\documentclass[journal]{IEEEtran}
\usepackage{amsmath,amsfonts}
\usepackage{algorithmic}
\usepackage{algorithm}
\usepackage{array}
\usepackage[caption=false,font=normalsize,labelfont=sf,textfont=sf]{subfig}
\usepackage{textcomp}
\usepackage{caption}
\usepackage{stfloats}
\usepackage{url}
\usepackage{verbatim}
\usepackage{graphicx}
\usepackage{cite}
\usepackage{graphicx}
\usepackage{accents}
\usepackage{ragged2e}
\usepackage{amssymb}
\usepackage{xcolor}
\usepackage{etoolbox}
\usepackage{xparse}
\usepackage{hyperref}
\usepackage{cleveref}
\usepackage{crossreftools}
\usepackage{comment}
\usepackage{bm}
\usepackage{bbm}
\usepackage{braket}
\usepackage{mathtools}
\allowdisplaybreaks

\newcommand{\indep}{\raisebox{0.05em}{\rotatebox[origin=c]{90}{$\models$}}}

\newcommand{\bU}{\mathbb{U}}
\newcommand{\cA}{\mathcal{A}}
\newcommand{\cB}{\mathcal{B}}
\newcommand{\cC}{\mathcal{C}}
\newcommand{\cD}{\mathcal{D}}
\newcommand{\cE}{\mathcal{E}}
\newcommand{\bE}{\mathbb{E}}
\newcommand{\bP}{\mathbb{P}}

\newcommand{\cH}{\mathcal{H}}
\newcommand{\cI}{\mathcal{I}}
\newcommand{\cM}{\mathcal{M}}
\newcommand{\cN}{\mathcal{N}}
\newcommand{\cO}{\mathcal{O}}
\newcommand{\cP}{\mathcal{P}}

\newcommand{\cR}{\mathcal{R}}
\newcommand{\cS}{\mathcal{S}}

\newcommand{\cU}{\mathcal{U}}
\newcommand{\cX}{\mathcal{X}}
\newcommand{\cY}{\mathcal{Y}}

\newtheorem{theorem}{Theorem}
\newtheorem{corollary}{Corollary}[theorem]
\newtheorem{lemma}{Lemma}[theorem]

\newtheorem{proposition}{Proposition}
\newtheorem{definition}{Definition}

\newtheorem{fact}{Fact}
\newtheorem{remark}{Remark}[theorem]
\renewcommand{\epsilon}{\varepsilon}
\newcommand{\eps}{\epsilon}

\newcommand{\norm}[1]{\left|\!\left| #1 \right|\!\right|}
\DeclareMathOperator{\Tr}{Tr}

\newcommand{\ketbra}[1]{\ket{#1}\!\!\bra{#1}}

\newcommand{\one}{\leavevmode\hbox{\small1\kern-3.8pt\normalsize1}}
\newcommand{\widesim}[2][1.5]{
  \mathrel{\overset{#2}{\scalebox{#1}[1]{$\sim$}}}
  }
  \newcommand\blfootnote[1]{%
  \begin{NoHyper}%
  \renewcommand\thefootnote{}\footnote{#1}%
  \addtocounter{footnote}{-1}%
  \end{NoHyper}%
}

\begin{document}

\title{One-shot Multiple Access Channel Simulation}

\author{
  \IEEEauthorblockN{Aditya Nema$^{1,2}$, 
  Sreejith Sreekumar$^{1}$,  
  Mario Berta$^{1}$}\\
  \IEEEauthorblockA{$^{1}$ Institute for Quantum Information,
                    RWTH Aachen University, Germany\\
                    $^{2}$ Department of Electrical Engineering, Indian Institute of Technology, Gandhinagar, India\\
                        Email: 
                        aditya.nema30@gmail.com
                   }
                    }

\maketitle
\begin{abstract}
We consider the problem of shared randomness-assisted multiple access channel (MAC) simulation for product inputs and characterize the one-shot communication cost region via almost-matching inner and outer bounds in terms of the smooth max-information of the channel, featuring auxiliary random variables of bounded size. The achievability relies on a rejection-sampling algorithm to simulate an auxiliary channel between each sender and the decoder, and producing the final output based on the output of these intermediate channels. The converse follows via information-spectrum based arguments. To bound the cardinality of the auxiliary random variables, we employ the perturbation method from [Anantharam {\it et al.}, IEEE Trans. Inf. Theory (2019)] in the one-shot setting. For the asymptotic setting and vanishing errors, our result expands to a tight single-letter rate characterization and consequently extends a special case of the simulation results of [Kurri {\it et al.}, IEEE Trans. Inf. Theory (2022)] for fixed, independent and identically distributed (iid) product inputs to universal simulation for any product inputs. \\
We broaden our discussion into the quantum realm by studying feedback simulation of quantum-to-classical (QC) MACs with product measurements [Atif {\it et al.}, IEEE Trans. Inf. Theory (2022)]. For fixed product inputs and with shared randomness assistance, we give a quasi tight one-shot communication cost region with corresponding single-letter asymptotic iid expansion.
\end{abstract}

\section{Motivation}

\blfootnote{This work was presented in part at the IEEE  International Symposium on Information Theory (ISIT), 2024\cite{Nema_Simulation}.}The channel simulation problem deals with the task of quantifying the minimum amount of communication required to establish correlation remotely, as dictated by the input-output joint distribution of the channel to be simulated. The most basic point-to-point channel simulation setup  consists of an encoder-decoder pair that with access to shared randomness and communication over a noiseless rate-limited link achieves the channel simulation task. More specifically, the encoder observes a random variable, say $X$  with distribution $q_X$, and based on the shared randomness,  sends a message to the decoder. Based on this message and shared randomness, the decoder outputs a random variable $Y$. The aim of the protocol is to ensure that the trace distance
between the joint distribution  $(X, Y)$ and the joint distribution induced by passing the source $X$ through a discrete memory-less channel $q(Y|X)$ is as small as possible. The channel simulation task is closely related to the task of creating a desired joint distribution between two distributed parties, also known as strong coordination \cite{Cuff_2010}.

Here, we consider the problem of simulating a two-sender classical and quantum to classical multiple-access channel (MAC). We assume that the respective encoders and decoder have access to unlimited shared randomness. This framework was first
investigated by Bennett {\it et al.} \cite{bennett_rev_Shannon}  to establish a so-called ‘reverse Shannon theorem’ to simulate a noisy channel from a noiseless channel in the asymptotic independent and identically distributed (iid) regime. They showed that the least communication cost for this purpose is equal to the mutual information, $I(X;Y)$,  between the input and output of the channel. The minimum one-shot rate for simulating a point-to-point classical channel was ascertained in \cite{Berta_simulation}. Extensions to broadcast channels were obtained in \cite{Berta_simulation}, and recently extended to the quantum setting \cite{Berta_broadcast}.

In both the point-to-point and broadcast channel simulation tasks, one may gain intuition from the scheme achieving the minimal communication rate as follows. Consider the case of point-to-point channel simulation:  since both the encoder, say Alice, and the decoder, Bob, know the channel to be simulated,  Alice can determine the channel output at her end and then compress it `optimally' and send it to Bob using the rate limited link. Bob then just outputs the target sequence after decompressing what he received from Alice. Similar intuition also works for the broadcast channel simulation problem. However, this approach breaks down for the MAC since there are two senders involved. More specifically, although each sender knows the MAC to be simulated, they cannot "locally" simulate the channel since the input of the other sender is unknown. Hence, novel schemes are required to circumvent this technical hurdle, which we address in this work.  While there is a comprehensive literature on simulating a point-to-point channel (both classical and quantum, in one-shot and asymptotic iid setting) and broadcast channel, only some more restricted results are known for MACs. In this regard, bounds on the asymptotic rate region for MAC simulation with fixed iid inputs were previously given in \cite{Kurri_MAC_simulation}. The inner bounds were derived by using the so-called OSRB technique of Yassaee {\it et al.}~\cite{Gohari_OSRB}. A matching outer bound for the case of fixed iid product inputs and shared randomness assistance was proven by identifying appropriate auxiliary random variables from the simulation code. 

Our main results are as follows:
\begin{itemize}
\item We obtain the one-shot cost region for simulating a MAC with two independent classical inputs $(X_1,X_2)$ and a single classical output $Y$, where the MAC is represented by the conditional probability distribution $q_{Y|X_1,X_2}$. We characterize the cost region, first for fixed product inputs in Theorem~\ref{thm:classical_MAC_main}, and then for universal simulation with arbitrary product inputs in Theorem~\ref{thm:one-shot_universal}. In order to simulate $q_{Y|X_1,X_2}$, for $j \in \{1,2\}$,  encoder $\cE_j$ of the sender $j$ sends a message $M_j \in [1:2^{R_j}]$ to the decoder $\cD$ over their respective noiseless links based on their individual observations and shared randomness with the decoder. We assume unlimited shared randomness $S_1~ (|\cS_1|=\infty)$ between $\cE_1$ and $\cD$ and $S_2~ (|\cS_2|=\infty)$ between $\cE_2$ and $\cD$. Since there is neither a one-shot nor a universal analogue of the Yassaee {\it et al.} \cite{Gohari_OSRB} OSRB techniques with the desired one-shot entropic quantity\,---\,which happens to be {\em smoothed max mutual information} in our case\,---\,this makes ours the first work towards simulating a MAC in the one-shot and universal regime.

\item We specialize our result to the asymptotic iid setting and show that it recovers \cite[Theorem~1]{Kurri_MAC_simulation} for fixed product iid inputs as Corollary~\ref{cor:classical_IID} of Theorem~\ref{thm:classical_MAC_main}. Further we also obtain a new single-letter formula for the universal case of arbitrary (not necessarily iid) product inputs in Corollary~\ref{cor:iid_universal}.

\item We tightly characterize the one-shot cost and asymptotic rate region for simulating classical scrambling quantum-inputs and classical output MACs with feedback in Theorem~\ref{thm:QC_MAC_main} and Corollary~\ref{cor:QC_iid}, respectively. This is referred to as classical scrambling QC-MAC with feedback, where feedback denotes the property that the classical inputs to the scrambler should also be available at the sender(s) after the simulation protocol has been executed.
\end{itemize}

{\bf Technical contributions:} A general recipe to obtain an inner bound on the rate region is applying the simple yet widely applicable technique of {\em rejection sampling}. One of the main technical hurdles, besides the unavailability of both the inputs at the encoders, preventing the import of earlier results is that this task cannot be seen as naively carrying out two point-to-point channel simulation. The main reason is that the output must be correlated with both the inputs. This is resolved by defining appropriate \textit{auxiliary} random variables, which are quantized versions of the respective inputs such that they approximately simulate the channel. The distribution of these auxiliary random variables can equivalently be viewed as point-to-point channel and hence we use these to decompose MAC into two point-to-point channels. We also give the bounds on the cardinality of these auxiliary random variables in the one-shot setting for smoothed max-mutual information, which is rarely studied like in \cite{auxiliary_Eric}, the only known work to the best of our knowledge. But their technique of the so-called generalized support lemma does not suffice for the task of MAC simulation due to an extra requirement of preserving the property that the output should be generated in correlation with auxiliary random variables. Hence we apply for the first time, the perturbation technique developed by Anantharam {\it et al.} \cite{Perturbation_first} to obtain the cardinality bounds on auxiliary random variables for the smoothed max-mutual information.

The paper is organized as follows. In Section \ref{sec:notation}, we introduce the preliminary notations and definitions of the relevant entropic quantities. In Section \ref{sec:Classical}, we present the one-shot cost region for fixed product input MAC simulation by proving its achievability and converse. In Section \ref{sec:oneshot-iid}, we give the asymptotic cost region by taking the asymptotic limit of our one-shot cost region with iid product inputs. In Section \ref{Sec:universal}, we state the one-shot universal MAC simulation cost region with product inputs. In Section \ref{sec:qc}, we establish the one-shot and asymptotic cost region for simulating QC MAC with feedback.  In Section \ref{Sec:conclusion}, we conclude the paper with a  discussion on some related open problems. Some of the detailed proofs are deferred to the Appendices.


\section{Notation}\label{sec:notation}
\subsection{Notation for classical setting}
The random variables are denoted by capital letters and their alphabets by scripted letters, for example, $X$ is a random variable with alphabet $\cX$, distributed according to $p_X$. $p_X$ is also the probability vector with the set of non-zero entries denoting its support represented by $supp(p_X)$. For brevity of notation, we use $\vec{X}$ to denote a finite length sequence of random variables $\{X_i\}_{i < \cI}$, where $\cI$ is any index set. The notation $[1:n]$ is used as a shorthand to denote the discrete set $\{1,2,\dots,n\}$. Expectation of a random variable is denoted by $\bE$.  We use the abbreviation p.m.f. to mean the probability mass function of the underlying discrete valued random variable. For any discrete random variable, one can define the notion of a sub-distribution which is nothing but a vector with all non-negative entries that sum up to less than $1$.  The set of all probability vectors is denoted by $\cP$ and sub-distribution vectors by $\cP_<$. For $p,q \in \cP$, the notation $p \ll q$ means that $supp(p) \subseteq supp(q)$. $h_2(\epsilon)$ is the binary entropy of distribution $\{\epsilon,1-\epsilon \}$, for $\epsilon \in (0,1)$. It is defined as $h_2(\epsilon):=-\epsilon \log_2 \epsilon -(1-\epsilon) \log_2 (1-\epsilon)$. We use the notation $\norm{\cdot}_1$ to denote the $\ell_1$ norm of a vector, which is the sum of absolute value of its components. We use $\norm{x-y}_{tvd}:=\frac{\norm{x-y}_1}{2}$ to denote the total variation distance between two vectors or operators. We denote an $n$-length sequence of random variables $\{X_1,X_2\ldots,X_n\}$ by the short hand $X_1^n$ taking values in alphabet $\cX^{\times n}$ and a random variable $U$ that takes values over an alphabet that depends on a sequence of length $n < \infty$ is denoted by $U^{(n)}$, that is, we do not necessarily enumerate $U^{(n)}$ as $U_1^n$. We define the joint distribution over a set of random variables by small case letters with the subscript denoting the random variables and the distribution restricted to a subset of random variables denotes their marginal. For example $p_{X_1,X_2,\dots,X_n}$ denotes a joint distribution on the random variables $X_1,X_2,\ldots,X_n$ and $p_{X_1,X_2}$ denotes the marginal on $X_1,X_2$ (by summing over the random variables $X_3,X_4,\ldots,X_n$). Similarly, if there is more than one random, the shorthand notation $\{X_i^n\}_{i=1}^d$ denotes the collection of random variables $\{X_{i1}, X_{i2}, \dots,X_{in}\}_{i=1}^d$. We will only be working with two random vectors in this work and thus the joint distribution over the random vectors $X_1^n,X_2^n \sim p_{X_1^n,X_2^n}$ denotes the  joint distribution $p_{X_{11},\ldots,X_{1n},X_{21},\ldots,X_{2n}}$. For brevity of notation we also define $p_{\vec{X}}:=p_{X_1,X_2,\dots,X_n}$. The notation $X \sim q$ indicates that the random variable $X$ is distributed according to $p$ or the p.m.f. of $X$ is $p$ and $X \overset{\epsilon}{\sim} q$ means that the p.m.f. $p$ of $X$ is $\epsilon$-close to the p.m.f. $q$ in the total variation distance. We also use the notation $p \overset{\epsilon}{\approx}q$ to denote that $\norm{p-q}_{tvd} \leq \epsilon$. We use the notation $p \ll q$, to mean that the p.m.f. $p$ is absolutely continuous with respect to the p.m.f. $q$, with emphasis on the property that the $supp(p) \subseteq supp(q)$.
The notation $\one_{A}$ denotes the indicator random variable which takes a value $1$ if event $A$ occurs and is zero otherwise. 
Scripted letters denotes the encoders, decoders and channels. $cl\{\cS\}$ denotes the closure of the set $\cS$. For conditional distributions $p_{Y|X} \text{ and }q_{Y|X}$, the total variation distance is defined as  $\norm{p_{Y|X}-q_{Y|X}}_{tvd}:=\max\limits_x \norm{p_{Y|X=x}-q_{Y|X=x}}_{tvd}$. We use $tvd$ as our distance measure unless stated otherwise. 

\subsection{Notation for quantum setting}
Analogous to the above classical notations, we denote any finite dimensional Hilbert space for the quantum setting by $\cH$. The set of all density operators in $\cH$ is denoted by $\cD$ or $\cD(\cH)$ and sub-states, that is, positive semi-definite operators with trace less than $1$ by $\cD_<$. When the argument of $\norm{\cdot}_1$ is an operator, it denotes the Schatten $1$-norm of the underlying operator, that is $\norm{A}_1:=\Tr[\sqrt{A^\dagger A}]$ for any operator $A \in \cH$, with $A^\dagger$ denoting the adjoint of $A$. The marginal or the reduced state of a quantum system is obtained taking the partial trace, e.g. the reduced state $\rho^B$ on the system $B$, of the joint bipartite state $\rho^{AB}$ is defined as $\rho^B:=\Tr_A(\rho^{AB})$. We also denote a sequence of quantum states admitting the tensor product form $\varphi_{x_{i,1}} \otimes \varphi_{x_{i,2}} \otimes \ldots$ with the short hand notation $\varphi_{x_i}^{\otimes n}$.\\
All the alphabets and the dimensions of classical and/or quantum systems are finite, unless stated otherwise. The notation $A \cong B$ is used to mean that the systems $A$ and $B$ are isomorphic to each other.

Abbreviations: \textbf{MAC} - \textbf{M}ultiple \textbf{A}ccess \textbf{C}hannel, \textbf{iid}- \textbf{i}ndependent and \textbf{i}dentically \textbf{d}istributed, \textbf{QC} - \textbf{Q}uantum to \textbf{C}lassical and \textbf{CS} - \textbf{C}lassical \textbf{S}crambling. 

We now give the definitions of the entropic quantities used in this work.

\begin{definition}
    The $\max$-divergence between any two probability distributions $p$ and $q$ on the support $\cX$ is defined as
    \begin{eqnarray*}
        D_{\max}(p||q) 
        & :=  &
        \log \max_{x \in \cX} \left[{p(x)}/{q(x)}\right].
    \end{eqnarray*}
\end{definition}

One then defines the notion of $\max$-mutual information of a joint distribution $p_{X,Y}$ from $D_{\max}$ as follows.

\begin{definition} \label{def:Imax}
    For a given bipartite distribution $p_{X,Y}$ the $\max$-mutual information is defined as \cite{Berta_rev_Shannon}
    \begin{align*}
        I_{\max}(X;Y)_p &:=\mathop{\inf}\limits_{q_Y \in \cP(\cY)} D_{\max}(p_{X,Y}||p_X \times q_Y)\\
        & =\mathop{\inf}\limits_{q_Y \in \cP(\cY)} \max_x D_{\max}(p_{Y|X=x}||q_Y)\;.
    \end{align*}
and the $\epsilon$-smoothed max-mutual information of $p_{X,Y}:=p_X p_{Y|X}$ for $\epsilon\geq0$ is defined as \cite{AEP}
    \begin{align} \label{eq:def_stateImax}
    I^\epsilon_{\max}(X;Y)_p
             &
        := \mathop{\inf}\limits_{ p'_{X,Y} \in \cB^\epsilon(p_{X,Y})}  I_{\max}(X;Y)_{p'} \notag\\
        &= \mathop{\inf}\limits_{ {p'_{X,Y} \in \cB^\epsilon(p_{X,Y})}}   \mathop{\inf}\limits_{q_Y \in \cP(\cY)} \max\limits_{x,y} \log \frac{p'_{Y|X}(y|x)}{q_Y(y)}
    \end{align}
            where  $\cB^\epsilon(p_{X,Y})
        := \{ p'_{X,Y} \in \cP: p'_X=p_X \mbox{ and } \mathop{\bE}\limits_{p_X} \lVert p_{Y|X} - p'_{Y|X} \rVert_{tvd} \leq \epsilon \}$.
\end{definition}

We now define the smoothed max mutual information of the channel using the notion of $\max$-mutual information defined above.

\begin{definition} \label{def:Imax_quantum}
    For a given bipartite distribution $p_{X,Y}$ the $\max$-mutual information is defined as \cite{Berta_rev_Shannon}
    \begin{align*}
        I_{\max}(A;B)_\tau &:=\mathop{\inf}\limits_{\sigma^B \in \cD(\cH^B)} D_{\max}(\tau^{AB}|| \tau^A \otimes \sigma^B)\\
        &= \log \norm{(\tau^A \otimes \sigma^B)^{-1/2} \tau^{AB} (\tau^A \otimes \sigma^B)^{-1/2}}_\infty,
    \end{align*}
    where $\norm{M}_\infty$ of an operator $M$ is the largest singular value or the largest magnitude of eigenvalue (if $M$  is Hermitian) of the operator $M$ and is also known as the operator norm or Schatten $\infty$-norm.
\end{definition}

\begin{definition}[Channel smoothed $\max$-mutual information] \label{def:smooth_Imax}
    Let $p_{Y|X}$ denote a channel with input $X$ and output $Y$. Let $X \sim p_X$ be a given input. Then for any given $\epsilon >0$,
the $\epsilon$-smoothed $\max$-mutual information of the channel $p_{Y|X}$ is defined as \cite{Berta_rej_sampling}:
    \begin{align} \label{eq:def_channelImax}
    I^\epsilon_{\max}(p_{Y|X})
             &
        := \hspace{-15 pt} \mathop{\inf}\limits_{ p'_{Y|X} \in \cB^\epsilon(p_{Y|X})}  \mathop{\inf}\limits_{q_Y \in \cP(\cY)} \max\limits_x D_{\max}(p'_{Y|X=x}||q_Y) \notag\\
        &= \mathop{\inf}\limits_{ p'_{Y|X} \in \cB^\epsilon(p_{Y|X})}  \mathop{\inf}\limits_{q_Y \in \cP(\cY)} \max\limits_{x,y} \log \frac{p'_{Y|X}(y|x)}{q_Y(y)},
    \end{align}
            where 
      \begin{align}
    \cB^\epsilon(p_{Y|X})
        := \{ p'_{Y|X} \in \cP:  \max\limits_x \lVert p_{Y|X=x} - p'_{Y|X=x} \rVert_{tvd} \leq \epsilon \}. \notag
      \end{align}      
\end{definition}

The analogous $\max$-mutual information and the smoothed $\max$-mutual information for bipartite quantum states is defined as follows \cite{AEP}:

\begin{definition}[State smoothed $\max$-mutual information]\label{def:smooth_Imax_quantum}
    The smoothed quantum $\max$-mutual information is defined as
    \begin{align} \label{eq:quantum_Imax}
         I^\epsilon_{\max}(A;B)_\rho 
        := \hspace{-15 pt} \inf_{\rho'^{AB} \in \cB^{\epsilon}(\rho^{AB})} \inf _{\sigma^B \in \cD(\cH^B)}D_{\max}(\rho'^{AB}|| \rho^A \otimes \sigma^B),
        \end{align}
        where 
        \[ D_{\max}(\rho||\sigma) := \inf\{ \lambda: \rho \leq 2^\lambda \sigma \} = \log \lVert \sigma^{-1/2} \rho \sigma^{-1/2}\rVert_\infty, 
        \]
      \begin{align*}
      \mbox{and }\cB^\epsilon(\rho^{AB})
        &:= \Big\{ \rho^{'AB} \in \cD(\cH) : \Tr_B(\rho')=\Tr_B(\rho),  \\
        &\qquad \quad  \lVert\rho - \rho' \rVert_{tvd} \leq \epsilon \Big\}.
        \end{align*}
\end{definition}

In the following, we use the term rate region or cost region interchangeably to mean the amount of classical communication used (or charged for) in the simulation protocol, and formally is the set of rate tuples $(R_1,R_2)$ that ensures MAC simulation.


\section{One-shot cost region for fixed product input MAC simulation} \label{sec:Classical}

\subsection{Task}

\begin{figure} 
    \centering
  \includegraphics[width=\linewidth]{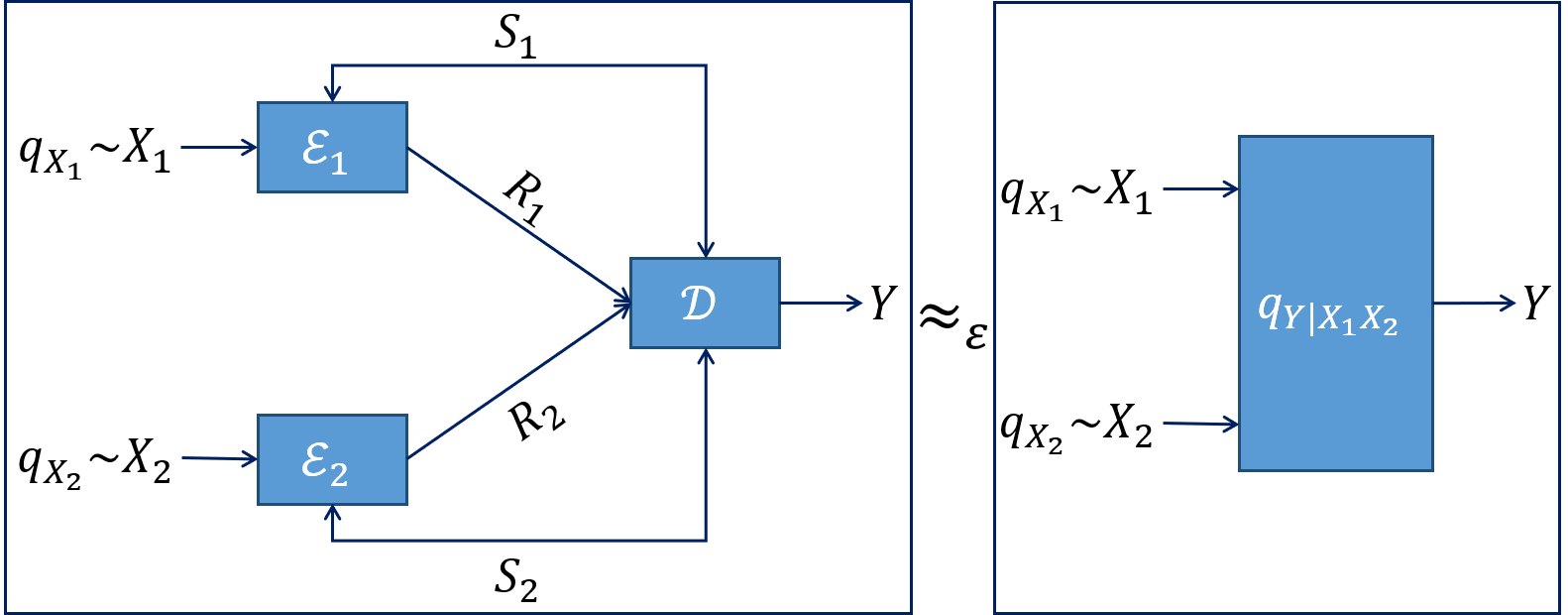}
  \caption{MAC Simulation: Encoders $\cE_j: (X_j,S_j) \xrightarrow[sampling]{rejection} M_j$, Decoder $\cD:(M_1,M_2,S_1,S_2) \mapsto Y$; $\approx_\varepsilon$ denotes closeness in tvd.}
  \label{fig:classical_MAC_simulation}
\end{figure}

We start by giving the formal definition of a $2$-user MAC simulation code, described in Figure~\ref{fig:classical_MAC_simulation}.

\begin{definition}[Classical MAC simulation with fixed input]\label{def:task_fixed_ip}
    An $(R_1,R_2, \epsilon)$ simulation protocol for a $2$-independent user MAC $q_{Y|X_1X_2}$ with inputs $q_{X_1} \times q_{X_2}$ and access to unlimited shared randomness between $\textrm{Sender}$1$ \overset{S_1}{\leftrightarrow} \textrm{Receiver}$ and $\textrm{Sender}$2$ \overset{S_2}{\leftrightarrow} \textrm{Receiver}$, consists of:
    \begin{itemize}
        \item A pair of encoders of form $\cE_1 \times \cE_2$, such that: $\cE_j:\cX_j \times \cS_j \to \cM_j:=\big[1:2^{R_j}\big]$ for $j \in \{ 1,2\}$;
        \item Two independent noiseless rate-limited links of rate $R_j$, $j \in \{1,2\}$ and;
        \item A decoder $\cD: \cM_1 \times \cS_1 \times \cM_2 \times \cS_2 \to \cY$;
        \item
        The overall joint distribution induced by the encoder-decoder pair   is given by 
        \begin{align*}
        &p_{X_1,X_2,S_1,S_2,M_1,M_2,Y}\\
        &=\left[\cD \circ \left(\mathop{\times}\limits_{j=1}^2 \cE_j \right) \right] \left\{\mathop{\times}\limits_{j=1}^2 \left(q_{X_j} \times p_{S_j} \right)\right\},
        \end{align*}
        such that
    \end{itemize}
     \begin{align} \label{eq:sim_protocol}
 &\norm{ p_{X_1,X_2,Y}-q_{X_1,X_2,Y} }_{tvd}  \notag \\
& =\mathop{\bE}\limits_{q_{X_1} \times q_{X_2}} \norm{p_{Y|X_1,X_2}-q_{Y|X_1,X_2}}_{tvd} \leq \epsilon\;.
     \end{align}
      The rate region $\cR(\epsilon)$ for simulating MAC is defined as the closure of the set of all rate pairs $(R_1,R_2)$ as given above satisfying \eqref{eq:sim_protocol}.  
\end{definition}
In this section, we henceforth consider  $p_{X_1,X_2,U_1,U_2,Y}$ to be a p.m.f. of the form:
\begin{align}
    p_{X_1,X_2,U_1,U_2,Y}=q_{X_1}q_{X_2}p_{U_1|X_1}p_{U_2|X_2} p_{Y|U_1,U_2}. \label{eq:jointpmf}
\end{align}
We now define the following regions that will  turn out to be inner and outer bounds for characterizing the rate region $\cR(\epsilon)$ for the task of one-shot MAC simulation given in Definition~\ref{def:task_fixed_ip}.
In these definitions,  the index  $j\in \{1,2\}$ unless specified otherwise.
\begin{definition}\label{def:inner_outer}
Let $\epsilon>0$ and $\epsilon_1, \epsilon_2,\delta \in (0,1)$ be such that  $\delta  < \min\{\epsilon_1,\epsilon_2\}$ and $\max\{\epsilon_1,\epsilon_2\} \leq \frac{\epsilon}{2}$. 
Let $\cR_{inner}(\epsilon_1,\epsilon_2,\delta)$ be the set of non-negative real numbers $(R_1,R_2)$ defined as:
   \begin{align} 
   & \cR_{inner}(\epsilon_1,\epsilon_2,\delta) \notag\\
   &=cl\Biggl\{{\bigcup} (R_1,R_2) \hspace{-2 pt}:  \hspace{-3 pt}
   R_j \geq   I_{\max}^{\epsilon_j-\delta} (X_j;U_j)_{q_{X_j}p_{U_j|X_j}} \hspace{-8 pt} +\log \log \frac{1}{\delta}\Biggr\},
   \label{eq:fixed_inner}
    \end{align} 
    where the union is over $\left(p_{U_1|X_1},p_{U_2|X_2}\right) \in \cA_{\epsilon}^{inner}$ and 
    \begin{align}
    \cA^{inner}&:= \big\{ \left(p_{U_1|X_1},p_{U_2|X_2}\right): \exists~ p_{Y|U_1, U_2}  \notag \\
    &  \qquad \quad \mbox{ satisfying } ~p_{X_1,X_2,Y}=q_{X_1,X_2,Y}\big\}. \label{eq:fixed_A_inner}
    \end{align}
    Similarly, let $\cR_{outer}(\epsilon_1,\epsilon_2,\epsilon)$ be the set of non-negative real numbers $(R_1,R_2)$ defined as:
    \begin{align}
    & \cR_{outer}(\epsilon_1,\epsilon_2, \epsilon) \notag \\
    &:=cl\Bigg\{\bigcup~ (R_1,R_2): R_j \geq   I_{\max}^{\epsilon_j} (X_j;U_j)_{q_{X_j}p_{U_j|X_j}} \Bigg\},
    \label{eq:fixed_outer}
    \end{align}
    where the union is over $\left(p_{U_1|X_1},p_{U_2|X_2}\right) \in \cA_{\epsilon}^{outer}$ and
\begin{align}\label{eq:fixed_A_outer}
    &\cA^{outer}_{\epsilon} := \left\{ \left(p_{U_1|X_1},p_{U_2|X_2}\right): \exists ~p_{Y|U_1, U_2} \mbox{ satisfying }  \right.\notag\\
    &~~ \left. \norm{  p_{X_1,X_2,Y}-q_{X_1,X_2,Y}}_{tvd} \leq 2\epsilon, |\cU_1|, |\cU_2|  \leq |\cX_1||\cX_2||\cY| \right\}. 
    \end{align}
\end{definition}

The following characterization is our main result in this section.

\begin{theorem} \label{thm:classical_MAC_main}
 Let $q_{Y|X_1,X_2}$ be a given $2$-sender, $1$-receiver MAC with input $q_{X_1} \times q_{X_2}$.
 For any $\epsilon \in (0,1)$, and  $\epsilon_1, \epsilon_2,\delta \in (0,1)$  such that  $\delta  < \min\{\epsilon_1,\epsilon_2\}$ and $\max\{\epsilon_1,\epsilon_2\} \leq \frac{\epsilon}{2} $,  the one-shot rate region for MAC simulation satisfies:
 \begin{align}
     \cR_{inner}(\epsilon_1,\epsilon_2,\delta) \subseteq \cR(\epsilon) \subseteq \cR_{outer}(\epsilon_1,\epsilon_2, \epsilon),
 \end{align}
 where  the  inner $(\cR_{inner}(\epsilon_1,\epsilon_2,\delta))$ and the outer $(\cR_{outer}(\epsilon_1,\epsilon_2))$ bounds are as defined in Definition~\ref{def:inner_outer}.
\end{theorem}

The proof comprises of two parts, direct part or achievability as shown in Lemma~\ref{lem:classical_achievability} and converse as proven in  Lemma~\ref{lem:classical_converse}.

    \begin{remark}
      Note that  $\cR_{inner}(\epsilon_1,\epsilon_2,\delta)$ and $\cR_{outer}(\epsilon_1,\epsilon_2)$ characterize the one-shot rate region $\cR(\epsilon)$ (for $\epsilon_1+\epsilon_2 \leq \epsilon$) up to a fudge factor that depends on $\delta$. Also, the characterization of $\cR(\epsilon)$ in terms of $ \cR_{inner}(\epsilon_1,\epsilon_2,\delta)$ and $
 \cR_{outer}(\epsilon_1,\epsilon_2)$ in Theorem \ref{thm:classical_MAC_main} does not involve any sum-rate constraints. This is due to the availability of infinite shared randomness between both the sender-receiver pairs. The output random variable $Y$ plays a role in this characterization through the Markov chain $(X_1,X_2) \rightarrow (U_1,U_2) \rightarrow Y$ under the distribution $p_{X_1,X_2,U_1,U_2,Y}$. 
    \end{remark}


\subsection{Achievability}

\begin{lemma}\label{lem:classical_achievability}
     For any given $\epsilon>0$, let $\epsilon_1, \epsilon_2 >0$ be such that  $\max\{\epsilon_1,\epsilon_2\} \leq \frac{\epsilon}{2}$ and  $\delta \in (0, \min\{\epsilon_1,\epsilon_2\})$. Then,  $ \cR_{inner}(\epsilon_1,\epsilon_2,\delta) \subseteq \cR(\epsilon)$.
      \end{lemma}

\begin{IEEEproof}
Fix  $(\epsilon_1,\epsilon_2,\delta)$ satisfying the conditions in the lemma and let $q_{X_1} \times q_{X_2}$ be the fixed input distribution. We need to show that for any $(R_1,R_2) \in  \cR_{inner}(\epsilon_1,\epsilon_2,\delta)$ (defined in \eqref{eq:fixed_inner}), there exists an $(R_1,R_2,\epsilon)$ one-shot MAC simulation protocol as mentioned in Definition \ref{def:task_fixed_ip}.
   
\underline{Idea:} We will use the point-to-point channel simulation algorithm of Fact~\ref{lem:rej_sampling}-(i) independently at the two senders.
    \begin{itemize}
         \item \textbf{Sender-$j$}: Let $s_{U_j}$ be a distribution with full support and choose $U_j \sim s_{U_j}$ as the shared randomness between the pair $(\cE_j,\cD)$. Using the rejection sampling algorithm stated in Fact \ref{fact:accept-reject}, sender $j$ sends the appropriately chosen index of the shared randomness using $R_j$ bits to perform point-to-point channel simulation for the auxiliary channel $p_{U_j|X_j}$.\\
         
         \item \textbf{Decoding:} 
        After receiving the transmitted index of shared randomness from both the encoders, the decoder first generates $\{U_j\}_{j=1}^2$ and  applies the stochastic map $p_{Y|U_1,U_2}$ to simulate $q_{Y|{X_1,X_2}}$.\\
        
        \item The output distribution of $U_j$ at $\cD$ is denoted by  $p^{algo}_{U_j|X_j}$, the joint distribution of $(X_j,U_j)$ by $p^{algo}_{X_j,U_j}:= q_{X_j} p^{algo}_{U_j|X_j}$ and satisfies (from Fact \ref{lem:rej_sampling}-(i)) :
        \begin{align}
       \norm{p^{algo}_{U_j,X_j}-p_{U_j,X_j}}_{tvd}=\underset{q_{X_j}}\bE\norm{p^{algo}_{U_j|X_j}-p_{U_j|X_j}}_{tvd} \leq \epsilon_j. \label{eq:sim1}       
        \end{align}
     
         The amount of classical communication required for this task is given by (see Fact \ref{lem:rej_sampling}-(i)): 
         \begin{align*}
             R_j \geq I_{\max}^{\epsilon_j-\delta} (X_j;U_j)_p + \log \log \frac{1}{\delta}.
         \end{align*}
         \end{itemize}
         Thus, our algorithm results in the overall distribution
         \begin{equation} \label{eq:algo_output}
         p^{algo}_{X_1,X_2,U_1,U_2,Y} = q_{X_1} \times q_{X_2} \times p^{algo}_{U_1|X_1} \times p^{algo}_{U_2|X_2} \times p_{Y|U_1,U_2}\;. 
         \end{equation}
       To complete the proof,  we need to show
       \begin{align}
           \underset{q_{X_1} \times q_{X_2}}{\bE}\lVert p^{algo}_{Y|X_1,X_2}-q_{Y|X_1,X_2}\rVert_{tvd}\leq \epsilon_1+\epsilon_2. \notag
       \end{align}
     This follows by the following chain of inequalities:
         \begin{align*}
   & \underset{q_{X_1}\times q_{X_2}}{\bE}\lVert p^{\text{algo}}_{Y|X_1,X_2}-q_{Y|X_1,X_2}\rVert_{tvd}\\
    &\overset{(a)}{\leq}
     \underset{q_{X_1}\times q_{X_2}}{\bE}\lVert p^{\text{algo}}_{Y|X_1,X_2}-p_{Y|X_1,X_2}\rVert_{tvd} \\
     & \quad +  \underset{q_{X_1}\times q_{X_2}}{\bE}\lVert p_{Y|X_1,X_2}-q_{Y|X_1,X_2}\rVert_{tvd}  \\
    & \overset{(b)}{=} \underset{q_{X_1}\times q_{X_2}}{\bE} \left\lVert \mathop{\Sigma}\limits_{u_1,u_2}p_{Y|U_1=u_1,U_2=u_2}\left(p^{\text{algo}}_{U_1|X_1}(u_1) \times \right.\right. \\
    &\qquad \qquad \quad \left.\left. p^{\text{algo}}_{U_2|X_2}(u_2)-p_{U_1|X_1}(u_1)p_{U_2|X_2}(u_2)\right) \right\rVert_{tvd}\\
    &= \underset{q_{X_1}\times q_{X_2}}{\bE}\mathop{\Sigma} \limits_{u_1,u_2}\mathop{\Sigma}\limits_y p_{Y|U_1,U_2}(y|u_1,u_2)\left\lvert \left(p^{\text{algo}}_{U_1|X_1}(u_1) \times \right.\right.\\
    &\qquad \qquad \quad \left.\left. p^{\text{algo}}_{U_2|X_2}(u_2) -p_{U_1|X_1}(u_1) p_{U_2|X_2}(u_2) \right)\right\rvert\\
    & \overset{(c)}{\leq} \underset{q_{X_1}\times q_{X_2}}{\bE} \lVert p^{\text{algo}}_{U_1|X_1} p^{algo}_{U_2|X_2}-p^{algo}_{U_1|X_1} p_{U_2|X_2}\rVert_{tvd} \\
    & \quad +
    \underset{q_{X_1}\times q_{X_2}}{\bE}\lVert p^{\text{algo}}_{U_1|X_1} p_{U_2|X_2}-p_{U_1|X_1} p_{U_2|X_2}\rVert_{tvd}\\
    &= \underset{q_{X_1}}{\bE}\lVert p^{\text{algo}}_{U_1|X_1}\rVert_{1} \underset{q_{X_2}}{\bE}\lVert p^{algo}_{U_2|X_2}-p_{U_2|X_2}\rVert_{tvd} \\
    & \quad +
     \underset{q_{X_2}}{\bE}\lVert p_{U_2|X_2}\rVert_1 \underset{q_{X_1}}{\bE}\lVert p^{\text{algo}}_{U_1|X_1}-p_{U_1|X_1}\rVert_{tvd}\\
    &\overset{(d)}{\leq} \epsilon_1+\epsilon_2 \leq \epsilon, 
         \end{align*}
         where $(a)$ and $(c)$ follow from  triangle inequality; $(b)$ follows from the definition of distribution induced by the code in \eqref{eq:algo_output}; 
        and  $(d)$ follows from  \eqref{eq:sim1}. Thus, we have shown that $\cR_{inner}(\epsilon_1,\epsilon_2) \subseteq \cR(\epsilon)$. 
\end{IEEEproof}
    Note that without the loss of generality, we can  assume $|\cU_j|$ to be of finite support since by continuity of $I_{\max}(X_j,U_j)$ \cite{D_max_continuity}  (see \eqref{eq:continuity}) and finiteness of $I_{\max}(X_j,U_j) \leq -\log p_{\min} $, where 
\begin{align}
   p_{\min} \coloneqq \mathop{\min}\limits_{x_j: p_{X_j}(x_j)>0}\{p_{X_j}(x_j)\}, \notag
\end{align}
  $|\cU_j|$ can be approximated arbitrarily well by $I_{\max}(X_j,U_j')$ for some $U_j'$ with $|\cU_j'|<\infty$. Hence, the above proof does not create any ambiguity in defining $I_{\max}^{\epsilon_j - \delta}(X_j;U_j)$, which is implicitly defined for $X_j, U_j$ with finite support.

\begin{remark}
We would like to remark that there is a one-shot version of OSRB \cite{NonIID_OSRB} and one can try and prove a one-shot achievability using this technique, as in the asymptotic iid case \cite{Kurri_MAC_simulation}. However, the one-shot rates will be in-terms of the empirical entropic quantities. Further, one-shot OSRB provides the analysis  only for the average-error criteria. Our rejection sampling analysis on the other hand gives the rate in terms of smoothed $\max$-mutual information, when also considering simulation with the $\max$-error criterion (in the classical setting). This is important for the universal MAC simulation. But, for fixed input asymptotic iid simulation under average error criterion, one-shot OSRB achievability will also give the same achievable region. 
The rejection sampling type technique can also be generalized in the fully quantum setting (which can be pursued as a potential future work), which is essentially the convex-split lemma for fully quantum setting. It is not immediately clear whether the one-shot OSRB with empirical entropies can be lifted as it is for the fully quantum setting.
Finally, we can prove a converse (which we do in Lemma~\ref{lem:classical_converse}) with smoothed $\max$-mutual information as the leading term and not with the empirical entropy.
\end{remark}


\subsection{Converse}

\begin{lemma}\label{lem:classical_converse}
For any given $\epsilon \in (0,1)$, let $\epsilon_1, \epsilon_2 > 0$ be such that $\max\{\epsilon_1, \epsilon_2\} \leq \frac{\epsilon}{2}$. Then, $\cR(\epsilon) \subseteq \cR_{outer}(\epsilon_1,\epsilon_2,\epsilon)$.
\end{lemma}

\begin{IEEEproof} 
Let $(\epsilon_1,\epsilon_2)$ and $\epsilon$ satisfy the conditions of the lemma. We need to show that any $(R_1,R_2,\epsilon)$ MAC simulation protocol according to Definition \ref{def:task_fixed_ip} has $(R_1,R_2) \in \cR_{outer}(\epsilon_1,\epsilon_2,\epsilon)$ (defined in \eqref{eq:fixed_outer}).

Consider a MAC simulation protocol with the overall distribution as $\mathop{\bigotimes}\limits_{j=1}^2 \left( q_{X_j}q_{S_j}p'_{M_j|S_j,X_j} \right)p'_{Y|\vec{M},\vec{S}}$. The encoders are specified by ${p'}_{M_1|X_1,S_1}$ and ${p'}_{M_2|X_2,S_2}$, and the decoder is specified by ${p'}_{Y|M_1,M_2,S_1,S_2}$. Since, the code is a faithful simulation code, we have from Definition~\ref{def:task_fixed_ip}:
     \begin{align} \label{eq:tilde_constraint_classical}
        \norm{p'_{X_1,X_2,Y}-q_{X_1,X_2,Y}}_{tvd} &= \hspace{-10 pt}\underset{q_{X_1}\times q_{X_2}}{\bE} \hspace{-5 pt}\norm{{p'}_{Y|X_1,X_2} - q_{Y|X_1,X_2}}_{tvd} \notag \\
        & \leq \epsilon. 
     \end{align} 
    One of the difficulties in using the standard converse based on information non-locking property of $I_{\max}$ (see e.g. \cite[Theorem~5]{Berta_rej_sampling}) is the identification of the auxiliary random variables $(U_1,U_2)$ that are essential for characterizing $\cR(\epsilon)$. Hence, we give a proof inspired from the information spectrum approach (e.g. \cite{Tomamichel_book,Inf_spectrum}). The main element of the proof is eliminating a small probability subset of the message and shared randomness for every input symbol. Then, we identify the auxiliary $U_j$ (for $j \in \{1,2\}$) as the tuple of message and the shared randomness restricted to the complement of the above eliminated set. The same intuition applies to the proof of outer bounds of Lemmas~\ref{lem:classical_converse_universal} and \ref{lem:QCMAC_converse}. The formal description now follows.\\ 
  We now define the following set for every $2$-tuple $\vec{x}=(x_1,x_2)$
\begin{align} \label{eq:goodset_classical}
    \bar{\cC}_{\vec{x}}:=\left\{(\vec{m},\vec{s}) : p'_{M_j|S_j,X_j}(m_j|s_j,x_j) \geq \frac{\epsilon_j}{|\cM_j|}, j=1,2 \right\}.
\end{align}
We henceforth denote the projection of $\cC_{\vec{x}}$ onto $(M_j,S_j,X_j)$ (or the $j^{th}$ user) as $\cC_{x_j}$ and we make the similar identification for their respective complements.\\ 
    Note that by union bound, we have
    \begin{align}
        \bP_{p'}(\cC_{\vec{x}}) &\leq  \sum_{j=1}^2\bP\left(\left\{p'_{M_j|S_j,X_j}(m_j|s_j,x_j) \leq \frac{\epsilon_j}{|\cM_j|}\right\}\right) \notag\\
        &\leq  \epsilon_1+\epsilon_2, \label{eq:unionbnd}
    \end{align}
    where we have used:
    \begin{align} \label{eq:P(c_j)_classical}
     &\bP_{p'}(\cC_{x_j}) \notag\\
     &:=\bP_{p'}\left(\left\{(m_j,s_j):p'_{M_j|S_j,X_j}(m_j|s_j,x_j) \leq \frac{\epsilon_j}{|\cM_j|}\right\}\right) \notag\\
     &=\sum_{(m_j,s_j):p'_{M_j|S_j,X_j}(m_j|s_j,x_j) \leq \frac{\epsilon_j}{|\cM_j|} } \hspace{-15 pt}p'_{S_j}(s_j)p'_{M_j|S_j,X_j}(m_j|s_j,x_j)  \notag \\
     &\leq \sum_{(m_j,s_j)}\frac{\epsilon_j}{|\cM_j|}q_{S_j} \leq \epsilon_j.
    \end{align}
    Hence, $\bP_{p'}(\bar \cC_{\vec{x}}) \geq 1-\epsilon_1 - \epsilon_2$, for all $\vec{x}$.\\     
    Consider the distribution defined as follows:
    \begin{align} \label{eq:p'}
         & p_{M_j,S_j|X_j}(m_j,s_j|x_j) \notag\\
         &:=
         \begin{cases}
         \frac{p'_{S_j}(s_j)p'_{M_j|S_j,X_j}(m_j|s_j,x_j)}{\bP_{p'}(\bar \cC_{x_j})}, & \mbox{ if }(m_j,s_j) \in \bar \cC_{x_j} \\
         0 & else\;.
         \end{cases}
    \end{align}
    We have thus identified the auxiliary random variable $\{U_j\}_{j=1}^2$ for each $x_j$ as:    
    \begin{align*}
        U_j&:=(M_j,S_j) \one_{\bar \cC_{x_j}}\\
        &\sim p_{U_j|X_j}(u_j|x_j):=p_{M_j,S_j|X_j}(m_j,s_j|x_j)\\
        &=\frac{p'_{S_j}(s_j)p'_{M_j|S_j,X_j}(m_j|s_j,x_j) \one_{(m_j,s_j) \in \bar \cC_{x_j}}}{\bP_{p'}(\bar \cC_{x_j})}.
    \end{align*}
    Using this we identify the conditional distribution $p_{\vec{U},Y|\vec{X}}$ (for every $\vec{x}$) as:
    \begin{align} \label{eq:p-jointfix'}
         p_{\vec{U},Y|\vec{X}}(\vec{u},y|\vec{x}) &:= \mathop{\bigotimes}\limits_{j=1}^2 \left[\frac{p'_{S_j}(s_j)p'_{M_j|S_j,X_j}(m_j|s_j,x_j)}{\bP_{p'}(\bar \cC_{x_j})}\right] \notag \\
         & \quad \times p'_{Y|\vec{S},\vec{M}}(y|\vec{s},\vec{m})\one_{\{m_j,s_j \in \bar \cC_{x_j}\}}
    \end{align}
     Now, we identify the complete joint distribution $p$ defined as follows:
     \begin{align} \label{eq:p-joint}
         & p_{\vec{X},\vec{U},Y}(\vec{x},\vec{u},y) \notag\\
         &:=
         \begin{cases}
         \mathop{\bigotimes}\limits_{j=1}^2 \left[ \frac{q_{X_j}(x_j)p'_{U_j|X_j}(u_j|x_j)}{\bP_{p'}(\bar \cC_{x_j})} \right] p'_{Y|U_1,U_2}(y|u_1,u_2), & \vec{u} \in \bar{\cC}_{\vec{x}} \\
         0, & else
         \end{cases}
    \end{align}
    Note that \eqref{eq:P(c_j)_classical} also gives: 
    \begin{align}
        &\mathop{\bE}\limits_{q_{\vec{X}}}\norm{p_{Y|\vec{X}}-p'_{Y|\vec{X}}}_{tvd} \notag \\ &=\mathop{\bE}\limits_{q_{\vec{X}}}\norm{\mathop{\Sigma}\limits_{\vec{m},\vec{s}} \left( p_{\vec{M}=\vec{m},\vec{S}=\vec{s}|\vec{X}}- p'_{\vec{M}=\vec{m},\vec{S}=\vec{s}|\vec{X}}\right)p'_{Y|\vec{M}=\vec{m},\vec{S}=\vec{s}}}_{tvd} \notag\\
        &\leq \mathop{\sum}\limits_{\vec{x}}  \frac{q_{X_1}(x_1)q_{X_2}(x_2)}{2} \left[\mathop{\sum}\limits_{(\vec{m},\vec{s})} \bigg\{ p(m_1,s_1|x_1) \times \right.\notag\\
        & \qquad \left| p(m_2,s_2|x_2) - p'(m_2,s_2|x_2)\right| +  p'(m_2,s_2|x_2) \times \notag\\
        & \qquad \left|p(m_1,s_1|x_1)- p'(m_1,s_1|x_1)\right| \bigg\} \bigg] \notag \\
        & \leq \mathop{\sum}\limits_{\vec{x}} \frac{q_{X_1}(x_1)q_{X_2}(x_2)}{2} \left[ \mathop{\sum}\limits_{(m_1,s_1) }p(m_1,s_1|x_1) \times  \right. \notag\\
        & \qquad \bigg\{ \mathop{\sum}\limits_{(m_2,s_2) \in \bar \cC_{x_2}}\left| {p(m_2,s_2|x_2)-p'(m_2,s_2|x_2)}\right| + \notag\\
        & \qquad \left. \mathop{\sum}\limits_{(m_2,s_2) \in \cC_{x_2}}\left| {p(m_2,s_2|x_2)-p'(m_2,s_2|x_2)}\right| \right\}+ \notag\\
        &\qquad \mathop{\sum}\limits_{(m_2,s_2)} p'(m_2,s_2|x_2) \times \notag\\
        & \qquad \bigg\{ \mathop{\sum}\limits_{(m_1,s_1) \in \cC_{x_1}} \left| p(m_1,s_1|x_1)- p'(m_1,s_1|x_1)\right| + \notag\\ & \mathop{\sum}\limits_{(m_1,s_1) \in \bar \cC_{x_1}} \left|p(m_1,s_1|x_1)-  p'(m_1,s_1|x_1)\right| \bigg\} \bigg]\notag \\
        &= \mathop{\sum}\limits_{x_2} \frac{q_{X_2}(x_2)}{2} \left[ \mathop{\sum}\limits_{(m_2,s_2) \in \bar \cC_{x_2}} p'(m_2,s_2|x_2) \left| \frac{1}{\bP_{p'}(\bar \cC_{x_2})}-1 \right| \right. \notag\\
        &  \qquad \qquad \qquad \qquad \qquad \qquad  + \bP_{p'}(\cC_{x_2})\bigg] \notag\\
        & + \mathop{\sum}\limits_{x_1}  \frac{q_{X_1}(x_1)}{2} \left[\mathop{\sum}\limits_{(m_1,s_1) \in \bar \cC_{x_1}} p'(m_1,s_1|x_1) \left| \frac{1}{\bP_{p'}(\bar \cC_{x_1})}-1 \right| \right.\notag \\
        & \qquad \qquad \qquad \qquad \qquad \qquad  + \bP_{p'}(\cC_{x_1})\bigg] \notag\\
        &= \frac{2\mathop{\sum}\limits_{x_1} q_{X_1}(x_1)\bP_{p'}(\cC_{x_1})+2 \mathop{\sum}\limits_{x_2} q_{X_2}(x_2)\bP_{p'}(\cC_{x_2})}{2} \notag \\
        & \leq \epsilon_1+\epsilon_2\; .\label{eq:tvddistbnd_classical}
    \end{align}
Finally,  we define the following distribution on the random variable $U_j (=(M_j,S_j))$ that will be used to evaluate the quantity $I_{\max}^\epsilon(X_j;U_j)_p$ for $j \in \{1,2\}$:
    \begin{align}
        r_{U_j}(u_j):= q_{S_j}(s_j)\frac{1}{|\cM_j|} 
    \end{align}
   These identifications lead to the following implications on the rate of the protocol:
    \begin{align}\label{eq:rate_classical}
        2^{I_{\max}^{\epsilon_j}(X_j;U_j)_{p}} & \overset{(a)}{\leq} 2^{D_{\max}(p'_{X_j,U_j}||p'_{X_j} \times r_{U_j})} \notag\\
         &=\max_{x_j} \max_{u_j}\frac{p'_{X_j,U_j}(x_j,u_j)}{p'_{X_j}(x_j)r_{U_j}(u_j)} \notag\\
        &\overset{(b)}{=} \max_{x_j} \max_{(m_j, s_j)} \frac{q_{S_j}(s_j)p'_{M_j|S_jX_j}(m_j|s_j,x_j)}{q_{S_j}(s_j)/|\cM_j|} \notag\\
        &\overset{(c)}{\leq } |\cM_j|,
    \end{align}
    where $(a)$ follows from the definition of smoothed $I_{\max}$ in Definition~\ref{def:smooth_Imax} and  observing that distribution $p_{U_j|X_j=x_j}=p_{M_j,S_j|X_j=x_j} \in \cB^{\epsilon_j}(p'_{M_j,S_j|X_j=x_j})$ because:
    \begin{align*}
        &\mathop{\bE}\limits_{q_{X_j}} \norm{p_{U_j|X_j}-p'_{U_j|X_j}}_{tvd} \\
        &= \sum_{x_j}\frac{q_{X_j}(x_j)}{2}  \sum_{m_j,s_j} \left| p'_{M_j,S_j|X_j}(m_j,s_j|x_j) \right.\\
        & \qquad \qquad \qquad \qquad \quad- \left. p_{M_j,S_j|X_j}(m_j,s_j|x_j)\right|\\
        &=\frac{1}{2}\sum_{x_j}q_{X_j}(x_j) \left[ \sum_{(m_j,s_j) \in \bar \cC_{x_j}} p'_{M_j,S_j|X_j}(m_j,s_j|x_j) \times \right.\\
        & \left. \left(\frac{1}{\bP_{p'}(\bar \cC_{x_j})}-1 \right)+ \sum_{(m_j,s_j) \in \cC_{x_j}} p'_{M_j,S_j|X_j}(m_j,s_j|x_j) \right] \\
        &=\bP_{p'}(\cC_{x_j}) \leq \epsilon_j \; (\mbox{from \eqref{eq:P(c_j)_classical}});
    \end{align*}
    $(b)$ follows from the identification of $U_j=(M_j,S_j)$ for all $p'$, the definition of $r_{U_j}(u_j)$ and the Bayes rule and \\
    $(c)$ follows since $p'_{M_j,S_j|X_j}(m_j,s_j|x_j) \leq 1$.\\
    We thus have from \eqref{eq:rate_classical}, the rate of the code is lower bounded by:
    \[
        R_j=\log|\cM_j| \geq I^{\epsilon_j}_{\max}(X_j;U_j)_{p} \; \text{ for }j\in\{1,2\}.
    \]
From \eqref{eq:tvddistbnd_classical} we have that $p_{Y|\vec{X}=\vec{x}} \in \cB^{\epsilon_1+\epsilon_2}(p'_{Y|\vec{X}=\vec{x}})$.
This along with the simulation constraint of \eqref{eq:tilde_constraint_classical} yields by the triangle inequality:
    \begin{align*}
         \norm{p_{X_1,X_2,Y}-q_{X_1,X_2,Y}}_{tvd} \leq (\epsilon_1+\epsilon_2)+\epsilon \leq 2\epsilon.
    \end{align*}
  Thus, the conditional distribution of the identified auxiliary random variables satisfy $(p_{U_1|X_1},p_{U_2|X_2}) \in \cA_{\epsilon}^{outer}$ (defined in \eqref{eq:fixed_A_outer}). Finally, we also need to combine Lemma~\ref{lem:cardinality-fixed_ip} to obtain this result. Hence, we have shown that for any $(R_1,R_2,\epsilon)$-simulation code, the rate of the code is bounded below by
    \begin{align}
        R_j &\geq I^{\epsilon_j}_{\max}(X_j;U_j)_{p}. \notag 
    \end{align}  
To complete the proof, we state the bound on the cardinalities of $\cU_1,\; \cU_2$ as Lemma~\ref{lem:cardinality-fixed_ip} below.
    \begin{lemma} \label{lem:cardinality-fixed_ip}
    The cardinalities of $\{\cU_1,\cU_2\}$ for the region $\cR_{outer}(\epsilon_1,\epsilon_2,\epsilon)$ can be upper bounded as:
    \begin{align} \label{eq:cardinality}
        |\cU_j| \leq |\cX_1||\cX_2||\cY|; \quad \text{ for } j\in \{1,2\}\;.
    \end{align}
\end{lemma}
The proof of Lemma \ref{lem:cardinality-fixed_ip} is shown  in Appendix~\ref{app:cardinality}.
    \end{IEEEproof}

\section{Asymptotic iid expansion}\label{sec:oneshot-iid}

We now evaluate the asymptotic limit of the iid expansion of our one-shot simulation result and show that the cost region  single-letterizes in this regime. This recovers a special case of  \cite[Theorem~1 and Theorem~4]{Kurri_MAC_simulation} with independent inputs and no side information at the decoder. We now state a general way to obtain asymptotic bounds from the extension of one-shot bounds. For the one-shot setting, our regime of interest is fixed $\epsilon_1 \text{ and }\epsilon_2$ and $\delta$ chosen independently as a fixed parameter subject to $ < \min\{\epsilon_1, \epsilon_2\}$. 
For asymptotic analysis, limit is first taken as $n \to \infty$ for fixed $\epsilon_1,\epsilon_2, \delta$
which leads to invoking Fact~\ref{fact:AEP_fixed_ip}. Subsequently, the limits  $\epsilon_j \to 0$ and $\delta \to 0$ are taken subject to $\delta < \min\{\epsilon_1, \epsilon_2\}$. Note that, this does not encounter the issue of  fudge terms blowing in the asymptotic iid setting and we get a matching converse.

For the sake of clarity we start by giving the formal definition of an $n$-letter MAC simulation code.
\begin{definition}[Classical MAC simulation with fixed input]\label{def:iid_task}
    An $(nR_1,nR_2, \epsilon)$ simulation protocol for simulating $q^{\otimes n}_{Y|X_1X_2}$ with inputs $q^{\otimes n}_{X_1} \times q^{\otimes n}_{X_2}$ and access to unlimited shared randomness between $\textrm{Sender}$1$ \overset{S_1}{\leftrightarrow} \textrm{Receiver}$ and $\textrm{Sender}$2$ \overset{S_2}{\leftrightarrow} \textrm{Receiver}$, consists of:
    \begin{itemize}
        \item A pair of encoders of form $\cE_1^{(n)} \times \cE_2^{(n)}$, such that: $\cE^{(n)}_j:\cX_j^{(n)} \times \cS_j^{(n)} \to \cM_j:=\big[1:2^{nR_j}\big]$ for $j \in \{ 1,2\}$;
        \item Two independent noiseless rate-limited links of rate $R_j$, $j \in \{1,2\}$ and;
        \item A decoder $\cD^{(n)}: \cM_1 \times \cS_1^{(n)} \times \cM_2 \times \cS_2^{(n)} \to \cY^{(n)}$;
        \item
        The overall joint distribution induced by the encoder-decoder pair   is given by 
        \begin{align*}
        &p_{X_1^n,X_2^n,S_1^n,S_2^n,M_1,M_2,Y^n}= \\
        &\left[\cD^{(n)} \circ \left(\mathop{\times}\limits_{j=1}^2 \cE_j^{(n)} \right) \right] \left\{\mathop{\times}\limits_{j=1}^2 \left(q_{X_j}^{\otimes n} \times p_{S_j}^{(n)} \right)\right\}
        \end{align*}
        such that
    \end{itemize}
     \begin{equation} \label{eq:sim_protocol_iid}
 \norm{ p_{X_1^n,X_2^n,Y^n}-q^{\otimes n}_{X_1,X_2,Y} }_{tvd} \leq \epsilon\;.
     \end{equation}
      The asymptotic iid rate region $\cR^{iid}$ for simulating MAC is defined as the closure of the set of all rate pairs $(R_1,R_2)$ such that for any $\epsilon>0$, there exists a simulation protocol  as given above  satisfying \eqref{eq:sim_protocol_iid} for all $n$ sufficiently large.
\end{definition}
Henceforth, in this section, we consider  $p_{X_1,X_2,U_1,U_2,Y}$ to be a p.m.f. of the form given in \eqref{eq:jointpmf}.
\begin{corollary} \label{cor:classical_IID}\cite[Theorem~1 and Theorem~4]{Kurri_MAC_simulation}
The cost region for simulating a MAC $q_{Y|X_1,X_2}$ with fixed inputs $q_{X_1} \times q_{X_2}$, using rate-limited links of rate $(R_1,R_2)$ and infinite shared randomness between each sender-receiver pair, in the asymptotic iid limit is given by:
    \begin{align} \label{eq:classical_iid_region}
   & \cR^{iid}=cl\Bigg\{\bigcup (R_1,R_2):    R_j \geq   I (X_j;U_j)_{q_{X_j}p_{U_j|X_j}}\Bigg\},
    \end{align} 
    
    where the union is taken over all the joint distribution $p_{X_1,X_2,U_1,U_2,Y}= q_{X_1}q_{X_2} p_{U_1|X_1}p_{U_2|X_2}p_{Y|U_1,U_2}$ such that $ p_{X_1,X_2,Y}=q_{X_1,X_2,Y}$ and $|\cU_1|,|\cU_2| \leq |\cX_1||\cX_2||\cY|$.
\end{corollary}

\begin{IEEEproof}
\textbf{Asymptotic iid Inner Bound: }The  one-shot inner bound can be straight away extended to obtain the optimal asymptotic iid rate region. Let $(R_1,R_2) \in  \cR^{iid}$  be such that for any $\eta>0$, 
\begin{align} \label{eq:limit_fixip}
    R_j \geq I(X_j;U_j)_p +\eta,
\end{align}
for some $ p_{X_1,X_2,U_1,U_2,Y} = q_{X_1}q_{X_2} p_{U_1|X_1} p_{U_2|X_2} p_{Y|U_1,U_2}$ satisfying $p_{X_1,X_2,Y} = q_{X_1,X_2,Y}$.\\
    Consider \begin{align}
    p_{X_1^n,X_2^n,U_1^{(n)},U_2^{(n)},Y^n}= q_{X_1}^{\otimes n}q_{X_2}^{\otimes n}p_{U_1|X_1}^{\otimes n}p_{U_2|X_2}^{\otimes n}
 p_{Y|U_1,U_2}^{\otimes n}. \label{eq:jointdist_iid}
\end{align}  
The AEP for the smoothed $\max$-mutual information (see \eqref{eq:AEP_classical} of Fact~\ref{fact:AEP_fixed_ip})  yields
    \[
        \lim_{n \to \infty} \frac{1}{n} \left[ I_{\max}^{\epsilon_j-\delta}(X_j^n,U_j^{(n)})_{p^n} + \log \log \left( \frac{1}{\delta} \right) \right]=I(X_j;U_j)_p,
    \]
    which by lower bound on $R_j$ given by \eqref{eq:limit_fixip} implies 
    \begin{align}
        nR_j \geq I_{\max}^{\epsilon_j-\delta}(X_j^n,U_j^{(n)})_{p^n} + \log \log  \frac{1}{\delta}, 
    \end{align}
    for all sufficiently large $n$ (depending on $\eta$). 
    This implies that 
$\cR^{iid} \subseteq \cR_{inner}^{(n)}(\epsilon_1,\epsilon_2)$, where
 
\begin{align}
 &\cR_{inner}^{(n)}(\epsilon_1,\epsilon_2) \notag\\
 &=\Bigg\{(R_1,R_2):  nR_j \geq I_{\max}^{\epsilon_j-\delta} (X_j^n;U_j^{(n)})_p +  \log \log \frac{1}{\delta}\Bigg\}.  
\end{align}
\textbf{Asymptotic iid Outer Bound: }
First note that obtaining the asymptotically optimal outer bound is not so straight forward as the $n$-fold extension of the random variable $U$ need not be iid. So, we prove a \textit{weak converse}. In order to do so, for any $\epsilon \in (0,1)$ we first define the following so-called $\epsilon$-approximate iid region  as follows:
\begin{align}\label{eq:delta_iid_converse}
    \cR^{iid}(\epsilon)&:= \Bigg\{ (R_1,R_2): R_j \geq I(X_j;U_j)_p, \notag\\
    & \quad \left. \forall \; p_{X_1,X_2,U_1,U_2,Y}=q_{X_1}q_{X_2}p_{U_1|X_1}p_{U_2|X_2}p_{Y|U_1,U_2} \right. \notag\\
    & \qquad \mbox{ such that }  \norm{p_{X_1,X_2,Y}- q_{X_1,X_2,Y}}_{tvd} \leq 2\epsilon \Bigg\}
\end{align}
For any $\epsilon \in (0,1)$ and $\epsilon_1, \epsilon_2 > 0$ such that $\max\{\epsilon_1,\epsilon_2 \} \leq \epsilon/2$, let $\cR_{outer}^{(n)}(\epsilon_1,\epsilon_2, \epsilon)$ be the $n$-fold extension of the region $\cR_{outer}(\epsilon_1,\epsilon_2, \epsilon)$ with respect to the input and auxiliary random variables   $(X_j^n,U_j^{(n)}) \sim q^{\otimes n}_{X_j} p_{U_j^{(n)}|X_j^n}$, i.e.
\begin{align}
 &\cR_{outer}^{(n)}(\epsilon_1,\epsilon_2,\epsilon) \notag\\
 &=\bigg\{(R_1,R_2):  nR_j \geq I_{\max}^{\epsilon_j} (X_j^n;U_j^{(n)})_p \bigg\},  
\end{align}
where \\
$p_{X_1^n,U_1^{(n)},X_2^n,U_2^{(n)},Y^n}\hspace{-3 pt}:=q_{X_1}^{\otimes n} q_{X_2}^{\otimes n}p_{U_1^{(n)}|X_1^n}p_{U_2^{(n)}|X_2^n}p_{Y^n|U_1^{(n)},U_2^{(n)}}$ is such that 
\begin{align}
    \norm{p^n_{X_1,X_2,Y}-q^{\otimes n}_{X_1,X_2,Y}}_{tvd} \leq 2\epsilon. \label{eq:simcrit}
\end{align}
 Suppose $(R_1,R_2) \in \cR_{outer}^{(n)}(\epsilon_1,\epsilon_2, \epsilon)$. 
Then: 
     \begin{align*}
         &nR_j \geq I_{\max}^{\epsilon_j}(X_j^n;U_{j}^{(n)})_{p^n}\\
         & \qquad \overset{(a)}{=} I_{\max} (X_j^n;U_{j}^{(n)})_{p^{'n}}\\
         & \qquad \overset{(b)}{\geq} I(X_j^n;U_{j}^{(n)})_{p^{'n}}\\ 
         &\qquad \overset{(c)}{\geq}I(X_j^n;U_{j}^{(n)})_{p^n}-2\epsilon_j \log |\cX_j|^n-2h_2\left(\frac{\epsilon_j}{1+\epsilon_j}\right)\\
         & \qquad \overset{(d)}{\geq}nI(X_j;U_j)_p-2 \epsilon_j \log |\cX_j|^n-2h_2\left(\frac{\epsilon_j}{1+\epsilon_j}\right)\\
         &\Rightarrow R_j \geq \lim_{\epsilon_j \to 0} \lim_{n \to \infty}  \left[\frac{nI(X_j;U_j)_p-2 \epsilon_j \log |\cX_j|^n}{n} \right.\\
         & \hspace{5 cm} \left.-\frac{2h_2\left(\frac{\epsilon_j}{1+\epsilon_j}\right)}{n}\right] \\
         & \Rightarrow R_j \geq I(X_j;U_j)_p,
         \end{align*}
     where $(a)$ holds by taking $p_{X_j^n,U_j^{(n)}}' \in \cB^{\epsilon_j}(p_{X_j^n, U_j^{(n)}})$ to be the optimizer for $I_{\max}^{\epsilon}$; $(b)$ holds by the fact the $I_{\max}(X;Y)_p \geq I(X;Y)_p$ for any joint distribution $p_{X,Y}$; $(c)$ follows due to continuity of mutual information from Fact~\ref{fact:continuity_I};
     $(d)$ follows by Proposition~\ref{prop:single_letter} shown in Appendix \ref{subsec:single_letter} for some $p_{X_j,U_j}=q_{X_j}p_{U_j|X_j}$ and finite $|\cU_j|$ from Lemma~\ref{lem:cardinality-fixed_ip}. Note that \eqref{eq:simcrit} and  monotonicity of trace distance implies that $\norm{p_{X_1,X_2,Y}-q_{X_1,X_2,Y}}_{tvd} \leq 2\epsilon$. 
     Hence, we have shown that in the asymptotic iid limit:
    \begin{equation} \label{eq:R_outsubset_R}
    \lim_{\epsilon_1,\epsilon_2 \to 0}\lim_{n \to \infty} \cR_{outer}^{(n)}(\epsilon_1,\epsilon_2,\epsilon) \subseteq \cR^{iid}(\epsilon).
    \end{equation}
    We have thus recovered the asymptotically optimal region of \cite[Theorem~1, Theorem~3]{Kurri_MAC_simulation} up to $\epsilon$.
     Since, in our setting we have bounded cardinalities of the auxiliary random variables, we can directly apply \cite[Lemma~6]{Gohari_continuity} (see Fact~\ref{fact:continuity} for a detailed analysis) and get a matching outer bound for $\epsilon \to 0$, as we argue in \eqref{eq:matching_ob1} and \eqref{eq:matching_ob2}.  We thus finally recover the asymptotically optimal region of \cite[Theorem~1, Theorem~3]{Kurri_MAC_simulation} in our setting of independent and fixed inputs and no side information at the decoder, to get 
     \begin{align} \label{eq:matching_ob1}
     \cR_{outer}:=    \mathop{\lim}\limits_{\epsilon \to 0}  \lim_{n \to \infty} \cR_{outer}^{(n)}(\epsilon_1,\epsilon_2,\epsilon) \subseteq \mathop{\lim}\limits_{\epsilon \to 0} \cR^{iid}(\epsilon)  = \cR^{iid}.
     \end{align}
     Thus we have shown that 
     \begin{align}\label{eq:matching_ob2}
        \cR_{outer} \subseteq \cR^{iid} \subseteq   \lim_{n \to \infty}\cR_{inner}^{(n)}(\epsilon_1,\epsilon_2) \subseteq \cR_{outer} \notag \\
        \Rightarrow  \cR_{inner}:=\lim_{\epsilon_1,\epsilon_2 \to 0}\lim_{n \to \infty}\cR_{inner}^{(n)}(\epsilon_1,\epsilon_2) = \cR^{iid} = \cR_{outer}.
     \end{align}
\end{IEEEproof}

\section{Universal MAC simulation} \label{Sec:universal}

In this section, we consider the task of universal channel simulation, where the   protocol should simulate the channel $q_{Y|X_1,X_2}$ irrespective of any particular choice of input distribution $q_{X_1} \times q_{X_2}$. Note the the inputs of the two senders are still independent, but arbitrary.
In the next proposition we show that Lemma~\ref{lem:classical_achievability} and Lemma~\ref{lem:classical_converse} of our MAC simulation protocol can be extended to achieve universal simulation with appropriate modifications.  These modifications refer to the simulation error being replaced by the maximum over the input samples $x_1,x_2$ (sampled according to any input distribution) in contrast to the average over a fixed input distribution.  
Before stating our result, we 
define  {\em universal} protocol for MAC simulation formally.
\begin{definition}[Universal  MAC simulation]\label{def:task-univ}
    An $(R_1,R_2, \epsilon)$ simulation protocol for a $2$-independent user MAC $q_{Y|X_1X_2}$ with inputs $q_{X_1} \times q_{X_2}$ and access to unlimited shared randomness between $\textrm{Sender}$1$ \overset{S_1}{\leftrightarrow} \textrm{Receiver}$ and $\textrm{Sender}$2$ \overset{S_2}{\leftrightarrow} \textrm{Receiver}$, consists of:
    \begin{itemize}
        \item A pair of encoders of form $\cE_1 \times \cE_2$, such that: $\cE_j:\cX_j \times \cS_j \to [1:2^{R_j}]$, for $j \in \{ 1,2\}$;
        \item Two independent noiseless rate-limited links of rate $R_j$, $j \in \{1,2\}$ and;
        \item A decoder $\cD: [1:2^{R_1}] \times \cS_1 \times [1:2^{R_2}] \times \cS_2 \to \cY$;
        \item
        The overall joint distribution induced by the encoder-decoder pair  final output is given by \begin{align*}&p_{X_1,X_2,S_1,S_2,M_1,M_2,Y}\\
        &=\left[\cD \circ \left(\mathop{\times}\limits_{j=1}^2 \cE_j \right) \right] \left\{\mathop{\times}\limits_{j=1}^2 \left(q_{X_j} \times p_{S_j} \right)\right\}
        \end{align*}
    \end{itemize}
     \begin{align} \label{eq:sim_protocoluniv}
      &\mbox{s.t. }\max_{x_1,x_2} \norm{ p_{Y|X_1=x_1,X_2=x_2}-q_{Y|X_1=x_1,X_2=x_2} }_{tvd} \leq \epsilon \notag\\
      &\text{ and }R_1=\log |\cM_1|,\; R_2=\log |\cM_2| \;.
     \end{align}
      The rate region $\cR_{\bU}(\epsilon)$ for universal simulation of a MAC is defined as the closure of the set of all rate pairs $(R_1,R_2)$ as given above satisfying \eqref{eq:sim_protocoluniv}. 
\end{definition}
    We say that the simulation protocol of Definition \ref{def:task-univ} is {\em universal} in the sense that it can simulate the given MAC  $q_{Y|X_1,X_2}$ for any input distribution $q_{X_1} \times q_{X_2}$, without being dependent on $q_{X_1}$ and $q_{X_2}$. This is ensured by the $\max$-error criterion in \eqref{eq:sim_protocoluniv}. 
    
In this section,  we henceforth consider  $p_{U_1,U_2,Y|X_1,X_2}$ to be a conditional p.m.f. of the form:
\begin{align}
p_{U_1,U_2,Y|X_1,X_2}=p_{U_1|X_1}p_{U_2|X_2} p_{Y|U_1,U_2}.  \label{eq:condpmf}
\end{align}

\subsection{One-shot Setting}
We first introduce the regions $\cR_{\bU}^{inner}$ and $\cR_{\bU}^{outer}$ which we will prove are the respective inner and outer bounds for the task of one-shot universal MAC simulation given by Definition~\ref{def:task-univ}.
\begin{definition}\label{def:one-shot_uni_in_out}
Let $\epsilon \in (0,1)$, and $\epsilon_1, \epsilon_2,\delta \in (0,1)$ be such that $\delta  < \min\{\epsilon_1,\epsilon_2\}$ and $\max\{\epsilon_1,\epsilon_2\} \leq \frac{\epsilon}{2} $. 
Let $\cR_{\bU}^{inner}(\epsilon_1,\epsilon_2,\delta)$ be the set of non-negative real numbers $(R_1,R_2)$ defined as:
   \begin{align} 
   & \cR_{\bU}^{inner}(\epsilon_1,\epsilon_2,\delta) \notag\\&=cl\Bigg\{\bigcup(R_1,R_2): R_j \geq   I_{\max}^{\epsilon_j-\delta} (p_{U_j|X_j})+\log \log \frac{1}{\delta}\Bigg\},
   \label{eq:universal_inner}
    \end{align} 
    where the union is taken over $\left(p_{U_1|X_1},p_{U_2|X_2}\right) \in \cA^{inner}$ and $\cA^{inner}$ is the set of all feasible distributions for evaluating $\cR_{\bU}^{inner}$ and is given by
    \begin{align}
    \cA^{inner}&:= \left\{ \left(p_{U_1|X_1},p_{U_2|X_2}\right):~ \exists ~p_{Y|U_1, U_2} \right.\notag\\
    & \qquad \left. \mbox{ satisfying } p_{Y|X_1,X_2}=q_{Y|X_1,X_2}\right\}. \label{eq:A_inner}
    \end{align}
     Similarly, let $\cR_{\bU}^{outer}(\epsilon_1,\epsilon_2,\epsilon)$ be the set of non-negative real numbers $(R_1,R_2)$ defined as:
    \begin{align} \label{eq:universal_outer}
     &\cR_{\bU}^{outer}(\epsilon_1,\epsilon_2,\epsilon) \notag\\
     &=  cl\Bigg\{\bigcup (R_1,R_2): R_j \geq   I_{\max}^{\epsilon_j} (p_{U_j|X_j})\Bigg\},
    \end{align}
    where the union is taken over $\left(p_{U_1|X_1},p_{U_2|X_2}\right) \in \cA_{\epsilon}^{outer} $ and $\cA_\epsilon^{outer}$ is the set of all feasible distributions for evaluating $\cR_{\bU}^{outer}$ and is given by
    \begin{align}
    & \cA^{outer}_\epsilon :=  \bigg\{\left(p_{U_1|X_1},p_{U_2|X_2}\right):~ \exists ~p_{Y|U_1, U_2} \mbox{ satisfying } \notag\\
&\quad  \mathop{\max}\limits_{x_1,x_2} \left\lVert p_{Y|X_1=x_1,X_2=x_2}-q_{Y|X_1=x_1,X_2=x_2} \right\rVert_{tvd} \leq 2\epsilon, \notag\\
& \quad~|\cU_1|, |\cU_2|  \leq |\cX_1||\cX_2||\cY|  \bigg\}. \label{eq:A_univ}
    \end{align}
\end{definition}
\begin{theorem} \label{thm:one-shot_universal}
 Let $q_{Y|X_1,X_2}$ be a given $2$-sender, $1$-receiver MAC. For any $\epsilon \in (0,1)$, and  $\epsilon_1, \epsilon_2,\delta \in (0,1)$ be such that  $\delta  < \min\{\epsilon_1,\epsilon_2\})$ and $\max\{\epsilon_1,\epsilon_2\} \leq \frac{\epsilon}{2} $,  one-shot rate region for  {\em universal} simulation of  MAC satisfies:
    \begin{align}\label{eq:universal}
        \cR_{\bU}^{inner}(\epsilon_1,\epsilon_2,\delta) \subseteq \cR_{\bU}(\epsilon) \subseteq \cR_{\bU}^{outer}(\epsilon_1,\epsilon_2,\epsilon)\;,
    \end{align}
    where  $\cR_{\bU}^{inner}(\epsilon_1,\epsilon_2,\delta)$ and $\cR_{\bU}^{outer}(\epsilon_1,\epsilon_2,\epsilon)$  are as defined in Definition~\ref{def:one-shot_uni_in_out}.
\end{theorem}
    The proof comprises of two parts:
    \begin{itemize}
        \item Direct part or the achievability as shown in Lemma~\ref{lem:classical_achievability_universal}; and
        \item Converse as proven in  Lemma~\ref{lem:classical_converse_universal}.
    \end{itemize}

\subsection{Achievability}

\begin{lemma}\label{lem:classical_achievability_universal}
For any given $\epsilon>0$, let $\epsilon_1, \epsilon_2 >0$ be such that  $\max\{\epsilon_1,\epsilon_2\} \leq \frac{\epsilon}{2}$ and  $\delta \in (0, \min\{\epsilon_1,\epsilon_2\})$. Then,  $ \cR_{\bU}^{inner}(\epsilon_1,\epsilon_2,\delta) \subseteq \cR_{\bU}(\epsilon)$.
\end{lemma}

The proof of this lemma (see  Appendix~\ref{sec:achievability_universal}) is very similar to that of the fixed input simulation case of Lemma~\ref{lem:classical_achievability}, with the difference being that the simulation error criterion is changed from average to maximum.


\subsection{Converse}

\begin{lemma}\label{lem:classical_converse_universal}
For any given $\epsilon \in (0,1)$, let $\epsilon_1, \epsilon_2 > 0$ be such that $\max\{\epsilon_1,\epsilon_2\} \leq \frac{\epsilon}{2}$. Then, $\cR_{\bU}(\epsilon) \subseteq \cR_{\bU}^{outer}(\epsilon_1,\epsilon_2,\epsilon)$.
\end{lemma}
The proof has minor technical changes compared to that of Lemma~\ref{lem:classical_converse} due to the average simulation error being replaced by the maximum simulation error criterion and is given in Appendix~\ref{sec:converse_universal}.

We now extend the one-shot result to the asymptotic iid setting.


\subsection{Asymptotic Expansion}

In this section, we consider a universal MAC simulation protocol that simulates $n$-iid copies of the channel, that is, $q^{\otimes n}_{Y|X_1,X_2}$ with general $n$-letter inputs denoted by $q_{X_1^n} \times q_{X_2^n}$.  
We note that this is in contrast to the generic usage of the term asymptotic iid, as used in previous sections, which refers to iid inputs $q_{X_1}^{\otimes n} \times q_{X_2}^{\otimes n}$.  
This leads to non-trivialities in extending the one-shot result to asymptotic iid as neither the inputs nor the auxiliary random variables that characterize the rate region are iid. Nevertheless we prove the following single-letter characterization even for this case. \\
Henceforth, in this section, we consider   $p_{U_1,U_2,Y|X_1,X_2}$ to be a conditional p.m.f. of the form given in \eqref{eq:condpmf}.
\begin{corollary} \label{cor:iid_universal}
    The rate region for  {\em universal} asymptotic iid simulation of MAC $q_{Y|X_1,X_2}$ is given by
    \begin{align} \label{eq:universal_iid}
     &\cR^{iid}_{\bU} \notag\\
     & =cl\Bigg\{\bigcup (R_1,R_2):R_j \geq \max_{q_{X_j}} I (X_j;U_j)_{q_{X_j}p_{U_j|X_j}}\Bigg\}.
    \end{align}
    where the union is taken over all the joint distributions $p_{U_1,U_2,Y|X_1,X_2}$ such that the Markov chain $(X_1,X_2) \to (U_1,U_2) \to Y$ holds and $p_{Y|X_1,X_2}=q_{Y|X_1,X_2}$ with $|\cU_1|,|\cU_2| \leq |\cX_1||\cX_2||\cY|$.
\end{corollary}
We prove this proposition in Appendix~\ref{sec:universal-iid}.
    \begin{remark} 
Note that the point-to-point channel is a special case of the MAC, where one of the inputs, say $X_2$, is redundant.  Then, setting  $U_1=Y$, $U_2=X_2=1$  with probability one, and  $\epsilon_2=0$ in the definitions of $\cR_{inner}(\epsilon_1,\epsilon_2,\delta)$ and $\cR_{outer}(\epsilon_1,\epsilon_2,\epsilon)$, we recover the one-shot point-to-point channel simulation result of \cite{Berta_simulation} stated in Fact \ref{lem:rej_sampling} (see Appendix \ref{app:classical}). Furthermore, this also recovers the asymptotically optimal point-point channel simulation rate $I(X;Y)_{q_{X,Y}}$ as shown in \cite{Cuff_distributed_synthesis, Berta_simulation}.  Moreover, for the point-to-point case our technique can be straight away extended to obtain the universal channel simulation by identifying optimal $U=Y$. 
\end{remark}
\begin{remark}
    We remark that the universal simulation protocol can also be used for the fixed input protocol of Section~\ref{sec:Classical}. The main difference is that the communication rates of the universal protocol are higher than necessary rate for the fixed input case because 
    \[
        I^\epsilon_{\max}(p_{U|X}) \geq I^\epsilon_{\max}(X;U)_{q_Xp_{U|X}}
    \]
    due to the fact that the minimization in the definition of the channel smoothed $\max$-mutual information (left hand side term above) is over a small set since the simulation error criterion is stronger. 
    Hence, we gave a separate analysis for the fixed input case. 
\end{remark}


\section{Quantum-classical MAC simulation}\label{sec:qc}

In this section, we take a step towards generalizing our simulation protocol in the quantum regime. To this end, we consider a MAC with two independent quantum inputs and one classical output. Generally, a channel that takes a quantum state as an input and outputs a probability distribution (or classical state as the output random variable) is modeled as a measurement device (or a measurement channel). Hence we refer to the $2$-quantum input and $1$-classical output as a QC MAC and think of it as a measurement channel.\\

 \textbf{Classical Scrambling QC MAC (CS-QC MAC):} Channel first does a product measurement on two inputs with classical outcomes $X_1,X_2$ and then scrambles them according to the conditional probability distribution $q_{Y|X
          _1,X_2}$. We refer to such channels as "\textit{classical scrambling}" (CS) channels, denoted as
         \begin{align} \label{eq:QC_classical_nofb}
                \cN_{CS}^{A_1A_2 \to Y}:= q_{Y|X_1,X_2} \circ \left( \Lambda^{A_1 \to X_1} \otimes \Gamma^{A_2 \to X_2} \right), 
          \end{align} 
          where $\Lambda^{A_1 \to X_1}$ and $\Gamma^{A_2 \to X_2}$ are  measurements with POVM elements $\{\Lambda_{x_1}\}_{x_1}$ and $\{\Gamma_{x_2}\}_{x_2}$, respectively. This is a special case of the model of a QC-channel proposed in \cite{Distributed_measurement},  termed as  \textit{distributed measurement channel having a separable decomposition with stochastic integration}.

          We characterize the cost of simulating the CS-QC MAC with feedback defined as follows:\\
          
\begin{figure*}
  \centering
  \includegraphics[width=\textwidth]{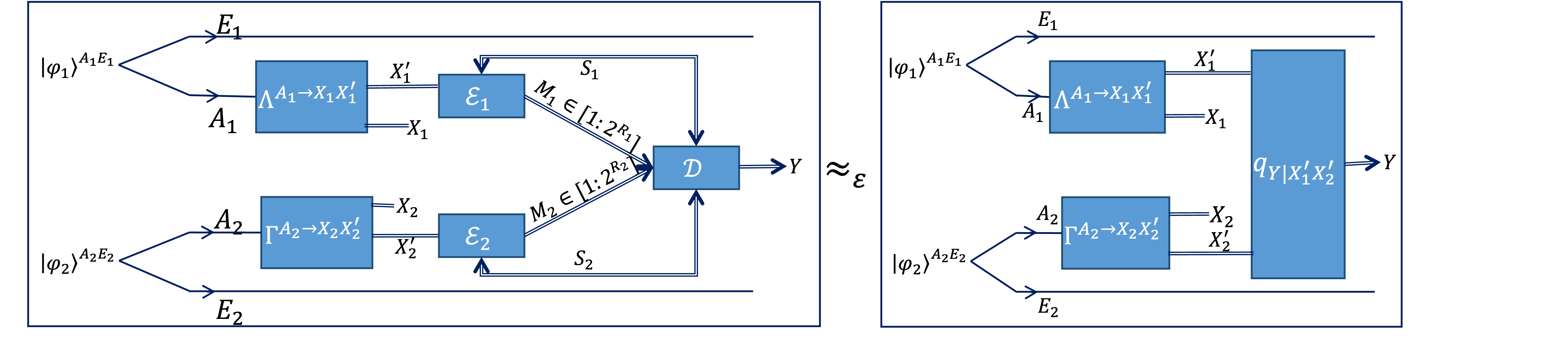}
  \caption{CS-QC MAC with feedback: Encoders $\cE_j: (X'_j,S_j) \xrightarrow[split]{convex} M_j \in [1:2^{R_j}]$, Decoder $\cD:(M_1,M_2,S_1,S_2) \mapsto Y$; $\approx_\varepsilon$ denotes closeness in tvd.}
  \label{fig:QC_MAC}
\end{figure*}

           \textbf{CS-QC MAC with Feedback:} Channel first does a product measurement on two inputs, creates two copies of the classical outputs and then scrambles one of the copies according to $q_{Y|X
          _1,X_2}$, while keeping the other untouched. We refer to such channels as "\textit{classical scrambling channels with feedback}", shown in the right hand side of Figure~\ref{fig:QC_MAC} and is denoted as:
          \[
            \cN_{CS}^{A_1A_2 \to YX_1X_2}:= q_{Y|X_1,X_2} \circ \left( \Lambda^{A_1 \to X_1X_1'} \otimes \Gamma^{A_2 \to X_2X_2'} \right),
          \]
          where the output of the measurement operators for fixed inputs are defined as:
          \begin{align*}
                &\cI^{E_1} \otimes \Lambda^{A_1 \to X_1X_1'}(\ketbra{\varphi_1}^{E_1A_1})\\
                &:= \sum_{x_1} p_{X_1}(x_1)\ketbra{x_1}^{X_1} \otimes \ketbra{x_1}^{X_1'} \otimes \varphi_{x_1}^{E_1}: \\
                & \qquad \Tr(\varphi_{x_1}^{E_1})=1;\\
                &\cI^{E_2} \otimes \Gamma^{A_2 \to X_2X_2'}(\ketbra{\varphi_2}^{E_2A_2})\\
                &:= \sum_{x_2} p_{X_2}(x_2)\ketbra{x_2}^{X_2} \otimes \ketbra{x_2}^{X_2'} \otimes \varphi_{x_2}^{E_2}:\\ 
                & \qquad \Tr(\varphi_{x_2}^{E_2})=1.
          \end{align*}
          Note that $(X_1',X_2')$ are just the classical copies of $(X_1,X_2)$. The conditional distribution $q_{Y|X_1,X_2}$ is a probability measure on $\cY$ conditioned on random variables $(X_1,X_2)$ taking values in $\cX_1 \times \cX_2$. Henceforth, we consistently use the notation $q_{Y|X_1,X_2}$ (instead of $q_{Y|X_1',X_2'}$) to represent the classical scrambling map to mean that random variables $(X_1',X_2')$ are stochastically mapped to the output random variable $Y$.
          Thus, the actual channel outcome is given as
         \begin{align} \label{eq:QC_classical}
                &\cN_{CS}^{A_1A_2 \to YX_1X_2}(\rho_1^{A_1} \otimes \rho_2^{A_2}) \notag\\
                &=\mathop{\Sigma}\limits_{x_1,x_2,y} \left\{ q_{Y|X_1,X_2}(y|x_1,x_2)\ketbra{y}^Y \otimes \Tr[\Lambda_{x_1}\rho_1] \ketbra{x_1}^{X_1} \right. \notag\\
                & \qquad \qquad \left. \otimes \Tr[\Gamma_{x_2}\rho_2]\ketbra{x_2}^{X_2} \right\}. 
          \end{align}  
   Here, we focus on the task of simulating CS-QC MAC with feedback represented by \eqref{eq:QC_classical}.


\subsection{Feedback Simulation for Fixed Product Input} \label{sec:qtask}

\begin{definition}[CS-QC MAC simulation with feedback]\label{def:qtask}
    An $(R_1,R_2, \epsilon)$ simulation code for a $2$-independent user CS-QC MAC with feedback given in \eqref{eq:QC_classical}  and access to unlimited shared randomness between $\textrm{Sender}1 \overset{S_1}{\leftrightarrow} \textrm{Receiver}$ and $\textrm{Sender}2 \overset{S_2}{\leftrightarrow} \textrm{Receiver}$, consists of:
    \begin{itemize}
    \item Inputs to the two encoders are ${\varphi_1}^{E_1X_1X_1'} \otimes {\varphi_2}^{E_2X_2X_2'}$, where 
    \begin{align*}
    {\varphi_j}^{E_jX_jX_j'} & :=\sum_{x_j}p_{X_j(x_j)}\ketbra{x_j}^{X_j} \otimes \ketbra{x_j}^{X_j'} \otimes {\varphi_{x_j}}^{E_j};\\
    & p_{X_j}(x_j):=\Tr[\Lambda_{x_j}\rho_j],
    \end{align*}
   and $\varphi_{x_1}^{E_1} \otimes \varphi_{x_2}^{E_2}$ is the normalized post-measurement state of the measurement $\Lambda \otimes \Gamma$. Note that $X'$ is just a classical copy of $X$, to perform the simulation with feedback;
        \item A pair of encoders $\cE_1 \otimes \cE_2$ with inputs as the measurement outcomes $X_1,\; X_2$ and shared randomness $S_1 \; S_2$, denoted by: $\cE_j:\cX_j \otimes \cS_j \to [1:2^{R_j}]$, for $j \in \{ 1,2\}$;
        \item Two separate noiseless rate-limited classical links of rate $R_j$, $j \in \{1,2\}$ and;
        \item A decoder $\cD: [1:2^{R_1}] \times \cS_1 \times [1:2^{R_2}] \times \cS_2 \to \cY$;
    \item The simulation algorithm produces the overall state as 
    \begin{align}
    \tau^{YX_1X_2E_1E_2}&:=\left\{\cD \circ \left(\mathop{\otimes}\limits_{j=1}^2 \cE_j \right) \right\}\bigg[\ketbra{\varphi_1} \otimes \ketbra{\varphi_2}   \notag\\
    & \qquad \qquad \qquad \qquad \qquad  \otimes \left(\mathop{\otimes}\limits_{j=1}^2  S_j \right)\bigg], \notag    
    \end{align}
   such that
     \begin{equation} \label{eq:QC_sim}
     \norm{ \tau^{YX_1X_2E_1E_2}-\eta^{YX_1X_2E_1E_2} }_1 \leq \epsilon,
     \end{equation}
     where $\eta^{YX_1X_2E_1E_2}:=\cN^{A_1A_2\to YX_1X_2}(\ketbra{\varphi_1}^{A_1E_1}\otimes \ketbra{\varphi_2}^{A_2E_2})$, is the output state of CS-QC MAC $\cN$ to be simulated.
     \end{itemize}
           The rate region $\cR(\epsilon)$ for simulating MAC is defined as the closure of the set of all rate pairs $(R_1,R_2)$ as given above. 
\end{definition}
    The classical MAC simulation described in Section~\ref{sec:Classical} is the non-feedback simulation. However, in order to extend the classical proof technique of the converse to CS-QC MAC, we require the encoders to have access to the classical outcomes of the measurement channels $\Lambda^{A_1 \to X_1} \otimes \Gamma^{A_2 \to X_2}$. 
    Observe that simulation criteria in \eqref{eq:QC_sim} is an average error criterion similar to  the classical MAC simulation criteria with fixed inputs given  in \eqref{eq:sim_protocol}. \\
In what follows, let $\tau^{E_1E_2X_1X_2U_1U_2Y}$ and $\eta^{E_1E_2X_1X_2Y}$ be the following classical-quantum (CQ) states: 
           \begin{align}  
     &\tau^{E_1E_2 U_1U_2X_1X_2Y} \notag\\
         &  := 
          \mathop{\Sigma} \limits_{\vec{u},\vec{x},y} \bigg[p_{Y|U_1,U_2}(y|u_1,u_2)p_{U_1|X_1}(u_1|x_1)p_{X_1}(x_1) \notag \\
          & p_{U_2|X_2}(u_2|x_2) p_{X_2}(x_2) \ketbra{y}^Y \otimes  \ketbra{x_1}^{X_1} \otimes \ketbra{u_1}^{U_1}    \notag\\
         &   \otimes \ketbra{x_2}^{X_2} \otimes \ketbra{u_2}^{U_2} \otimes \frac{\Tr_{A_1}\big[\left\{\cI^{E_1} \otimes \Lambda^{A_1}_{x_1}\right\}(\varphi_1 ^{E_1A_1}) \big]}{p_{X_1}(x_1)} \notag \\ 
         &  \left. \otimes  \frac{\Tr_{A_2}\big[\left\{\cI^{E_2} \otimes\Gamma^{A_2}_{x_2}\right\}(\varphi_2^{E_2A_2})\big]}{p_{X_2}(x_2)} \right];\mbox{ and }\label{eq:evaluating_state}\\
     &\eta^{YE_1E_2 X_1X_2} \notag\\
          & :=  \mathop{\Sigma} \limits_{\vec{x},y} \bigg[ q_{Y|X_1,X_2}(y|x_1,x_2) p_{X_1}(x_1) \otimes p_{X_2}(x_2) \ketbra{y}^Y \otimes \notag\\
         & \ketbra{x_1}^{X_1} \otimes \ketbra{x_2}^{X_2} \otimes  \frac{\Tr_{A_1}\big[\left\{\cI^{E_1} \otimes \Lambda^{A_1}_{x_1}\right\}(\varphi_1 ^{E_1A_1}) \big]}{p_{X_1}(x_1)} \notag \\
         & \left. \otimes \frac{\Tr_{A_2}\big[\left\{\cI^{E_2} \otimes \Gamma^{A_2}_{x_2}\right\}(\varphi_2^{E_2A_2})\big]}{p_{X_2}(x_2)} \right]. \label{eq:target_state}
     \end{align}
We now define the regions analogous to the classical case given by Definition~\ref{def:inner_outer} and \ref{def:one-shot_uni_in_out}, which will be proven as the {\em inner and outer bound} regions for the task of QC-MAC simulation {\em with feedback} analogous to Definition~\ref{def:inner_outer}.

\begin{definition}[Inner and Outer bounds] \label{def:QC_in_out}
Let $\epsilon \in (0,1)$, and  $\epsilon_1, \epsilon_2,\delta \in (0,1)$ be such that  $\delta  < \min\{\epsilon_1,\epsilon_2\}$ and $\max\{\epsilon_1,\epsilon_2\} \leq \frac{\epsilon}{2}$.
Let $\cR^{QC-fb}_{inner}(\epsilon_1,\epsilon_2,\delta)$ be the set of non-negative real numbers $(R_1,R_2)$ defined as:
   \begin{align} 
   & \cR^{QC-fb}_{inner}(\epsilon_1,\epsilon_2,\delta) \notag\\
   & =cl\Bigg\{\bigcup (R_1,R_2): R_j \geq   I_{\max}^{\epsilon_j-\delta} (E_j;U_j)_{\tau}+2\log \frac{1}{\delta}\Bigg\},
   \label{eq:fixed_inner_QC}
    \end{align} 
    where union is taken over $\left(\tau^{E_1X_1U_1},\tau^{E_2X_2U_2}\right) \in \cA^{inner}$ and $\cA^{inner}$ is the set of all feasible states for evaluating $\cR_{inner}^{QC-fb}$ and is given by
    \begin{align}
    \cA^{inner}&:= \bigg\{ \left(\tau^{E_1X_1U_1},\tau^{E_2X_2U_2}\right):~ \forall\tau \notag \\
    &  \qquad \qquad \mbox{ s.t. } \tau^{E_1E_2X_1X_2Y}=\eta^{E_1E_2X_1X_2Y}\bigg\}. \label{eq:fixed_A_inner_QC}
    \end{align}
    Similarly, let $\cR^{QC-fb}_{outer}(\epsilon_1,\epsilon_2)$ be the set of non-negative real numbers $(R_1,R_2)$ defined as:
    \begin{align}
     &\cR^{QC-fb}_{outer}(\epsilon_1,\epsilon_2,\epsilon) \notag\\
    & =cl\Bigg\{\bigcup (R_1,R_2):  R_j \geq   I_{\max}^{\epsilon_j} (E_j;U_j)_{\tau} \Bigg\},
    \label{eq:fixed_outer_QC}
    \end{align}
    where union is taken over $\left(\tau^{E_1X_1U_1},\tau^{E_2X_2U_2}\right) \in \cA_{\epsilon}^{outer}$ and $\cA_\epsilon^{outer}$ is the set of all feasible states for evaluating $\cR_{outer}^{QC-fb}$ and is given by
    \begin{align}
    &\cA^{outer}_\epsilon := \bigg\{ \left(\tau^{E_1X_1U_1},\tau^{E_2X_2U_2}\right):~\forall~ \tau \mbox{ s.t. }  \notag\\
    & \qquad \qquad \quad \left\lVert \tau^{E_1E_2X_1X_2Y}-\eta^{E_1E_2X_1X_2Y}\right\rVert_{tvd} \leq 2\epsilon,  \notag\\
    & \qquad \qquad \quad |\cU_1|,|\cU_2| \leq |\cX_1||\cX_2||\cY| \bigg\}. \label{eq:fixed_A_outer_QC}
\end{align}
\end{definition}

The following characterization is our main result in this section.

\begin{theorem} \label{thm:QC_MAC_main}
 Let $\cN_{CS}^{AB \to YX_1X_2}$ be a given $2$-sender, $1$-receiver MAC with input $\rho^{A_1} \otimes \rho^{A_2}$ and their respective purifications denoted by $\ket{\varphi}^{E_1A_1} \otimes \ket{\varphi}^{E_2A_2}$. For $\epsilon \in (0,1)$ and $\epsilon_1, \epsilon_2 >0 $ with $\max\{\epsilon_1,\epsilon_2\} \leq \frac{\epsilon}{2} $, and $\delta \in (0,\min\{\epsilon_1,\epsilon_2 \})$, we have
 \begin{align}
     \cR^{QC-fb}_{inner}(\epsilon_1,\epsilon_2,\delta) \subseteq \cR(\epsilon) \subseteq \cR^{QC-fb}_{outer}(\epsilon_1,\epsilon_2,\epsilon),
 \end{align}
  where  $\cR^{QC-fb}_{inner}(\epsilon_1,\epsilon_2,\delta)$ and  $\cR^{QC-fb}_{outer}(\epsilon_1,\epsilon_2, \epsilon)$  are given in Definition~\ref{def:QC_in_out}.
\end{theorem}

The proof comprises of two parts:
    \begin{itemize}
        \item Direct part or the achievability, which we prove in Lemma~\ref{lem:CS_QC_MAC}; and
        \item Converse shown in Lemma~\ref{lem:QCMAC_converse}.
    \end{itemize}

Note that $\cR^{QC-fb}_{inner}(\epsilon_1,\epsilon_2,\delta)$ and $\cR^{QC-fb}_{outer}(\epsilon_1,\epsilon_2,\epsilon)$ characterize the one-shot rate region $\cR(\epsilon)$ up to a fudge factor that depends on $\delta, \epsilon_1 \text{ and }\epsilon_2$.
\subsection{Achievability}
\begin{lemma} \label{lem:CS_QC_MAC}
For any given $\epsilon>0$, let $\epsilon_1, \epsilon_2 >0 $ be such that  $\max\{\epsilon_1,\epsilon_2\} \leq \frac{\epsilon}{2}$ and $\delta \in (0, \min\{\epsilon_1,\epsilon_2\})$. Then $ \cR^{QC-fb}_{inner}(\epsilon_1,\epsilon_2,\delta) \subseteq \cR(\epsilon)$.
\end{lemma}

\begin{IEEEproof}
    Fix  $(\epsilon_1,\epsilon_2,\delta)$ satisfying the conditions in the lemma and let $\ket{\varphi_1}^{E_1A_1} \otimes \ket{\varphi_2}^{E_2A_2}$ be the purifications of the fixed quantum inputs $\rho_1^{A_1} \otimes \rho_2^{A_2}$, with $E_1,E_2$ denoting the purifying reference (or the environment) systems. We need to show that for any $(R_1,R_2) \in  \cR^{QC-fb}_{inner}(\epsilon_1,\epsilon_2,\delta)$ (defined in \eqref{eq:fixed_inner_QC}), there exists an $(R_1,R_2,\epsilon)$ one-shot MAC simulation  protocol as mentioned in Definition \ref{def:qtask}.\\
    We note that since the overall state used to evaluate $\cR_{inner}(\epsilon_1, \epsilon_2, \delta)$ is a CQ state $\tau$ given in \eqref{eq:code_p_achieve} satisfying the simulation criterion, we have:
    \begin{align} 
           &\eta^{E_1E_2X_1X_2Y} = \Tr_{U_1,U_2} \tau^{E_1E_2X_1X_2U_1U_2Y} \notag \\
        &\Rightarrow p_{X_1}(x_1)p_{X_2}(x_2)q_{Y|\vec{X}}(y|\vec{x}) =\notag \\
        &~\sum_{\vec{u}} p_{X_1}(x_1)p_{X_2}(x_2) p_{U_1|X_1}(u_1|x_1)p_{U_2|X_2}(u_2|x_2) p_{Y|\vec{U}}(y|\vec{u})\;. \label{eq:code_p_achieve}
           \end{align}
    We henceforth consider the joint distribution $p_{X_1,X_2,U_1,U_2,Y}$ in \eqref{eq:code_p_achieve}  throughout the proof. We will show that $(R_1,R_2)$ is achievable by constructing a protocol that uses the convex split lemma from Fact~\ref{fact:convex_split}, for each sender. 
     \begin{itemize}
             \item \textbf{Encoding: }The input to the encoders $\cE_1$ and $\cE_2$ are the post-measurement states ${\varphi}_1^{E_1 X_1X_1'}$ and $\varphi_2^{E_2  X_2X_2'}$, respectively. 
             The encoders then generate the classical auxiliary random variables $U_j$ by post processing $X_j'$ with the dephasing map $\cI^{E_j} \otimes\; \cC_{j}^{X_j' \to U_{j}}$. This map is essentially a measurement channel that measures the state on $X_j'$ in an orthonormal basis $\{\ket{u_j}\}^{U_j}$ and outputs the classical variable $U_j$ distributed according to the conditional distribution $p_{U_j|X_j}$, formally defined as follows:
            \begin{align*}
                &\cC_j^{X_j' \to U_j}: \mathop{\Sigma}\limits_{x_j}p_{X_j'}(x_j) \ketbra{x_j}^{X_j'}\\
                & \qquad \qquad \mapsto  \mathop{\Sigma}\limits_{u_j,x_j}p_{X_j'}(x_j)p_{U_j|X_j}(u_j|x_j) \ketbra{u_j}^{U_j} 
            \end{align*}
          Thus, for $j\in \{1,2\}$, the overall states $ \varphi_1^{E_1X_1U_1}$ and $ \varphi_2^{E_2X_2U_2}$ are given by:
        \begin{align} \label{eq:phi-EU}
         &\varphi_j^{E_jX_jU_j}:= \notag\\
         &\sum_{u_j,x_j} \hspace{-5 pt} p_{X_j}(x_j) p_{U_j|X_j}(u_j|x_j) \ketbra{x_j}^{X_j} \otimes \ketbra{u_j}^{U_j} \otimes \varphi_{x_j}^{E_j}  \notag\\
       &\Rightarrow  \varphi_j^{E_jU_j} = \hspace{-5 pt}
       \sum_{u_j,x_j} p_{U_j}(u_j) \ketbra{u_j}^{U_j}  \otimes p_{X_j|U_j}(x_j|u_j)   \varphi_{x_j}^{E_j} \notag\\
       & \qquad \qquad = \sum_{u_j} p_{U_j}(u_j) \ketbra{u_j}^{U_j}  \otimes \varphi_{u_j}^{'E_j}, 
        \end{align}
        where $\varphi_{u_j}^{'E_j}:=\sum_{x_j}p_{X_j|U_j}(x_j|u_j)   \varphi_{x_j}^{E_j}$. 
         Note that these auxiliary random variables $U_j \sim p_{U_j|X_j}$ will satisfy the condition of \eqref{eq:code_p_achieve}. 
         The resultant state to be further encoded or compressed to achieve lower rates is the purification of ${ \varphi_1}^{E_1 X_1U_1} \otimes {\varphi_2}^{E_2 X_2U_2}$, which is given by:
         \begin{align}\label{eq:tilde_phi}
            &\mathop{\otimes}_{j=1}^2 \left(\ket{\varphi_j}^{E_j E_j' X_jX_j'' U_j \tilde U_j}\right):= \notag \\
            &\mathop{\otimes}_{j=1}^2 \left[ \sum_{x_j,u_j} \sqrt{p_{X_j}(x_j)p_{U_j|X_j}(u_j|x_j)} \ket{\varphi_{x_j}}^{E_j E_j'} \otimes\right. \notag\\
            & \qquad \qquad\left. \ket{x_jx_j}^{X_jX_j''}  \otimes \ket{u_ju_j}^{U_j \tilde U_j}\right].
         \end{align}
         Sender $j$ holds the registers $E_j',X_j, X_j'', U_j \mbox{ and }\tilde U_j$. Now we use the {\em encoders} $\cE_{j, \textrm{meas. comp.}}:{\cS_j \times \tilde \cU_j \to [1:2^{R_j}]}$ of the measurement compression protocol with feedback (see Definition~\ref{def:meas_comp} and the proof in Appendix~\ref{appendix:CS-QC_achievability}). These measurement compression encoders compress  $\tilde U_j$ using one half of the available shared randomness to a message $M_j$ described by $R_j$ bits. 
         We denote the overall encoder $\cE_j=\cE_{j, \textrm{meas. comp.}} \circ \cC^{X_j' \to U_j}$ (see (Appendix~\ref{appendix:CS-QC_achievability} for details of $\cE_{j, \textrm{meas. comp.}}$).\\
        \item \textbf{Decoding: }The decoding is composed of the following two steps:\\
        \subitem{(i). }The receiver first recovers $U_j$ from the received message index and the available shared randomness. This is accomplished by using the decoders
        $\cD_j:[1:2^{R_j}] \otimes \cS_j \to \bar \cU_j$ of the measurement compression theorem ($\cD_{\textrm{meas. comp.}}$) from Definition~\ref{def:meas_comp} (see Appendix~\ref{appendix:CS-QC_achievability} for exact details). Essentially these are the isometries that are guaranteed by Uhlmann's theorem (Fact~\ref{fact:Uhlmann}) in the convex split lemma. The recovered pairs are denoted by $\bar U_1,\bar U_2$ and in effect the  correlations with $E_1,E_2$ are "preserved". \\
        Let the overall state after the application of $ (\cD_j \circ \cE_j)$ be denoted as:
        \begin{align} \label{eq:post_meas_comp_state}
            &\ket{\tilde \varphi_j}^{E_j E_j' X_jX_j'  U_j \bar U_j} := \notag \\
            &\sum_{x_j,u_j} \sqrt{\tilde p_{X_j,\bar U_j}(x_j,u_j)} \ket{\varphi_{x_j}}^{E_j E_j'} \ket{x_jx_j}^{X_jX_j'} \ket{u_ju_j}^{U_j \bar U_j} \\
            &\Rightarrow \tilde \varphi_j^{E_j \bar U_j}=\sum_{x_j,u_j} \tilde p_{X_j,\bar U_j}(x_j,u_j) \varphi_{x_j}^{E_j} \otimes \ketbra{u_j}^{\bar U_j} \mbox{ and } \label{eq:rate_ineq}\\
            & \epsilon_j \overset{(a)}{\geq} \norm{\tilde \varphi_j-\varphi_j}_{tvd} \geq \norm{\tilde p_{X_j,\bar U_j}-p_{X_j,U_j}}_{tvd} \label{eq:appx_comp}
        \end{align}
        where $(a)$ follows from Fact~\ref{fact:convex_split} for
       \begin{align}
          R_j \geq I_{\max}^{\epsilon_j-\delta}(E_j;U_j)_\tau + 2  \log  \frac{1}{\delta} ; \; j \in \{ 1,2 \}. \label{eq:rate_qc}   
       \end{align}
    
        \subitem{(ii). } The decoder $\cD$ use these recovered classical states $\bar U_j$ and finally outputs $Y \sim p_{Y|\bar U_1,\bar U_2}$.\\

        \item \textbf{Analysis of the Protocol: }We now show that the code defined above using $\cE_1 \otimes \cE_2$ as the encoder and $p_{Y|U_1,U_2} \circ (\cD_1 \otimes \cD_2)$ as the decoder satisfies the simulation constraint and hence is a valid simulation code for CS-QC MAC with feedback. And we finally evaluate the rate of this code. For this, we first recall that the actual channel output  $\eta^{YE_1E_2X_1X_2}$ from \eqref{eq:QC_classical} can be written as:
        \begin{align} \label{eq:channel_op}
         &\eta^{YE_1E_2 X_1X_2} := \sum_{\vec{x},y} \left[q_{Y|\vec{X}}(y|\vec{x}) p_{X_1}(x_1)p_{X_2}(x_2)\ketbra{y}^Y \right.\notag\\
         & \qquad \qquad \qquad \left. \otimes \varphi_{x_1}^{E_1} \otimes \varphi_{x_2}^{E_2} \otimes \ketbra{x_1}^{X_1} \otimes \ketbra{x_2}^{X_2} \right]
         \end{align}  
         with an extension:
         \begin{align}
         &\tau^{YE_1E_2 X_1X_2U_1U_2}:= \notag\\
         &\sum_{\vec{x},\vec{u},y} \left[p_{Y|\vec{U}}(y|\vec{u}) p_{X_1}(x_1)p_{U_1|X_1}(u_1|x_1)p_{X_2}(x_2) \right.\notag\\
         & \quad p_{U_2|X_2}(u_2|x_2) \ketbra{y}^Y \otimes \varphi_{x_1}^{E_1} \otimes \varphi_{x_2}^{E_2} \otimes \ketbra{x_1}^{X_1} \otimes  \notag\\
         & \quad \left. \ketbra{x_2}^{X_2} \otimes \ketbra{u_1}^{U_1} \otimes \ketbra{u_2}^{U_2}\right] \mbox{ such that } \notag \\
    &\tau^{E_1E_2X_1X_2Y}= \eta^{E_1E_2X_1X_2Y} \notag\\
         &\Rightarrow p_{X_1,X_2,Y}=p_{X_1}p_{X_2}q_{Y|X_1,X_2}\;. \label{eq:simul_eqaulity}
         \end{align}
         where \eqref{eq:simul_eqaulity} holds due to the block diagonal structure of the CQ states $\eta$ and $\tau$.
         Validity of the simulation constraint:
         Let the overall final state of the protocol be: \begin{align} \label{eq:tau'}
         &\tilde \tau^{YE_1E_2 X_1X_2U_1U_2}:=\sum_{\vec{x},\vec{u},y} \left[p_{Y|\vec{U}}(y|\vec{u}) \tilde p_{X_1, U_1}(x_1,u_1) \right. \notag\\
         & \tilde p_{X_2,U_2}(x_2,u_2) \ketbra{y}^Y \otimes \varphi_{x_1}^{E_1} \otimes \varphi_{x_2}^{E_2} \otimes \ketbra{x_1}^{X_1} \otimes\notag \\
         & \qquad \qquad ~\left. \ketbra{x_2}^{X_2} \otimes \ketbra{u_1}^{U_1} \otimes \ketbra{u_2}^{U_2} \right].
         \end{align}  
         We can now apply triangle inequality to obtain the following bound:
         \begin{align} \label{eq:qc_simul}
         &\norm{\tilde \tau^{\;E_1E_2X_1X_2Y}-\eta^{E_1E_2X_1X_2Y}}_{tvd} \notag\\
         &\leq \norm{\tilde \tau^{E_1E_2X_1X_2Y}-\tau^{E_1E_2X_1X_2Y}}_{tvd} \notag\\
         & \quad+\norm{\tau^{E_1E_2X_1X_2Y}-\eta^{E_1E_2X_1X_2Y}}_{tvd}  \notag \\
         & \overset{(i)}{=} \norm{\tilde \tau^{E_1E_2X_1X_2Y}-\tau^{E_1E_2X_1X_2Y}}_{tvd} \notag\\
         &\overset{(ii)}{\leq} \epsilon_1+\epsilon_2\;,
         \end{align}
         where (i) follows from \eqref{eq:simul_eqaulity} and $(ii)$
         holds due to the following analysis:
         \begin{align*}
            &\norm{\tilde \tau-\tau}_{tvd} = \notag\\
            &\bigg\lVert \mathop{\Sigma}\limits_{y,\vec{u},\vec{x}} p_{Y|\vec{U}}(y|\vec{u}) \bigg[ \tilde p_{X_1,U_1}(x_1,u_1) \tilde p_{X_2,U_2}(x_2,u_2)- \notag\\
            & \quad p_{X_1,U_1}(x_1,u_1)p_{X_2,U_2}(x_2,u_2) \bigg]  \ketbra{y} \otimes \ketbra{u_1}\\
             & \quad \otimes \ketbra{u_2} \otimes \ketbra{x_1} \otimes \ketbra{x_2} \otimes \varphi_{x_1}^{E_1} \otimes \varphi_{x_2}^{X_2}\bigg\rVert_{tvd}\\
        &=\frac{1}{2} \mathop{\Sigma}\limits_{y,\Vec{u},\vec{x}} p_{Y|\Vec{U}}(y|\vec{u}) \left|\bigg[ \tilde p_{X_1,U_1}(x_1,u_1) \tilde p_{X_2,U_2}(x_2,u_2)- \right. \\
        & \qquad \qquad \qquad \left. p_{X_1,U_1}(x_1,u_1)p_{X_2,U_2}(x_2,u_2) \bigg]\right| \\
        & =\frac{1}{2} \mathop{\Sigma}\limits_{\Vec{u},\vec{x}} \left|\bigg[ \tilde p_{X_1, U_1}(x_1,u_1) \tilde p_{X_2,U_2}(x_2,u_2)- \right. \notag \\
        & \qquad \qquad \qquad  \left. p_{X_1,U_1}(x_1,u_1)p_{X_2,U_2}(x_2,u_2) \bigg]\right| \\
        & \leq \frac{1}{2} \mathop{\Sigma}\limits_{\Vec{u},\vec{x}} \left[\left|\tilde p_{X_1,U_1}(x_1,u_1) \tilde p_{X_2,U_2}(x_2,u_2)- \right. \right. \notag\\
        & \quad \left. \left. p_{X_1,U_1}(x_1,u_1) \tilde p_{X_2,U_2}(x_2,u_2)\right| + \left|p_{X_1,U_1}(x_1,u_1) \times \right. \right. \\
        & \quad \left. \left. \tilde p_{X_2,U_2}(x_2,u_2)-p_{X_1,U_1}(x_1,u_1) p_{X_2,U_2}(x_2,u_2)\right| \right] \\
        & \overset{(a)}{\leq} \frac{1}{2} \mathop{\Sigma}\limits_{\Vec{u},\vec{x}} \left[|\tilde p_{X_1,U_1}(x_1,u_1)-p_{X_1,U_1}(x_1,u_1)| \times \right.\\
        & \qquad \qquad  |\tilde p_{X_2,U_2}(x_2,u_2)| + |p_{X_1,U_1}(x_1,u_1)| \times\\
        & \qquad \qquad  \left. | \tilde p_{X_2,U_2}(x_2,u_2)-p_{X_2,U_2}(x_2,u_2)| \right] \\
        &\overset{(b)}{\leq}  \epsilon_1+ \epsilon_2 \leq \epsilon,
         \end{align*}
         where $(a)$ holds by triangle inequality and $(b)$ holds by using \eqref{eq:appx_comp}.
        
         We have thus shown that the protocol with $\cE_1 \otimes \cE_2=$ $(\cE_{1,\textrm{meas. comp.}} \circ \cC^{X_1 \to U_1}) \otimes (\cE_{2,\textrm{meas. comp.}} \circ \cC^{X_2 \to U_2})$ as encoders and $\cD^{X_1X_2 \to Y}:=p_{Y|U_1,U_2} \circ (\cD_{1,\textrm{meas. comp.}} \otimes \cD_{2,\textrm{meas. comp.}})$ as the decoder, is an $(R_1,R_2,\epsilon_1+\epsilon_2)$-simulation code for CS-QC MAC with feedback provided \eqref{eq:rate_qc} holds.
     \end{itemize}
    \end{IEEEproof}
\subsection{Converse}
\begin{lemma} \label{lem:QCMAC_converse}
Let $\epsilon_1, \epsilon_2 \in (0,1)$ and $\max\{\epsilon_1,\epsilon_2\} \leq \frac{\epsilon}{2}$. Then, $\cR(\epsilon) \subseteq \cR^{QC-fb}_{outer}(\epsilon_1,\epsilon_2,\epsilon)$.
\end{lemma}
\begin{IEEEproof}
          Let $(\epsilon_1,\epsilon_2)$ and $\epsilon$ satisfy the conditions of the lemma. We need to show that any $(R_1,R_2,\epsilon)$ CS-QC MAC simulation protocol according to Definition \ref{def:qtask} has $(R_1,R_2) \in \cR^{QC-fb}_{outer}(\epsilon_1,\epsilon_2)$  (defined in \eqref{eq:fixed_outer_QC}). 
          Let the  output state of the channel with feedback be $\eta^{E_1E_2X_1X_2Y}$ and let $(R_1,R_2,\epsilon)$ simulation code produce the state $\tau^{\prime\;E_1E_2X_1X_2Y}$ having the form of \eqref{eq:evaluating_state} and satisfying the following simulation constraint:
          \[
            \norm{\tau^{\prime\;E_1E_2X_1X_2Y}-\eta^{E_1E_2X_1X_2Y}}_{tvd} \leq \epsilon.
          \]
          Let the respective purifications of the post-measurement state after  measurement $(\Lambda \otimes \Gamma)$ be given by $\ket{\varphi}^{E_1E'_1X_1X_1'} \otimes \ket{\psi}^{E_2E'_2X_2X_2'}$ and the shared randomness be denoted by $S_1^{A_1'\mathring A_1' } \otimes S_2^{A_2'\mathring A_2'}$. The Stinespring isometry of the encoders $V_{\cE_1}^{X'_1A_1' \to M_1 M_1' \bar{X}_1 \bar{A}_1} \otimes V_{\cE_2}^{X'_2A_2' \to M_2 M_2' \bar{X}_2 \bar{A}_2}$ creates the pure states (where $\bar{A}_j \cong A'_j,\; \bar{X}_j \cong X_j$):
          \begin{align} \label{eq:pure_encoded}
            &\ket{\nu_1'}^{E_1E'_1M_1M_1'\bar A_1 \accentset{\circ}{A'_1} A''_1\accentset{\circ}{A''_1}X_1\bar{X}_1}:= \notag \\  &\mathop{\Sigma} \limits_{x_1,m_1,s_1} \hspace{-10 pt} \sqrt{p_{M_1|X_1,S_1}(m_1|x_1,s_1)p_{X_1}(x_1)p_{S_1}(s_1)} \ket{m_1m_1}^{M_1M_1'} \notag\\
            & \quad \otimes \ket{x_1x_1}^{X_1\bar{X}_1} \otimes  \ket{s_1s_1}^{\bar A_1A_1''} \ket{s_1s_1}^{\accentset{\circ}{A'_1}\accentset{\circ}{A''_1}} \ket{\varphi_{x_1}}^{E_1E'_1}\\
             & \ket{\nu_2'}^{E_2E'_2M_2M_2'\bar A_2 \accentset{\circ}{A'_2} A_2''\accentset{\circ}{A_2''}X_2\bar{X}_2}:= \notag\\ &\mathop{\Sigma} \limits_{x_2,m_2,s_2} \hspace{-10 pt} \sqrt{p_{M_2|X_2,S_2}(m_2|x_2,s_2)p_{X_2}(x_2)p_{S_2}(s_2)} \ket{m_2m_2}^{M_2M'_2} \notag\\
             & \quad \otimes \ket{x_2x_2}^{X_2\bar{X}_2} \otimes \ket{s_2s_2}^{\bar A_2A_2''}\ket{s_2s_2}^{\accentset{\circ}{A'_2}\accentset{\circ}{A''_2}}  \ket{\varphi_{x_2}}^{E_2E_2'}. 
        \end{align}
        In summary, the encoders produce the following output states:
              \begin{align} \label{eq:tx}
            &{\nu_1'}^{E_1M_1 \accentset{\circ}{A'_1}} = \notag\\ 
            &\mathop{\Sigma} \limits_{x_1,m_1,s_1} {p_{M_1|X_1,S_1}(m_1|x_1,s_1)p_{X_1}(x_1)p_{S_1}(s_1)} \ketbra{m_1}^{M_1} \notag\\
            & \quad \otimes \ketbra{s_1}^{\accentset{ \circ}{A'_1}} \otimes {\varphi}_{x_1}^{E_1},\\
             &{\nu_2'}^{E_2M_1 \accentset{\circ}{A'_2}}= \notag \\
             &\mathop{\Sigma} \limits_{x_2,m_2,s_2} {p_{M_2|X_2,S_2}(m_2|x_2,s_2)p_{X_2}(x_2)p_{S_2}(s_2)} \ketbra{m_2}^{M_2} \notag\\
             & \quad \otimes 
             \ketbra{s_2}^{\accentset{ \circ}{A'_2}} \otimes {\varphi}_{x_2}^{E_2}.
        \end{align}
        Now, similar to the converse part of Theorem~\ref{thm:classical_MAC_main} in Lemma~\ref{lem:classical_converse}, we will identify the auxiliary random variables $U_j$ from the classical message $M_j$ and parts of shared randomness. In order to do so, we will curtail the states $\nu_j'$ in their eigenbasis, so that the resultant state has a similar CQ form and is appropriately close to $\nu'$. We now give the formal description of this intuition.\\
        For every $\vec{x}$, we now define the  `bad' set $\cC_{\vec{x}}$ as the complement of the following set:
        \begin{align} \label{eq:badset_QC}
         \bar{\cC}_{\vec{x}}:=\left\{(\vec{m},\vec{s}) :p'_{M_j|S_j,X_j}(m_j|s_j,x_j) \geq \frac{\epsilon_j}{|\cM_j|}, j=1,2 \right\}.
     \end{align}
     Note that this is very similar to the set defined in \eqref{eq:goodset_classical} for the classical case. Henceforth, the analysis is almost fully classical from this step, except that the rates will be evaluated with respect to the CQ state $\tau$ to be identified in \eqref{eq:CQ_tau}.\\ 
    Consider the following joint distribution on $\vec{X},\vec{M},\vec{S},Y$:
     \begin{align} \label{eq:p'_tau}
         &p'_{\vec{X},\vec{S},\vec{M},Y}(\vec{x},\vec{s},\vec{m},y):= \notag\\
         &\begin{cases}
         \mathop{\bigotimes} \limits_{j=1}^2 \frac{p_{X_j}(x_j)p_{M_j,S_j|X_j}(m_j,s_j|x_j)}{\bP_{p}(\bar{\cC}_{x_j})}P_{Y|\vec{M},\vec{S}}(y|\vec{m}\vec{s}), & (\vec{m},\vec{s}) \in \bar \cC_{\vec{x}} \\
         0, & else\;.
         \end{cases}
    \end{align}
        Now using the definitions from equations~\eqref{eq:badset_QC}, \eqref{eq:p'_tau} we identify the classical auxiliary random variable for every $x_1$ as $U_1=(M_1,S_1)\one_{\bar{\cC}_{x_1}} \sim \frac{p_{U_1|X_1}(u_1|x_1)\one_{\bar \cC_{x_1}}}{\bP_{p}(\bar \cC_{x_1})}$ and $U_2=(M_2,S_2)\one_{\bar{\cC}_{x_2}} \sim \frac{p_{U_2|X_2}(u_2|x_2)\one_{\bar \cC_{x_2}}}{\bP_{p}(\bar \cC_{x_2})}$. We thus define the encoded states $\nu_1, \nu_2$ with $U_1,U_2$ as:  
        \begin{align*}
            &\nu_1^{ E_1U_1A_1''X_1} := \\
            &\mathop{\Sigma} \limits_{\genfrac{}{}{0 pt}{2}{x_1 \in \cX_1,}{ (m_1,s_1) \in \bar{\cC}_{x_1}}} \hspace{-0.3cm} p_{X_1}(x_1)p_{S_1}(s_1) p'_{M_1|S_1,X_1}(m_1|s_1,x_1) \ketbra{x_1}^{X_1} \otimes \\
            & \quad \ketbra{m_1,s_1}^{U_1} \otimes 
             \ketbra{s_1}^{A_1''} 
             \otimes \varphi_{x_1}^{E_1}\\
             &=\mathop{\Sigma} \limits_{x_1, u_1} p_{X_1}(x_1)p'_{U_1|X_1}(u_1|x_1) \ketbra{x_1}^{X_1} \otimes  \ketbra{u_1}^{U_1} \otimes \\
             & \quad \ketbra{s_1}^{A_1''} 
             \otimes \varphi_{x_1}^{E_1}\\
             &\nu_2^{ E_2U_2A_2''X_2} := \\
             &\mathop{\Sigma} \limits_{\genfrac{}{}{0 pt}{2}{x_2 \in \cX_2,} {(m_2,s_2) \in \bar{\cC}_{x_2}}} \hspace{-0.3cm} p_{X_2}(x_2)p_{S_2}(s_2) p_{M_2|S_2,X_2}(m_2|s_2,x_2)  \ketbra{x_2}^{X_2} \\
             & \quad \otimes  \ketbra{m_2,s_2} ^{U_2} \otimes \ketbra{s_2}^{B_2''}
              \otimes \psi_{x_2}^{E_2}\\
              &=\mathop{\Sigma} \limits_{x_2, u_2}  p_{X_2}(x_2)p'_{U_2|X_2}(u_2|x_2) \ketbra{x_2}^{X_2} \otimes  \ketbra{u_2}^{U_2} \otimes \\ 
             & \quad \ketbra{s_2}^{A_2''} 
             \otimes \varphi_{x_2}^{E_2}
             \end{align*}
             The probability of the set $\cC_{\vec{x}}$  is upper bounded  by $\epsilon_1+\epsilon_2$ similar to \eqref{eq:unionbnd}.
    Hence, $\bP(\bar \cC_{\vec{x}}) \geq 1-(\epsilon_1+\epsilon_2)$.
             We now show that $\norm{\nu_j-\nu'_j}_{tvd} \leq \epsilon_j$ for $j \in \{1,2\}$. For this, first note that for every $\vec{x}$ the set $\bar \cC_{\vec{x}}$ can be seen as Cartesian product of the sets $\bar \cC_{x_j} := \left\{ (m_j,s_j): p_{M_j,S_j}(m_j,s_j) > \frac{\epsilon_j}{|\cM_j|} \right\}$. Also, $\bP_p(\bar \cC_{x_j}) \geq 1-\epsilon_j$. Thus:
             \begin{align} \label{eq:nu_prime-nu}
             &\norm{\nu'_j-\nu_j}_1= \notag\\
             & \sum_{x_j} \sum_{(m_j,s_j) \in \bar \cC_{x_j}} \bigg\lvert p_{X_j}(x_j)\bigg[\frac{p_{M_j,S_j|X_j}(m_j,s_j|x_j)}{\bP_p(\bar \cC_{x_j})} -  \notag\\
             & \quad p_{M_j,S_j|X_j}(m_j,s_j|x_j)\bigg]\bigg\lvert+ \notag \\
             &\quad \sum_{x_j} \sum_{m_j,s_j \in  \cC_{x_j}} p_{X_j}(x_j)\left[p_{M_j,S_j|X_j}(m_j,s_j|x_j)\right] \notag\\
            &= \sum_{x_j} p_{X_j}(x_j) \left[\bP_{p}(\bar \cC_{x_j}) \left( \frac{1}{\bP_{{p}}(\bar \cC_{x_j})}-1\right) + \bP_{p}(\cC_{x_j}) \right]\notag \\
            & = 2 \sum_{x_j} p_{X_j}(x_j) \bP_{p}(\cC_{x_j}) \leq  2\epsilon_j, \notag \;\\
            & \Rightarrow \norm{\nu'_j-\nu_j}_{tvd} \leq \epsilon_j \;. 
    \end{align}
            Similar to \eqref{eq:pure_encoded}, we  define the  purifications of the states $\nu_j'$  as follows:
             \begin{align}
                 &\ket{\nu_j'}^{E_jE'_j M_jM_j'\bar{A}_j \accentset{\circ}{A'_j} A''_j\accentset{\circ}{A''_j}X_j\bar{X}_j}:= \notag\\
                 & \sum_{x_j}\sum_{(m_j,s_j) \in \bar{\cC}_{x_j}} \left[ \sqrt{p'_{M_j|X_j,S_j}(m_j|x_j,s_j)p_{X_j}(x_j)p_{S_j}(s_j)}  \right. \notag\\
             & \quad \ket{m_jm_j}^{M_jM_j'}  \otimes \ket{x_jx_j}^{X_j \bar{X}_j} \otimes \ket{s_js_j}^{\bar{A}_j A''_j} \otimes \notag\\
             & \quad \left. \quad \ket{s_js_j}^{\accentset{\circ}{A'_j}\accentset{\circ}{A''_j}} \otimes \ket{\varphi_{x_j}}^{E_jE'_j}  \right],\notag\\
              &  \Rightarrow \ket{\nu_j'}^{E_jE'_jU_jU_j' \accentset{\circ}{A'_j}\accentset{\circ}{A''_j} X_j\bar{X}_j}:= \notag\\ &\sum_{x_j}\sum_{( m_j,s_j) \in \bar{\cC}_{x_j}} \sqrt{p'_{U_j|X_j}(u_j|x_j)p_{X_j}(x_j)}\ket{u_ju_j}^{U_jU_j'} \notag\\ 
             & \quad \otimes \ket{x_jx_j}^{X_j \bar{X}_j} \ket{s_js_j}^{\accentset{ \circ}{A'_j}\accentset{\circ}{A''_j}} \ket{\varphi_{x_j}}^{E_jE'_j}
             \end{align}
             The overall state   after the action of the decoder   acting on $\ket{\nu_1}\otimes \ket{\nu_2}$ is denoted (after tracing out all but $E_1E_2U_1U_2X_1X_2Y$ subsystems) by  $
             \tau^{ E_1E_2U_1U_2X_1X_2Y}$:
             \begin{align} \label{eq:CQ_tau}
            &\tau^{E_1E_2U_1U_2X_1X_2Y}  := \notag\\
            & \cD^{U_1U_2 \to Y} (\nu_1 \otimes \nu_2) \notag\\
            &= \sum_{\vec{x},y} \sum_{\vec{m},\vec{s} \in \bar{\cC}} \left[ p_{\vec{X}}(\vec{x}) \frac{p_{\vec{M},\vec{S}|\vec{X}}(\vec{m},\vec{s}|\vec{x})}{\bP_p(\bar{\cC})}p_{Y|\vec{M},\vec{S}}(y|\vec{m},\vec{s}) \ketbra{y}^Y \right.\notag\\
            & \quad \otimes \ketbra{x_1}^{X_1} \otimes \ketbra{x_2}^{X_2} \otimes \otimes \ketbra{m_1,s_1}^{U_1} \notag\\ 
            & \quad  \otimes \ketbra{m_2,s_2}^{U_2} \otimes \varphi_{x_1}^{E_1} \otimes \varphi_{x_2}^{E_2} \bigg]\notag
             \end{align}
             We thus have that:
             \begin{align}
                 &\norm{\tau^{' E_1E_2X_1X_2Y}-\tau^{E_1E_2X_1X_2Y}}_{tvd}\\
                 &= \left\lVert \cD \left(\nu_1^{'E_1X_1M_1M_1'\accentset{\circ}{A'_1}\accentset{\circ}{A''_1}} \otimes \nu_2^{'E_2X_2M_2M_2'\accentset{\circ}{A'_2}\accentset{\circ}{A''_2}} \right. \right. \notag\\
                 & \quad \left. \left.-\nu_1^{E_1X_1M_1M_1'\accentset{\circ}{A'_1}\accentset{\circ}{A''_1}} \otimes \nu_2^{E_2X_2M_2M_2'\accentset{\circ}{A'_2}\accentset{\circ}{A''_2}} \right) \right\rVert_{tvd} \notag\\
                 &\leq \norm{\nu'_1 \otimes \nu'_2-\nu_1 \otimes \nu_2}_{tvd} \notag\\
                 &\overset{(a)}{\leq} \sum_{j=1}^2 \norm{\nu'_j-\nu_j}_{tvd} \notag\\
                 &\overset{(b)}{\leq} \epsilon_1+ \epsilon_2,
             \end{align}
             where $(a)$ holds because of triangle inequality and  $(b)$ follows from \eqref{eq:nu_prime-nu}.
             Finally, using the simulation constraint\\
             $\norm{\tau^{\prime\;E_1E_2X_1X_2Y}-\eta^{E_1E_2X_1X_2Y}}_{tvd} \leq \epsilon$ and the triangle inequality again, we get that
             \[
                \norm{ \tau^{E_1E_2X_1X_2Y}-\eta^{E_1E_2X_1X_2Y}}_{tvd} \leq  (\epsilon_1+\epsilon_2) + \epsilon \leq 2\epsilon.
             \]
Thus, the identified auxiliary states satisfy $(\tau^{E_1X_1U_1},\tau^{E_2X_2U_2}) \in \cA_\epsilon^{outer}
$ (defined in \eqref{eq:fixed_A_outer_QC}).

We now evaluate the rate of the code. First we notice that $\tau^{E_jU_j}=\tau^{E_jA''_jM_j}=\nu_j^{E_jA''_jM_j}$ and $\tau^{\prime\;E_jA_j''M_j}=\nu_j^{' E_jA''_jM_j}$ for $j \in \{1,2\}$. Hence, we have 
\begin{align}
    \tau^{E_jU_j} = \nu_j^{E_jA''_jM_j} &\overset{(a)}{\leq} |\cM_j| \left(\nu_j^{E_j} \otimes \tilde \nu_j^{A''_jM_j}\right), \label{eq:op_ineq}
\end{align}
where $(a)$ holds due to
\begin{align*}
 &\nu_j^{E_jM_jA_j''}= \\
 &\sum_{x_j}p_{X_j}(x_j) \varphi_{x_j}^{E_j} \otimes   \sum_{(m_j,s_j) \in \bar \cC_{x_j}}p_{S_j}(s_j)p_{M_j|X_j,S_j}(m_j|x_j,s_j) \\
 & \quad \ketbra{m_j}^{M_j} \otimes \ketbra{s_j}^{A_j''}  \notag \\
 & \leq \sum_{x_j}p_{X_j}(x_j) \varphi_{x_j}^{E_j} \otimes   \sum_{m_j,s_j }p_{S_j}(s_j) \ketbra{m_j}^{M_j} \otimes \ketbra{s_j}^{A_j''}.
\end{align*}
In the above, we used 
\begin{align}
   &  p_{M_j|S_j,X_j}(m_j|s_j,x_j) \leq 1, ~\forall (x_j,s_j,m_j), \notag \\
   \mbox{ and } & \tilde \nu_j^{A''_jM_j} := \sum_{m_j,s_j}\frac{p_{S_j}(s_j)}{|\cM_j|} \ketbra{m_j}^{M_j} \otimes \ketbra{s_j}^{A_j''}. \notag
\end{align}
Hence, from \eqref{eq:op_ineq} and Definition~\ref{def:smooth_Imax_quantum} of the quantum smoothed $\max$-mutual information, we have 
$$ R_j= \log |\cM_j| \geq I_{\max}^{\epsilon_j}(E_j;U_j)_{\tau} .$$ 
We also need to combine Lemma~\ref{lem:cardinality_QC} to get this result. We now finally state the cardinality bounds of $\cU_1, \cU_2$ as Lemma~\ref{lem:cardinality_QC} below. This completes the proof of the converse.
\end{IEEEproof}

\begin{lemma} \label{lem:cardinality_QC}
    The cardinalities of $\{\cU_1,\cU_2\}$ for $\cR_{outer}^{QC-fb}$ can be upper bounded as:
    \begin{align} \label{eq:cardinality_QC}
        |\cU_j| \leq |\cX_1||\cX_2||\cY|; \quad \text{ for } j\in \{1,2\}\;.
    \end{align}
\end{lemma}
The proof  is very similar to that of Lemma~\ref{lem:cardinality-fixed_ip}, as $U_1, U_2$ are classical, and is given in Appendix~\ref{app:cardinality_QC}.

\subsection{Asymptotic iid Expansion}

    We now give the asymptotic iid characterization of the rate region for simulating a CS-QC MAC with feedback.
\begin{corollary} \label{cor:QC_iid}
         Consider the classical scrambling QC-MAC $\cN^{A_1A_2 \to YX_1X_2}$ with feedback, given by \eqref{eq:QC_classical} and inputs $\rho_1^{A_1} \otimes \rho_2^{A_2}$ with their respective purifications of form $\ket{\varphi_1}^{E_1 A_1}\otimes \ket{\varphi_2}^{E_2 A_2}$. 
         The rate region for simulating $\cN^{A_1A_2 \to YX_1X_2}$ using infinite shared randomness between each sender-receiver pair and classical communication over links of $(R_1,R_2)$ is given by:
             \begin{align} \label{eq:classical_QC_converseIID}
     &\cR_{iid}^{QC-fb}=cl\Biggl\{\bigcup (R_1,R_2):    R_j \geq I(E_j;U_j)_\tau\Biggr\},
    \end{align}
    where the union is taken over all the states $\tau^{E_1E_2X_1X_2U_1U_2Y}$ satisfying the Markov condition $(E_1,E_2) \to (X_1,X_2) \to (U_1,U_2) \to Y$  and $\tau^{E_1E_2X_1X_2Y}=\eta^{E_1E_2X_1X_2Y}$ with $|\cU_1|,|\cU_2| \leq |\cX_1||\cX_2||\cY|$.
\end{corollary}

\begin{IEEEproof}
\textbf{Asymptotic iid Inner Bound: }The  one-shot inner bound can be straight away extended to obtain the optimal asymptotic iid rate region. Let $(R_1,R_2) \in  \cR_{iid}^{QC-fb}$  be such that for any $\zeta>0$, 
\begin{align} \label{eq:limit_CQ}
    R_j \geq I(E_j;U_j)_\tau +\zeta, 
\end{align}
    for some $\tau_{E_1,E_2,X_1,X_2,U_1,U_2,Y}$ having the form described by \eqref{eq:evaluating_state} and satisfying $ \tau_{E_1,E_2,Y} = \eta_{E_1,E_2,Y}$. Consider the following $n$-letter iid extension of $\tau$, defined as $\tau^{(n)}_{E_1 E_2 X_1 X_2 U_1 U_2 Y}:=\tau^{\otimes n}_{E_1 E_2 X_1 X_2 U_1 U_2 Y}$. Note that 
\begin{align}\label{eq:sim_iid_inner}
    \tau^{(n)}_{E_1E_2Y} = \tau_{E_1E_2Y}^{\otimes n}=\eta_{E_1E_2Y}^{\otimes n}.
\end{align}
Now, the AEP for the smoothed $\max$-mutual information (see \eqref{eq:AEP_quantum} of Fact~\ref{fact:AEP_fixed_ip})  yields
    \[
        \lim_{n \to \infty} \frac{1}{n} \left[ I_{\max}^{\epsilon_j-\delta}(E_j^n,U_j^{(n)})_{\tau^{(n)}} +2 \log\left( \frac{1}{\delta} \right)  \right]=I(E_j;U_j)_\tau,
    \]
    which by \eqref{eq:limit_CQ} means that 
    \begin{align}
        nR_j \geq I_{\max}^{\epsilon_j-\delta}(E_j^n,U_j^{(n)})_{\tau^{(n)}} + 2\log  \frac{1}{\delta}, 
    \end{align}
    for all sufficiently large $n$ (depending on $\zeta$). 
    This along with \eqref{eq:sim_iid_inner} implies
$\cR^{QC-fb}_{iid} \subseteq \left(\cR_{inner}^{QC-fb}\right)^{(n)}(\epsilon_1,\epsilon_2)$, where
 
\begin{align}
 &\left(\cR_{inner}^{QC-fb}\right)^{(n)}(\epsilon_1,\epsilon_2)= \left\{(R_1,R_2): \right.\notag\\
 &\quad \left.  nR_j \geq I_{\max}^{\epsilon_j-\delta} (E_j^n;U_j^{(n)})_{\tau^{(n)}}  + 2 \log \frac{1}{\delta}; \mbox{ for } j\in \{1,2\}\right\}.  
\end{align}       

\textbf{Asymptotic iid Outer Bound: }
In order to prove the converse, for any $\epsilon \in (0,1)$ we first define the following $\epsilon$-approximate iid region as follows:
\begin{align}\label{eq:delta_iid_converse_QC}
    \cR^{QC-fb}_{iid}(\epsilon)&:= \left\{ (R_1,R_2): R_j \geq I(E_j;U_j)_\tau, \right. \notag\\
    & \left. \forall \; \tau_{E_1,E_2,X_1,X_2,U_1,U_2,Y} \mbox{ is of  form given in } \eqref{eq:evaluating_state} \right. \notag\\
    & \quad \mbox{ s.t. } \left. \norm{\tau^{E_1,E_2,Y}- \eta^{E_1,E_2,Y}}_{tvd} \leq 2\epsilon \right\}.
\end{align}
We will now use the extension of the one-shot converse of Lemma~\ref{lem:QCMAC_converse}. In order to do so, for any $\epsilon_1, \epsilon_2, \epsilon \in (0,1)$ such that $\max\{\epsilon_1,\epsilon_2 \} \leq \epsilon/2$, let $\left(\cR_{outer}^{QC-fb}\right)^{(n)}(\epsilon_1,\epsilon_2, \epsilon)$ be the $n$-fold extension of the region $\cR_{outer}^{QC-fb}(\epsilon_1,\epsilon_2, \epsilon)$ with respect to the iid inputs $\ket{\varphi_j}_{E_jA_j}^{\otimes n}$ (for $j \in \{1,2\}$) and general auxiliary random variables   $U_j^{(n)} \sim  p_{U_j^{(n)}|X_j^n}$. Suppose $(R_1,R_2) \in \left(\cR_{outer}^{QC-fb}\right)^{(n)}(\epsilon_1,\epsilon_2, \epsilon)$ with 
\begin{align*} 
    &\tau^{(n)}_{E_1 E_2 X_1 X_2 U_1 U_2 Y}:= \notag\\ &\mathop{\Sigma}\limits_{\substack{{x_1^n,x_2^n}\\{u_1^{(n)},u_{2}^{(n)},y^n}}} \left[ p^{\otimes n}_{X_1}(x_1^n)p_{U^{(n)}_{1}|X^n_1}(u^{(n)}_1|x^n_1) \ketbra{x_1^n}_{X_1^n} \otimes \right.\\ 
    & \ketbra{u_1^{(n)}}_{U_1^{(n)}} \otimes \left(\mathop{\bigotimes}\limits_{i=1}^n\varphi_{x_{1,i}}^{E_{1,i}}\right)\otimes p^{\otimes n}_{X_2}(x_2^n) p_{U^{(n)}_{2}|X^n_2}(u^{(n)}_2|x^n_2) \\ 
    & \ketbra{x_2^n}_{X_2^n} \otimes \ketbra{u_2^{(n)}}_{U_2^{(n)}} \otimes \left(\mathop{\bigotimes}\limits_{i=1}^n \varphi_{x_{2,i}}^{E_{2,i}}\right) \otimes  \\
    & \left. p_{Y^n|U_1^{(n)},U_2^{(n)}}(y^n|u^{(n)}_1,u^{(n)}_2) \ketbra{y^n}_{Y^n} \right]
\end{align*}
be the state induced by any $n$-fold simulation code. Suppose $(R_1,R_2) \in \left(\cR_{outer}^{QC-fb}\right)^{(n)}(\epsilon_1,\epsilon_2, \epsilon)$ satisfying
\begin{align} \label{eq:tau_simcrit}
    \norm{\left({\tau^{(n)}}\right)^{E_1^nE_2^nX_1^nX_2^nY^n}- \left(\eta^{E_1E_2X_1X_2Y}\right)^{\otimes n}}_{tvd} \leq 2\epsilon.
\end{align}
Then, 
   \begin{align*}
     & nR_j \geq     I_{\max}^{\epsilon_j}(E_j^n;U_j^{(n)})_{\tau^{(n)}}  \\ &\overset{(a)}{=} I_{\max} (E_j^n;U_j^{(n)})_{\tau^{'(n)}}\\
         &\overset{(b)}{\geq} I(E_j^n;U_j^{(n)})_{\tau^{'(n)}}\\ 
         &\overset{(c)}{\geq}I(E_j^n;U_j^{(n)})_{\tau^{(n)}}-2\epsilon_j \log |\cE_j|^n-2h_2\left(\frac{\epsilon_j}{1+\epsilon_j}\right)\\
         &\overset{(d)}{\geq}nI(E_j;U_j)_{\tau_{E_jU_j}}-2 \epsilon_j \log |\cE_j|^n-2h_2\left(\frac{\epsilon_j}{1+\epsilon_j}\right),\\
         &\Rightarrow \lim_{\epsilon_j \to 0} \lim_{n \to \infty} \frac{I_{\max}^{\epsilon_j} (E^n_j;U^{(n)}_j)_{\tau^{(n)}}}{n} \geq \\
         &\lim_{\epsilon_j \to 0} \lim_{n \to \infty}  \left[\frac{nI(E_j;U_j)_{\tau_{E_jU_j}}-2 \epsilon_j \log |\cE_j|^n-g(\epsilon_j)}{n}\right] \\
         & = I(E_j;U_j)_{\tau_{E_jU_j}},
         \end{align*}
     where $(a)$ holds by choosing $\tau^{'(n)} \in \cB^{\epsilon_j}(\tau^{(n)}_{E_jU_j})$ to be the optimizer for $I_{\max}^{\epsilon_j}(E_j^n;U_j^{(n)})_{\tau^{(n)}_{E_jU_j}}$; $(b)$ holds by the fact the $I_{\max}(E;U)_{\tau'} \geq I(E;U)_{\tau'}$ for any state $\tau'_{E,U}$; $(c)$ follows from continuity of mutual information from Fact~\ref{fact:continuity_I} with $g(\epsilon_j)=2h_2\left(\frac{\epsilon_j}{1+\epsilon_j}\right)$;  
     $(d)$ follows by Proposition~\ref{prop:single_letter-QC}. 
      This implies 
     \begin{align*}
          R_j \geq I(E_j;U_j)_\tau.
    \end{align*}
Note that \eqref{eq:tau_simcrit} and  monotonicity of trace distance implies that $\norm{\tau^{E_1E_2X_1X_2Y}-\eta^{E_1E_2X_1X_2Y}}_{tvd} \leq 2\epsilon$ (see also \eqref{eq:final_tau}). Hence, we have shown that in the asymptotic iid limit:
    \begin{equation} \label{eq:R_outsubset_R_QC}
  \lim_{\epsilon_1,\epsilon_2 \to 0}  \lim_{n \to \infty}\left(\cR_{outer}^{QC-fb}\right)^{(n)}(\epsilon_1,\epsilon_2, \epsilon) \subseteq \cR^{iid}(\epsilon).
    \end{equation}
    We have thus recovered the asymptotically optimal region of \cite[Theorem~1, Theorem~3]{Kurri_MAC_simulation} up to $\delta$. Since, we also have that  cardinalities of the auxiliary random variables are bounded, we apply Proposition~\ref{fact:continuity_QC} to recover the asymptotically optimal region of Corollary~\ref{cor:QC_iid} as follows:
     \begin{align*}
        \cR_{outer}^{QC-fb}&:=\lim_{\epsilon \to 0}\lim_{n \to \infty}\left(\cR_{outer}^{QC-fb}\right)^{(n)}(\epsilon_1,\epsilon_2, \epsilon) \\
        &\subseteq \cR^{QC-fb}_{iid}=\mathop{\lim}\limits_{\epsilon \to 0} \cR^{QC-fb}_{iid}(\epsilon).
     \end{align*}
     Thus we have shown that in the asymptotic iid limit:
     \begin{align*}
        &\cR_{outer}^{QC-fb} \subseteq \cR^{iid} \subseteq \lim_{n \to \infty}\left(\cR_{inner}^{QC-fb}\right)^{(n)}(\epsilon_1,\epsilon_2) \subseteq \cR_{outer}^{QC-fb} \\
        &\Rightarrow \cR_{inner}^{QC-fb}:=\lim_{\epsilon_1,\epsilon_2 \to 0}\lim_{n \to \infty}\left(\cR_{inner}^{QC-fb}\right)^{(n)}(\epsilon_1,\epsilon_2)\\
        &= \cR^{QC-fb}_{iid} = \cR_{outer}^{QC-fb}.
     \end{align*}
\end{IEEEproof}

\section{Conclusion}\label{Sec:conclusion}

In this work, we have provided one-shot inner and outer bounds for simulating a two-user classical MAC, with  unlimited shared randomness for fixed product inputs and also universally for two independent  arbitrary inputs. Further, we have derived the corresponding generalizations to the classical scrambling QC-MAC with feedback and fixed inputs, along with proving a tight asymptotic iid rate region. There are several interesting connected open problems, e.g., characterizing the rate region of  CS-QC MAC  without feedback as well its universal rate region. 
Yet, another challenging and immediate open problem is the fixed input and universal simulation of fully quantum MACs. 
This would call for deeper understanding and interpretation of the global Markov condition $(A_1,A_2) \to (U_1,U_2) \to Y$ and how that can aid in viewing the MAC as two point-point channels. Another technical issue would be to bound the cardinalities of the quantum auxiliary systems $U_1,U_2$ , both in one-shot and asymptotic iid settings and prove a matching single-letter asymptotic iid converse.

Potential applications of our simulation techniques are in one-shot quantum multi-user rate distortion theory\cite{Distributed_measurement}, characterizing the communication complexity of computing a function across a network \cite{Nayak_comm_complexity}, remotely preparing a target quantum state between several users and analyzing the related dynamical resource theory \cite{auxiliary_Eric}, as well as simulating other general network topologies.


\section*{Acknowledgment}

The authors acknowledge support from the Excellence Cluster - Matter and Light for Quantum Computing (ML4Q), and thank Michael X. Cao and Hao-Chung Cheng for discussions. AN would like to thank Pranab Sen for several helpful discussions and Gowtham Raghunath Kurri for pointers on single-letterization of the classical MAC simulation rate region. MB acknowledges funding by the European Research Council (ERC Grant Agreement No. 948139).


\appendices


\section*{Appendix\\ \textsc{Organization}}

In the following appendices we give the key technical tools used to prove our results for the one-shot classical and classical scrambling QC-MAC simulation, along with the proofs of some of the main lemmas of this work. Although these tools are not new and has a quite exhaustive literature, we state the versions that we use in our proofs. We also give proofs of some of our key lemmas in some of the sections of the following appendices. 

\section{Rejection Sampling and Convex split lemma}\label{app:classical}

We use the version of rejection sampling inspired from \cite{HJMR} and developed in  \cite{Berta_rej_sampling}.
We first state the accept-reject technique to shape one distribution to some other distribution of interest.

\begin{fact}\mbox{\cite[Lemma~1]{Berta_rej_sampling}}\label{fact:accept-reject}
Let $Y$ be sampled from the distribution $q_Y$ and $p_Y$ denote the target distribution for sampling $Y$. Assume $p,q$ satisfy $p\ll q$. Let $M \geq 1$ be an integer. Suppose $Y_1,Y_2, \ldots, Y_M \sim q_Y$ be iid random variables. Define $ \lambda := \max_{y \in \cY} \frac{p_Y(y)} {q_Y(y)}= 2^{D_{\max}(p_Y||q_Y)}$. Then there exists an algorithm, called accept-reject or rejection sampling that either outputs a random variable $\tilde{Y} \in \{Y_1,\ldots,Y_M\}$ such that for any $\epsilon \in (0,1)$ and a large enough $M$ satisfying:
\begin{align*}
2^{-\left\{\frac{M}{2^{D_{\max}(p_Y||q_Y)}}\right\}} &\leq \epsilon \text{ it holds that}\\    
\norm{p_Y-\tilde{p}_Y}_{tvd} &\leq \epsilon,
\end{align*}
or outputs an abort message at termination of the algorithm.
\end{fact}
\begin{remark} \label{rem:rej_sampling}
  The choice of $M$ ensuring that  the  probability that the above {\em accept-reject} method aborts  is upper bounded by:
    \[
        \left(1-2^{-D_{\max}(p_Y||q_Y)}\right)^{M} \leq 2^{-\left\{\frac{M}{2^{D_{\max}(p_Y||q_Y)}}\right\}} \leq \epsilon.
    \]
    Thus, the accept-reject method outputs a sample distributed according to $p_Y$ (under no abort with probability at least $1-\epsilon$)  using $\log M \geq D_{\max}(p_Y||q_Y) + \log \log (1/\epsilon)$ trials from shared randomness. This can easily be converted to a point-to-point channel simulation achievability protocol as stated in the Fact~\ref{lem:rej_sampling} and implemented in \cite{Berta_rej_sampling}.  
\end{remark}
Using this fact we now state the specific versions of the one-shot classical point-to-point channel simulation costs for fixed input and the universal simulation criterion as the following facts:
\begin{fact} \label{lem:rej_sampling}
    Let $\epsilon>0$ and $\delta \in (0,\epsilon)$. Now let $p_{X,Y}$ denote a bipartite probability distribution and $S \sim p_S$ shared randomness between sender Alice and the receiver Bob. Alice sends a message $m(x,S) \in \mathcal{M}$ to Bob, so that Bob can generate a sample $y(m,s) \overset{\epsilon}{\sim} p_{Y|X=x}$.
    \begin{enumerate}
        \item[(i)]  An achievable rate (with an almost matching converse) for the above task is given by (by fixing the input distribution in \cite[Theorem~2]{Berta_rej_sampling}):
    $$
        R:= \log |\mathcal{M}| \geq I_{\max}^{\epsilon-\delta} (X;Y)_p + \log \log{\frac{1}{\delta}},
    $$
    and the resulting distribution $\tilde{p}_{X,Y}$ satisfies  $\bE_{p_{X}}\lVert \tilde{p}_{Y|X=x}-p_{Y|X=x}\rVert_{tvd} \leq \epsilon$.
    \item[(ii)]  An achievable rate for the above task which works independent of $p_X$ (also called universal simulation, \cite[Theorem~2]{Berta_rej_sampling}) is:
    $$
        R:= \log |\mathcal{M}| \geq I_{\max}^{\epsilon-\delta} (p_{Y|X}) + \log \log{\frac{1}{\delta}},
    $$
    and the resulting distribution $\tilde{p}_{X,Y}$ satisfies $\mathop{\max}\limits_{x}\lVert \tilde{p}_{Y|X=x}-p_{Y|X=x}\rVert_{tvd} \leq \epsilon$.
    Further an almost matching converse for the universal task is given by \cite[Theorem~4]{Berta_rej_sampling}:
    $$
        R:= \log |\mathcal{M}| \geq I_{\max}^{\epsilon} (p_{Y|X}) \;.
    $$
    \end{enumerate}
   
\end{fact}

Finally, for deriving our results for the QC-MAC in Section \ref{sec:qc}, we need a quantum analogue of rejection sampling called coherent rejection sampling or the convex split lemma (first formulated in \cite{convex_split}). Again, this is the core idea used to prove the one-shot measurement compression theorem in \cite{Anshu_compression}, which can be modified to carry out the CS-QC MAC simulation task, as we do here. We remark that the additive fudge term in the convex split lemma is $2 \log 1/\delta$ in contrast with $\log \log 1/\delta$ in the classical setting. This is because the classical rejection sampling `fine tunes' the input to be correlated with only the accepted sample from the shared randomness whereas the convex split step correlates input with all the registers of the shared randomness. We state the convex split lemma used in our proofs as the following fact:
\begin{fact}\mbox{(\cite[Corollary~2]{Anshu_compression} and \cite[Lemma~12]{AEP})} \label{fact:convex_split}
    Let $\epsilon> 0$ and  $\delta \in (0, \epsilon)$. Consider the  following states:
    \begin{align*}
        \tau^{EU}:=\sum_u p(u) \tau_u^E \otimes \ketbra{u}^U \mbox{ and } \sigma^U := \sum_u q(u) \ketbra{u}^U
    \end{align*}
    such that the $supp(\tau^U) \subseteq supp(\sigma^U)$ and $\{p(u)\}_u,\{q(u)\}_u$ are probability distributions. Further,  suppose $q$ is the distribution achieving the infimum in the definition of $I_{\max}^{\epsilon-\delta}(E;U)_\tau$ (in equation~\eqref{eq:quantum_Imax}). Let
    \begin{align*}
    &\bar \sigma^{U_1 U_2 \ldots U_n} := \\
    &\sum_{u_1,u_2, \ldots, u_n} \hspace{-8 pt}\bar q(u_1, u_2, \ldots,u_n) \ketbra{u_1u_2 \dots u_n}^{U_1 U_2 \ldots U_n} 
    \end{align*}
    be a quantum state satisfying $\bar \sigma ^{U_i} = \sigma^U$ for all $i \in [1:n]$ and $\bar{q}$ be a pairwise independent probability distribution of $U_1^{(n)}$ with each $U_i \sim q$.
    For the following  states
    \begin{align*}
    \tau^{E U_1\ldots U_n}_i &:= \sum_u p(u) \tau_u^E \otimes \ketbra{u}^{U_i} \otimes  \left( \mathop{\bigotimes}\limits_{j \neq i} \bar \sigma^{U_j} \right),\\
    \tau^{E U_1 \ldots U_n} &:=  \sum_{i=1}^n \frac{1}{n} \tau_i^{EU_1 \ldots U_n}, 
    \end{align*}
    and the value of parameter $n$ satisfying
    \begin{align*} 
        \log n \geq I_{\max}^{\epsilon-\delta}(E;U)_\tau + 2 \log \frac{1}{\delta},
    \end{align*}
    then it holds that 
    \begin{align}\label{eq:CQ_convex_split}  \norm{\tau^{EU_1\ldots U_n}-\tau^E \mathop{\bigotimes}\limits_{i=1}^n \sigma^{U_i}}_{tvd} & \leq \epsilon.
    \end{align}
\end{fact}
The above fact in \cite[Corollary~2]{Anshu_compression} proves an upper bound of $2\epsilon+\delta$ 
in \eqref{eq:CQ_convex_split}  and $\log n$ is lower bounded by $I_{\max}^{\epsilon}(E;U)_\tau$. The $2\epsilon$ term in the distance is due to a different definition of $I_{\max}^\epsilon$. More precisely, the marginal $\tau^{'E}$ of the optimal state $\tau'^{EU} \in \cB^{\epsilon}(\tau)$ for evaluating $I_{\max}^{\epsilon}(E;U)_{\tau}$ need not be the same as $\tau^E$, which is unlike our Definition~\ref{def:smooth_Imax_quantum}. Thus, by using our Definition~\ref{def:smooth_Imax_quantum} we first reduce the aforementioned distance to $\epsilon + \delta$. In addition, we eliminate the additional $\delta$ in the distance by choosing the radius of the ball used for smoothing to be $\epsilon-\delta$ instead of $\epsilon$ (in \cite[Corollary~2]{Anshu_compression}), which gives us $\log n \geq I_{\max}^{\epsilon-\delta}(E;U)_{\tau}$. The same modifications were made in \cite[Lemma~12]{AEP} when the state $\tau^{EU}$ is fully-quantum,  resulting in the similar rate expression. Even with these minor differences, the asymptotic iid limit of smoothed-$I_{\max}$ (with any of the two definitions) is $I(E;U)_\tau$ (see e.g. \cite{Berta_splitting, Tomamichel_book}). We also note that a fully quantum version of the convex split lemma with slightly different definition of $I_{\max}^{\epsilon}$ was also given in \cite{Wilde_convex_split}. However this CQ version of the convex split lemma suffices for our purpose.

\section{Cardinality bounds and single-letterization} \label{app:cardinality}

\subsection{\texorpdfstring{Cardinality of $\cU_1, \cU_2$ for Asymptotic iid Simulation}{Cardinality of cU1, cU2 for Asymptotic iid Simulation}}

The cardinality bounds of $\cU_1, \cU_2$ were proven in \cite[Lemma~5]{Kurri_MAC_simulation} using the so-called perturbation method of \cite[Claim~1]{Perturbation_2} and the support lemma of \cite[Appendix~C]{El_Gamal_Kim}. We state this result as the following fact:
\begin{fact}\mbox{\cite[Lemma~5 and Theorem~3]{Kurri_MAC_simulation}}\label{fact:support_lemma}
    Given the MAC simulation task in Definition~\ref{def:task_fixed_ip} with fixed inputs $(X_1,X_2) \sim q_{X_1} \times q_{X_2}$, suppose the cost region for simulating the MAC $q_{Y|X_1,X_2}$ is given as:
    \[
        \cR^{iid}=\left\{(R_1,R_2): R_j \geq I(X_j,U'_j)_{p'}, \text{ for }j \in \{1,2\} \right\}
    \]
    for some overall distribution\\ 
    \begin{align*}
    &p'_{X_1,X_2,U'_1,U'_2,Y}(x_1,x_2,u'_1,u'_2,y)= q_{X_1}(x_1)q_{X_2}(x_2) \times\\
    & \quad p_{U'_1|X_1}(u'_1|x_1)p_{U'_2|X_2}(u'_2|x_2)p_{Y|U'_1,U'_2}(y|u'_1,u'_2)
    \end{align*}
    such that $p_{X_1,X_2,Y}(x_1,x_2,y)=q_{X_1}(x_1)q_{X_2}(x_2)q_{Y|X_1,X_2}(x_1,x_2)$. Then the cardinalities of random variables $\cU_1',\cU_2'$ can be restricted as follows:
    \begin{align} \label{eq:iid_cardinality}
        |\cU_j'| & \leq |\cX_1||\cX_2||\cY|, \text{ for } j \in \{1,2\}.
    \end{align}
    Moreover, one can show that there exists an overall distribution $p_{X_1,X_2,U_1,U_2,Y}$, with cardinalities of $\cU_j$ as bounded above in \eqref{eq:iid_cardinality} and $p_{X_j}=q_{X_j}$ (for $j \in \{1,2\}$) such that $p_{X_1,X_2,Y}=q_{X_1,X_2,Y}$ and the cost region can then be simplified as:
    \[
        \cR^{iid}= \left\{(R_1,R_2): R_j \geq I(X_j;U_j)_p, \text{ for }j \in \{1,2\} \right\} \;.
    \]
\end{fact}


\subsection{\texorpdfstring{Cardinality of $\cU_1, \cU_2$ for One-shot Simulation}{Cardinality of U1, U2 for one-shot MAC simulation}} \label{sec:cardinality_proof}

We first state the following fact on a characterization of the $\max$-mutual information of two correlated random variables. 
\begin{fact} \mbox{\cite[Equation~77]{Lapidoth_2019}}\label{Lem:opt_Imax}
   The $\max$-mutual information between a pair of jointly distributed random variables $(X,U) \sim p_{X,U}$ is given by:
    \begin{align*}
        I_{\max}(X;U)_p &= \max_{x,u} \log \left( \frac{p_{X,U}(x,u)}{p_X(x)q_U(u)}\right)\\
        &=\log \sum_u \max_{x}  p_{U|X}(u|x) \;,
    \end{align*}
    where $q_U(u):=\frac{\mathop{\max}\limits_{x}p_{U|X}(u|x)}{\mathop{\sum}\limits_{u} \mathop{\max}\limits_{x}p_{U|X}(u|x) }$ is the optimal distribution in the Definition~\ref{def:Imax} of $I_{\max}$. 
\end{fact}

We now prove Lemma~\ref{lem:cardinality-fixed_ip} that upper bounds the cardinalities of auxiliary alphabets $\cU_1, \cU_2$.
\begin{IEEEproof}[Proof of Lemma~\ref{lem:cardinality-fixed_ip}]
    The proof is similar to the proof \cite[Lemma~5]{Kurri_MAC_simulation}. 
We show that the cardinalities of the auxiliary alphabets can be brought down to $|\cU_j| \leq |\cX_1||\cX_2||\cY|$ for $j \in \{1,2\}$. 
First of all let the outer bound (without any cardinality bounds) be denoted by $\cC$ (just for the simplicity of notation in this proof). Then the proof of bounding cardinality follows in the following two steps (same as that of \cite[Appendix~E]{Kurri_MAC_simulation}:
\begin{enumerate}
    \item To prove that:
    \begin{align} \label{eq:finiteU}
        \cC = \mbox{Closure} \left(\mathop{\bigcup}\limits_{K_1,K_2 \in \mathbb{N}}  \cC^{K_1, K_2} \right),
    \end{align}
     where $\cC^{K_1, K_2}$ is the region given by the union of rate tuple $(R_1, R_2)$ satisfying $R_j \geq I^{\epsilon}_{\max}(X_j;U_j)_p$ for some distribution 
\[
    p(\vec{x}, \vec{u},y) \coloneqq q(x_1)q(x_2)p(u_1|x_1)p(u_2|x_2)p(y|u_1, u_2) 
\]
with  $|U_1| \leq K_1$, $|U_2| \leq K_2$, and  satisfying $ \norm{p(x_1, x_2, y) - q(x_1, x_2, y)} \leq \epsilon$. 
Once this is shown, we can start the perturbation argument with $U_j$s assumed to be finite without loss of generality. The proof of \eqref{eq:finiteU} when the rate region is defined in terms of mutual information based quantities was shown in 
\cite[Appendix~E]{Kurri_MAC_simulation} in the asymptotic iid setting. i.e., when the region is given by $R_j \geq I(X_j;U_j)_p$. The proof therein is completed by a continuity argument for $I(X_j;U_j)$, i.e.,  $p_m \to p$ implies $I(X_j;U_j)_{p_m} \xrightarrow{m \to \infty} I(X_j;U_j)_p$. The proof for our one-shot setting is exactly the same,  except that we require continuity of $I_{\max}(X_j;U_j)$. 
The required continuity follows directly from \cite[Corollary~5.16]{Continuity_Imax} specialized to our classical setting (with $X_j \leftarrow A$, $U_j \leftarrow  B$,  and $ I_{\max}(X_j;U_j) \leftarrow I_{\max}(A;B)_\rho$) as follows:  
\begin{align} \label{eq:continuity}
    &|I_{\max}(X_j;U_j)_{p'}-I_{\max}(X_j;U_j)_p| \leq \log \left( 1 + \epsilon \frac{|\cX|}{p_{\min}}\right),\notag\\
    &\mbox{ for }\norm{p'-p}_{tvd} \leq \epsilon, 
\end{align}
where $p_{\min}:= \min\limits_{x_j}p(x_j)$.  Now, given \eqref{eq:finiteU}, we can approximate $\cC$ by a finite cardinality region arbitrarily well. 
\item  Given step 1 above, we now obtain an upper bound on $K_j$ via the perturbation method, first developed in \cite[Lemma~1 and 2]{Perturbation_first} and then simplified in \cite[Claim~1]{Perturbation_2}. An essential property  required to bound $|\cU_j|$ using this method is that the relevant entropic quantity ($I_{\max}$ here) under consideration should be invariant under the perturbed distribution. Along the lines of the perturbation method (\cite[Claim~1]{Perturbation_2}), we consider an optimization of the
    weighted sum $\nu_1 I_{\max}(X_1; U_1) + \nu_2 I_{\max}(X_2; U_2)$ for non-negative real numbers $\nu_1, \nu_2$ and come up with new auxiliary random variables with reduced alphabet sizes and not increasing the weighted sum, along with preserving the constraints on $p_{X_1,X_2,U_1,U_2,Y}$ from Lemma~\ref{lem:classical_converse}.

    Thus, for a given $p(x_1, x_2, u_1, u_2, y)$, consider the Lyapunov perturbation
    $L(u_1)$ and the perturbed distribution $p_\epsilon$ defined by:
    \begin{align} \label{eq:perturb}
    p_\epsilon (x_1, x_2, u_1, u_2, y) := p(x_1, x_2, u_1, u_2, y) (1 + \epsilon L(u_1)),
    \end{align}
where $\epsilon \in \mathbb{R}$ and $L$ is chosen such that $p_\epsilon$ is a valid distribution. For this, it should hold that $(1+\epsilon L(u_1)) \geq 0$ for all $u_1$, and $\Sigma_{u_1} p(u_1)L(u_1)=0$. We will further consider perturbations $L(u_1)$ satisfying 
    \begin{align} \label{eq:perturb_constraint}
        &\mathop{\bE}\limits_{U_1 \sim p(u_1)} [L(U_1)|X_1 = x_1, X_2 = x_2, Y = y] = \notag\\ 
        & \qquad \sum_{u_1} L(u_1)p(u_1|x_1, x_2, y)= 0, \; \forall x_1, x_2, y\;.
    \end{align}
    Note that the marginal distribution of $X_1, X_2$ is unchanged under the above perturbation, that is, $p_\epsilon(x_j)=p(x_j)$ for $j \in \{1,2\}$  and that $p_\epsilon (u_1|x_1)=p(u_1|x_1)(1+\epsilon L(u_1))$. 

    Equation 
    \eqref{eq:perturb_constraint} can also be seen as a linear equation  $L^T P(u|x_1,x_2,y)=0$, where $L= \{L(u_1)\}_{u_1}$ is a vector and  $[P(u_1|x_1,x_2,y)]_{|\cU_1| \times |\cX_1||\cX_2||\cY|}$ is a stochastic matrix. Thus, by the rank-nullity theorem the range space of $[P(u_1|x_1,x_2,y)]^T_{|\cU_1| \times |\cX_1||\cX_2||\cY|}$ is of dimension at most $|\cX_1||\cX_2||\cY|$ and hence a non-trivial (non-zero) perturbation exists as long as $|\cU_1| > |\cX_1||\cX_2||\cY|$. Further, for sufficiently small values of $|\epsilon|$, we also have $(1 + \epsilon L(u_1)) \geq 0$ for all $u_1$. A simple check ensures that this perturbation preserves the distribution $p(x_1, x_2, y)$. Similarly, a straightforward marginalization of \eqref{eq:perturb} ensures that the global Markov constraint $(X_1,X_2) \to (U_1,U_2) \to Y$ holds for the perturbed distribution $p_\epsilon$. 

    Now if the distribution $p(x_1, x_2, u_1, u_2, y)$ minimize $\nu_1 I_{\max}(X_1; U_1) + \nu_2 I_{\max} (X_2; U_2)$, then for any valid perturbation, the following first derivative condition must hold:
    \begin{align}\label{eq:First_derivative}
    \left.\frac{d}{d\epsilon} \left( \nu_1 I_{\max}(X_1; U_1)_{p_{\epsilon}} + \nu_2 I_{\max} (X_2; U_2)_{p_\epsilon}\right) \right|_{\epsilon=0}=0. 
    \end{align}
    We now evaluate weighted sum term under the perturbed distribution $p_\epsilon(\cdot)$. 
     Observe that $p_{\epsilon}(x_2,u_2)=p(x_2,u_2)$ results in the derivative with respect to the second term above vanishing. Hence,
    using \eqref{eq:perturb} and Fact~\ref{Lem:opt_Imax} we get 
\begin{align}\label{eq:optimal_Imax_perturb}
    I_{\max}(X_1;U_1)_{p_\epsilon}=\sum_{u_1}\max_{x_1} p_{U_1|X_1}(u_1|x_1)(1+\epsilon L(u_1))
    \end{align}
    remains unaltered due to perturbation as
    \begin{align} \label{eq:perturb_condition}
        & 0 =\frac{d}{d\epsilon}I_{\max}(X_1;U_1)_{p_\epsilon} \bigg|_{\epsilon=0} \notag\\
        & =\frac{d}{d \epsilon}\log \hspace{-3 pt}\left( \hspace{-3 pt} \sum_{u_1}\max_{x_1} p_{U_1|X_1}(u_1|x_1)(1+\epsilon L(u_1)) \hspace{-5 pt}\right) \hspace{-3 pt} \bigg|_{\epsilon=0} \notag\\
        & =  \frac{\sum_{u_1} \max_{x_1} p_{U_1|X_1}(u_1|x_1)L(u_1)}{\sum_{u_1} \max_{x_1} p_{U_1|X_1}(u_1|x_1) } \notag \\
       \Rightarrow 0 &= \sum_{u_1} \max_{x_1} p_{U_1|X_1}(u_1|x_1)L(u_1),
    \end{align}
    and finally substituting \eqref{eq:perturb_condition} in \eqref{eq:optimal_Imax_perturb} gives 
    $I_{\max}(X_1;U_1)_{p_{\epsilon}}=I_{\max}(X_1;U_1)_p$.
    We also get from \eqref{eq:perturb} $p_{\epsilon}(x_2,u_2)=p(x_2,u_2)$ and thus $I_{\max}(X_2;U_2)_{p_\epsilon}=I_{\max}(X_2;U_2)_p$. Note that from Fact~\ref{Lem:opt_Imax} \eqref{eq:First_derivative} is automatically satisfied. Now, 
    we choose $\epsilon$ such that $\min_{u_1}(1 + \epsilon L(u_1)) = 0$ (such an $\epsilon$ always exists since $\epsilon$ can be negative) and let $u_1 = u^*_1$ attain this minimum. This implies $p_\epsilon(u_1^*) = 0$, and hence we can reduce the cardinality of $\cU_1$ by $1$ or equivalently there exists a $U'_1$ such that $|\cU'_1| \leq |\cU_1|-1$ and the minimum of $\nu_1 I_{\max}(X_1;U_1)+ \nu_2 I(X_2;U_2)$ is preserved. Finally, we can proceed by induction until $|\cU_1| = |\cX_1||\cX_2||\cY|$, beyond which we are no longer guaranteed the existence of a non-trivial perturbation $L(u_1)$ satisfying \eqref{eq:perturb_constraint}.
    Hence we can restrict the cardinality to $|\cU_1| \leq |\cX_1||\cX_2||\cY|$.
    A very similar argument can be carried out for bounding the cardinality $|\cU_2| \leq |\cX_1||\cX_2||\cY|$.\\

    The same analysis can be carried out for the rate region of Lemma~\ref{lem:classical_converse} and Lemma~\ref{lem:classical_converse_universal}, where the rates are governed by smoothed $\max$-mutual information $I_{\max}^\epsilon(X_j;U_j)_p$ and the channel smoothed $\max$-mutual information $I_{\max}^{\epsilon_j}(p_{U_j|X_j})=I_{\max}^{\epsilon_j}(X_j;U_j)_{p_{X_j}p_{U_j|X_j}}$. 
    We can repeat the arguments above  by replacing the distribution $p$ with $p'$, such that $p'_{X_j,U_j} \in \cB^{\eps}(p_{X_j,U_j})$ and $p'_{X_j}=p_{X_j}$, without disturbing the global Markov property. The only modification is to  start with the Lyapunov perturbation of the distribution $p_{X_1}p'_{U_1|X_1}p_{X_2}p'_{U_2|X_2}p_{Y|U_1,U_2}$ instead of $p_{X_1}p_{U_1|X_1}p_{X_2}p_{U_2|X_2}p_{Y|U_1,U_2}$. Then, the same arguments as above can be applied by first considering $p'_{\gamma}(u_1|x_1)=(1+\gamma L(u_1))$ and showing that $I_{\max}^\epsilon(X_1;U_1)_{q_{X_1}p'_\gamma(u_1|x_1)}=I_{\max}^\epsilon(X_1;U_1)_p$. Note that under this perturbation $I_{\max}^\epsilon(X_2;U_2)_p$ remains unchanged. A similar analysis can be done by considering perturbation of $q_{X_2}p'_{U_2|X_2}(u_2|x_2) \in \cB^\epsilon(q_{X_2}p_{U_2|X_2}(u_2|x_2))$, keeping $q_{X_2}$ fixed. This shows that it suffices to choose $U_1,U_2$ with cardinality $|\cU_j| \leq |\cX_1||\cX_2||\cY|$ and the region $\cR_{outer}$ stays the same.
    \end{enumerate}
\end{IEEEproof}

\begin{remark}
    Another technique to prove the cardinality bounds of auxiliary random variables in classical information theory is {\em support lemma} (a corollary of Fenchel-Eggleston-Carath\'eodory's theorem) \cite[Appendix~C]{El_Gamal_Kim} and recently a fully quantum version of it was developed for $I_{\max}$ by \cite{auxiliary_Eric}. However, it does not suffice for our purpose. The reason being that the global Markov chain $(X_1,X_2) \to (U_1,U_2) \to Y$ cannot be preserved using support lemma. Hence, we have to use the perturbation technique as in the proof above. It is for the same reason,  the cardinality bounds for the asymptotic iid case as shown in \cite[Lemma~5]{Kurri_MAC_simulation} are also proven using the perturbation method.
\end{remark}

\subsection{\texorpdfstring{Cardinality of $\cU_1, \cU_2$ for CS-QC MAC Simulation}{Cardinality of U1, U2 for CS-QC MAC simulation}} \label{app:cardinality_QC}
We now state the following fact on a characterization of the $\max$-mutual information for a CQ state of our interest. 
\begin{lemma}\label{Lem:opt_Imax-QC}
Let  $\tau^{EU}$ be a CQ state of the form
    \[
        \tau^{EU}:= \mathop{\sum}\limits_{x,u} p(x)p(u|x) \varphi_x^E \otimes \ketbra{u}^U.
    \]
   Then, for any $r^{U}=\sum_{u} r(u) \ket{u}\bra{u}$, we have 
    \begin{align}
      &D_{\max}\left(\tau^{EU}\| \tau^{E} \otimes r^{U}\right)= \notag \\   &D_{\max}\left(r^{\star U}\| r^U\right)
       +\log \left(\sum_{u}\norm{ \sum_{x}p(u|x) A_x }_{\infty}\right),  \notag
    \end{align}
    where 
    \begin{align}
       r^{\star U}(u) & = \frac{\norm{ \sum_{x}p(u|x) A_x  }_{\infty}}{\sum_{u}\norm{ \sum_{x} p(u|x) A_x }_{\infty}} \mbox{and} \notag\\
       A_{x} & = \left(\sum_{x'} p(x') \varphi_{x'}\right)^{-\frac 12} p(x) \varphi_x \left(\sum_{x''} p(x'') \varphi_{x''}\right)^{-\frac 12}. \notag
    \end{align}
    Consequently
 \begin{align*}
        I_{\max}(E;U)_\tau &= \log \left(\sum_{u}\norm{ \sum_{x}p(u|x) A_x }_{\infty}\right).  
    \end{align*}
\end{lemma}
\begin{IEEEproof}[Proof of Lemma \ref{Lem:opt_Imax-QC}]
  We have 
    \begin{align}
        & D_{\max}\left(\tau^{EU}\| \tau^{E} \otimes r^{U}\right)  \notag\\
        & \stackrel{(a)}{=} \log \norm{\sum_u  \sum_{x} A_x \otimes \frac{p(u|x)}{r(u)}\ketbra{u}}_{\infty} \notag \\
       & \stackrel{(b)}{=}\log \max_{u}\frac{\norm{ \sum_{x} p(u|x) A_x  }_{\infty}}{r(u)} \notag \\
       & \stackrel{(c)}{=}\log \max_{u}\left(\frac{\norm{ \sum_{x}p(u|x) A_x  }_{\infty}}{\sum_{u}\norm{ \sum_{x} p(u|x) A_x }_{\infty}r(u)}\right) \notag\\
       & \quad + \log \left(\sum_{u}\norm{ \sum_{x}p(u|x) A_x }_{\infty}\right) \notag \\
       & =\log \max_{u}\left(\frac{r^{\star}(u)}{r(u)}\right)+\log \left(\sum_{u}\norm{ \sum_{x}p(u|x) A_x }_{\infty}\right)  \notag \\
       &= D_{\max}\left(r^{\star U}\|r^U \right)+ \log \left(\sum_{u}\norm{ \sum_{x}p(u|x) A_x }_{\infty}\right), \notag
    \end{align}
    where $(a)$ uses the definitions of $D_{\max}$ by computing the marginal state $\tau^{E}$;  $(b)$ is because $\{\ket{u}\}_{u}$ form an orthonormal basis of system $U$; and $(c)$ is because the scalar $\sum_{u}\norm{ \sum_{x}p(u|x) A_x }_{\infty}$ is independent of  $u$. 
\end{IEEEproof}
\begin{IEEEproof}[Proof of Lemma~\ref{lem:cardinality_QC}]
    The proof is very similar to that of the proof of Lemma~\ref{lem:cardinality-fixed_ip} since the auxiliary random variables $U_1,\;U_2$ are classical and are generated conditioned on $X_j$. 
    Note that \eqref{eq:continuity} also holds for quantum states and hence the same continuity equation is true for CQ states required for our purpose. Hence, we can assume the cardinality of $U_1$ and $U_2$ to be  finite using the continuity of $I_{\max}^{\epsilon_j}(E_j;U_j)$ (similar to step 1 in the proof of Lemma~\ref{lem:cardinality-fixed_ip}).  Next we apply the perturbation argument to bound the cardinality of the auxiliaries, similar to step 2 of Lemma~\ref{lem:cardinality-fixed_ip}.  
    To this end, consider the  perturbed state
\begin{align}\label{eq:tau_eps}
        &\tau_\epsilon^{E_jU_j}:= \notag\\
        &\mathop{\sum}\limits_{x_j,u_j}p_{X_j}(x_j) \left[ p_{U_j|X_j}(u_j|x_j) \ketbra{u_j}^{U_j} (1+\epsilon L(u_j))\right] \otimes \varphi_{x_j}^{E_j},
    \end{align}
    where $p_{U_j|X_j}$ is the conditional distribution induced by the  state $\tau$ (see \eqref{eq:evaluating_state}) used in the definition of $\cR_{outer}^{QC-fb}$.
    From Lemma \ref{Lem:opt_Imax-QC}, we have    \begin{align}\label{eq:I_max_perturb}
    I_{\max}(E_j;U_j)_{\tau_{\epsilon}}=\log \sum_{u_j}\norm{ \sum_{x_j}p(u_j|x_j)(1+\epsilon L(u_j)) A_{x_j} }_{\infty}.  
    \end{align}
   Setting the derivative with respect to $\epsilon$ to zero in the above expression gives
    \begin{align} 
        0&=\sum_{u_j}\norm{ \sum_{x_j}p(u_j|x_j)L(u_j) A_{x_j} }_{\infty}, \notag
    \end{align}
where the derivatives can be computed via the variational characterization of operator norm. 
    Substituting this in \eqref{eq:I_max_perturb} gives $I_{\max}(E_j;U_j)_{\tau_\epsilon}=I_{\max}(E_j;U_j)_\tau$.
    The rest of the proof is exactly the same as that in Section~\ref{sec:cardinality_proof}.
\end{IEEEproof}

\subsection{\texorpdfstring{Single Letterization of Asymptotic Expansion for MAC Simulation}{Single letterization of asymptotic expansion}} \label{subsec:single_letter}

Here, we prove a single-letter characterization of $\cR^{(n)}_{outer}(\epsilon_1,\epsilon_2,\epsilon)$.
Although this is a well-known technique in classical information theory, yet an argument is always needed for the task under consideration. Hence, we give a self contained proof to show that our one-shot rate region can be lifted to the asymptotic iid setting and also can be single-letterized to match the outer bound of \cite[Theorem~4]{Kurri_MAC_simulation}. 

\begin{proposition} \label{prop:single_letter}
    Given any $\epsilon \geq 0$, consider any $n$-letter simulation code ($n < \infty$) that induces the joint distribution 
    \begin{align}\label{eq:sim_iid}
    & p'_{X_1^n,U_1^{(n)},X_2^n,U_2^{(n)},Y^n}(x_1^n,u_1^{(n)},x_2^n,u_2^{(n)},y^n):= \notag\\
    &q_{X_1}^{\otimes n}(x_1^n) q_{X_2}^{\otimes n}(x_2^n) p'_{U_1^{(n)}|X_1^n}(u^{(n)}_1|x^n_1)p'_{U_2^{(n)}|X_2^n}(u_2^{(n)}|x_2^n) \notag\\
    & \qquad \qquad \times p'_{Y^n|U_1^{(n)},U_2^{(n)}}(y^n|u_1^{(n)},u_2^{(n)}),
    \end{align}
    such that 
    \begin{align} \label{eq:appx_sim}
    \norm{p'_{X_1^n,X_2^n,Y^n}-q^{\otimes n}_{X_1}q_{X_2}^{\otimes n} q_{Y|X_1,X_2}^{\otimes n}}_{tvd} \leq \epsilon.
    \end{align}
    Suppose the rate pair $(R_1,R_2)$ satisfy:
    \[
        R_j \geq \frac{1}{n}I(X^n_{j},U_j^{(n)})_{p'}, \text{ for }j \in \{1,2\}.
    \]
Then,
    \[
        R_j \geq I(X_j;U_j)_p \text{ for }j \in \{1,2\}\;,
    \]
    for some distribution $p_{X_1,U_1,X_2,U_2,Y}:=q_{X_1} q_{X_2} p_{U_1|X_1}p_{U_2|X_2}p_{Y|U_1,U_2}$ such that
    \[
        \norm{p_{X_1,X_2,Y}-q_{X_1}q_{X_2}q_{Y|X_1,X_2}}_{tvd} \leq \epsilon.
    \]
\end{proposition}
\begin{IEEEproof}
    The proof follows mostly by standard arguments. 
    An important point is to ensure that the cardinalities of the auxiliaries $(U_1,U_2)$ are bounded, which is ascertained by Fact~\ref{fact:support_lemma}.

    The proof of single-letterization is as follows (for $j \in \{1,2\}$):
    \begin{align} \label{eq:single_letter}
    R_j &\geq \frac{1}{n}I(X_j^n;U_j^{(n)})_{p'} \notag\\
    &=\frac{1}{n} \left[ H(X_j^n)_{q_{X_j}^{\otimes n}}-H(X_{j}^n|U_j^{(n)})_{p'} \right]\notag\\
    &\overset{(a)}{=}\frac{1}{n}\left[ \sum_{i=1}^n H(X_{j,i})_{q_{X_j}}-H(X_{j}^n|U_j^{(n)})_{p'} \right]\notag\\
    &=\sum_{i=1}^n \frac{1}{n} \left[ H(X_{j,i})_{q_{X_j}}-H(X_{j,i}|X_{j,1}^{i-1},U_j^{(n)})_{p'} \right]\notag\\
    &\overset{(b)}{\geq} \sum_{i=1}^n \frac{1}{n} \left[ H(X_{j,i})_{q_{X_j}}-H(X_{j,i}|U_{j}^{(n)})_{p'} \right] \notag\\
    &= \sum_{i=1}^n \frac{1}{n} I(X_{j,i};U_{j}^{(n)})_{p'}\notag\\
    &\overset{(c)}{=} \sum_{i=1}^n \frac{1}{n} I(X_{j,i};U_{j}^{(n)}|T=i)_{p'}\notag\\
    &=I(X_{j,T};U_{j}^{(n)}|T)_{p'} \notag \\
    &=H(U_j^{(n)}|T)_{\pi_{T}p'}-H(U_j^{(n)}|X_{j,T},T)_{\pi_{T}p'} \notag\\
&=H(U_j^{(n)})_{p'}-H(U_j^{(n)}|X_{j,T},T)_{\pi_{T}p'} \notag\\
    &\overset{(d)}{\geq}H(U_j^{(n)})-H(U_j^{(n)}|X_{j,T})_{\pi_{T}p'} \notag \\
    &=I(X_{j,T};U_j^{(n)})_{\pi_{T}p'},
    \end{align}
    where $(a)$ follows by iid distribution of $X_j$; $(b)$ and $(d)$ hold  by the fact that conditioning reduces  entropy; $(c)$ follows by identifying the so-called time-sharing random variable $T$, uniformly distributed on $\{1,2,\ldots,n\}$ with p.m.f. $\pi_T(i)=\frac{1}{n}$ and independent of $X_1^n,X_2^n,U_1^{(n)},U_2^{(n)},Y^n$; 

    Now we define $X_j := X_{j,T}$ and $U_j:=(U_{j}^{(n)})$, $Y_j:=Y_{j,T}$ and the joint distribution 
    \[p_{X_1,X_2,U_1,U_2,Y}:=p'_{X_{1,T}}p'_{X_{2,T}} p'_{U_{1}^{(n)}|X_{1,T}}p'_{U_{2}^{(n)}|X_{2,T}}p'_{Y_T|U_1^{(n)},U_2^{(n)}},
    \]
    where 
    \begin{align}\label{eq:identified_p}
    &p'_{X_{j,T}}=q_{X_j}, \; p'_{U_j^{(n)}|X_{j,T}}:= \sum_{i=1}^n \frac{1}{n} p'_{U_j^{(n)}|X_{j,i}}, \notag\\
    & p'_{X_{j,T},U_j^{(n)}}:= \sum_{i=1}^n \frac{1}{n} q_{X_j}p'_{U_j^{(n)}|X_{j,i}};  \mbox{ for }j=\{1,2\}  \notag \\
    &\mbox{ and } p'_{Y_T|U_1^{(n)},U_2^{(n)}}:= \sum_{i=1}^n \frac{1}{n} p'_{Y_i|U_1^{(n)},U_2^{(n)}}.
    \end{align} 
    The above defined $p$ leads to the rate constraint as
    \begin{align} \label{eq:converse_rate}
        R_j \geq I(X_j;U_j)_p
    \end{align}
    and satisfy 
    \[
        p_{X_1,X_2,U_1,U_2,Y} =  q_{X_1}q_{X_2}p_{U_1|X_1}p_{U_2|X_2}p_{Y|X_1,X_2}.
    \]
    Further, this $p$ also satisfy simulation constraint as follows (with $q_{X_1,X_2,Y}(x_1,x_2,y)=q_{X_1}(x_1)q_{X_2}(x_2)q_{Y|X_1,X_2}(y|x_1,x_2)$):
    
    \begin{align*}
    &\norm{p_{X_1,X_2,Y}-q_{X_1,X_2,Y}}_{tvd}\\
    &= \left \lVert \sum_{u_1,u_2} \left[p'_{X_{1,T}}p'_{X_{2,T}}p'_{U_{1}^{(n)}=u_1|X_{1,T}}p'_{U_{2}^{(n)}=u_2|X_{2,T}} \right. \right.\\
    & \quad \quad  \left. p'_{Y_T|U_{1}^{(n)}=u_1,U_{2}^{(n)}=u_2}\right]-q_{X_1,X_2,Y} \bigg\rVert_{tvd}\\
     &\overset{(i)}{=} \left\lVert \sum_{u_1,u_2} \left[q_{X_{1}}q_{X_{2}}p'_{U_{1}^{(n)}=u_1|X_{1,T}}p'_{U_{2}^{(n)}=u_2|X_{2,T}} \right. \right. \\
     & \quad \quad \left.  p'_{Y_T|U_{1}^{(n)}=u_1,U_{2}^{(n)}=u_2}\right]-  q_{X_1}q_{X_2}q_{Y|X_1,X_2} \bigg\rVert_{tvd}\\
     &\overset{(ii)}{=} \left\lVert \sum_{u_1,u_2} \left[q_{X_{1}}q_{X_{2}}\left( \sum_{i=1}^n \frac{1}{n} p'_{U_{1}^{(n)}=u_1|X_{1,i}} \right) \right.\right.\\
     & \quad\left. \left. \left( \sum_{j=1}^n \frac{1}{n} p'_{U_{2}^{(n)}=u_2|X_{2,j}} \right) \left( \sum_{k=1}^n \frac{1}{n} p'_{Y_k|U_{1}^{(n)}=u_1,U_{2}^{(n)}=u_2} \right)\right] \right.\\
     & \qquad \qquad \qquad \qquad - q_{X_1}q_{X_2}q_{Y|X_1,X_2} \bigg\rVert_{tvd}\\
     &\overset{(iii)}{=}     \left\lVert \sum_{u_1,u_2} \left[\left( \mathop{\bE}\limits_{T_1 \sim [1:n]} q_{X_{T_1}} p'_{U_{1}^{(n)}=u_1|X_{1,T_1}} \right) \times  \right. \right. \\
     & \quad  \quad \left( \mathop{\bE}\limits_{T_2 \sim [1:n]} q_{X_{T_2}} p'_{U_{2}^{(n)}=u_2|X_{2,T_2}} \right) \times \\
     & ~~\left. \left. \left(\mathop{\bE}\limits_{T_3 \sim [1:n]} p'_{Y_{T_3}|U_{1}^{(n)}=u_1,U_{2}^{(n)}=u_2} \right)\right] - q_{X_1}q_{X_2}q_{Y|X_1,X_2}\right\rVert_{tvd}\\
     &\overset{(iv)}{=} \left\lVert \sum_{u_1,u_2} \left[\left( \mathop{\bE}\limits_{T_1 \sim [1:n]} q_{X_{T_1}} p'_{U_{1}^{(n)}=u_1|X_{1,T_1}} \right) \times \right. \right.\\
     & \quad \quad  \left( \mathop{\bE}\limits_{T_2 \sim [1:n]} q_{X_{2,T_2}} p'_{U_{2}^{(n)}=u_2|X_{2,T_2}} \right) \times \\
     &~~\left. \left( \mathop{\bE}\limits_{T_3 \sim [1:n]} p'_{Y_{T_3}|U_{1}^{(n)}=u_1,U_{2}^{(n)}=u_2} \right)\right] - \left(\mathop{\bE}\limits_{T_1 \sim [1:n]}q_{X_{1,T_1}}\right) \times\\
     & ~~ \left. \left( \mathop{\bE}\limits_{T_2 \sim [1:n]} q_{X_{2,T_2}} \right) \left( \mathop{\bE}\limits_{T_3 \sim [1:n]} q_{Y_{T_3}|X_{1,T_3},X_{2,T_3}} \right) \right\rVert_{tvd}\\
    &\overset{(v)}{\leq} \mathop{\bE}\limits_{T_1,T_2,T_3 \underset{\mathrm{iid}}{\sim} [1:n]} \left\lVert \sum_{u_1,u_2} \left[p'_{X_{1,T_1},U_{1}^{(n)}=u_1}p'_{X_{2,T_2}U_{2}^{(n)}=u_2} \right. \right.\\
    &~~\left. \left. p'_{Y_{T_3}|U_{1}^{(n)}=u_1,U_{2}^{(n)}=u_2}\right]-  q_{X_{1,T_1}}q_{X_{2,T_2}}q_{Y_{T_2}|X_{1,T_1},X_{2,T_2}} \right\rVert_{tvd}\\
    & \leq \max_{t_1,t_2,t_3} \left\lVert \sum_{u_1,u_2}p'_{X_{1,t_1},U_{1}^{(n)}=u_1} p'_{X_{2,t_2} U_{2}^{(n)}=u_{2}} \right.\\
    &~~ p'_{Y_{t_3}|U_{1}^{(n)}=u_1,U_{2}^{(n)}=u_2}
    -q_{X_{1,t_1}}q_{X_{2,t_2}}q_{Y_{t_3}|X_{1,t_3}, X_{2,t_3}}\bigg\rVert_{tvd}\\
    & \overset{(vi)}{\leq}  \left\lVert\sum_{u_1,u_2}p'_{X^n_{1},U_{1}^{(n)}=u_1} p'_{X^n_{2},U_{2}^{(n)}=u_{2}} p'_{Y^n|U_{1}^{(n)}=u_1,U_{2}^{(n)}=u_2} \right.\\
    & \qquad \qquad \left.
    -q_{X_{1}}^{\otimes n } q^{\otimes n}_{X_{2}}q^{\otimes n}_{Y|X_1, X_2}\right\rVert_{tvd}\\
    &\overset{(vii)}{=} \norm{p'_{X^n_{1},X^n_{2},Y^n}
    -q_{X_{1},X_{2},Y}^{\otimes n}}_{tvd} \leq \epsilon, 
    \end{align*}
    where $(i)$ and $(ii)$ follows from the identified single letter distribution $p$ in \eqref{eq:identified_p}; $(iii)$ and $(iv)$ holds since $q_{X_{j,T_j}}=q_{X_j}$ and choosing $T_1, T_2, T_3$ iid according to uniform distribution on $[1:n]$ and the fact that $q_{Y_{T_3}|X_{1,T_3},X_{2,T_3}}=q_{Y|{X_1,X_2}}$; $(v)$ holds due to convexity of $tvd$; $(vi)$ follows from the monotonicity of $tvd$ and $(vii)$ holds from \eqref{eq:appx_sim}.\\
    Finally, from  Fact~\ref{fact:support_lemma},  we have that it suffices to take $U_j$ with cardinality $|\cU_j|\leq |\cX_1||\cX_2||\cY|$. 
    We have thus identified $p$ that satisfies the simulation constraint and using this $p$  in  \eqref{eq:single_letter} and from \eqref{eq:converse_rate} we obtain
    \[
        R_j \geq I(X_j;U_j )_p \;.
    \]
\end{IEEEproof}

\subsection{\texorpdfstring{Single Letterization for Asymptotic Expansion of CS-QC MAC Simulation}{Single letterization for asymptotic expansion of CS-QC MAC simulation}}

We now state and prove a proposition that ensures that CS-QC MAC simulation with feedback has a single-letter characterization.
\begin{proposition} \label{prop:single_letter-QC}
    Given any $\epsilon \geq 0$, consider any $n$-letter simulation code ($n < \infty$), that induces the state 
    \begin{align}\label{eq:sim_iid-QC}
    &\tau^{' E_1^nX_1^nU_1^{(n)}E_2^nX_2^nU_2^{(n)}Y^n}:= \notag \\
    &\sum_{\substack{{x_1^n,x_2^n}\\{u_1,u_2,y^n}}} \left[p_{X_1}^{\otimes n}(x_1^n) p_{X_2}^{\otimes n}(x_2^n) p'_{U_1^{(n)}|X_1^n} (u_1|x^n_1) p'_{U_2^{(n)}|X_2^n}(u_2|x_2^n) \right. \notag\\
    &  \quad p'_{Y^n|U_1^{(n)},U_2^{(n)}}(y^n|u_1,u_2) \ketbra{x_1^n}_{X_1^n} \otimes \ketbra{u_1}_{U_1^{(n)}}\otimes  \notag\\
    & \quad \ketbra{x_2^n}_{X_2^n} \otimes\ketbra{u_2}_{U_2^{(n)}}  \otimes \left( \varphi_{x_1}^{E_1}\right)^{\otimes n} \otimes \left( \varphi_{x_2}^{E_2}\right)^{\otimes n}  \notag\\
    & \quad \left. \otimes  \ketbra{y^n}_{Y_1^n}  \right], 
    \end{align}
    such that
    \begin{align} 
    \norm{\tau^{' E_1^nE_2^nX_1^nX_2^nY^n}-\eta^{\otimes n}_{E_1E_2X_1X_2Y}}_{tvd} \leq \epsilon. \label{eq:sim-QC1}
    \end{align}
  Suppose  the rate pair $(R_1,R_2)$ satisfy:
    \[
        R_j \geq \frac{1}{n}I(E^n_{j},U_j^{(n)})_{\tau'}, \text{ for }j \in \{1,2\}.
    \]
    Then, 
    \[
        R_j \geq I(E_j;U_j)_\tau \text{ for }j \in \{1,2\}\;,
    \]
    for some state 
    \begin{align} 
    & \tau^{E_1X_1U_1E_2X_2U_2Y}:= \notag \\
    &\sum_{\substack{{x_1,x_2}\\{u_1,u_2,y}}} \bigg[p_{X_1}(x_1) p_{X_2}(x_2) p_{U_1|X_1}(u_1|x_1) p_{U_2|X_2}(u_2|x_2) \notag\\
    & p_{Y|U_1,U_2}(y|u_1,u_2) \varphi_{x_1}^{E_1} \otimes  \varphi_{x_2}^{E_2} \otimes \ketbra{x_1}^{X_1} \otimes \ketbra{u_1}^{U_1}    \notag \\
    & \quad \otimes \ketbra{x_2}^{X_2} \otimes  \ketbra{u_2}^{U_2} \otimes \ketbra{y}^Y \bigg], \notag \\
    &\mbox{such that } \norm{\tau^{E_1E_2X_1X_2Y}-\eta^{E_1E_2X_1X_2Y}}_{tvd} \leq \epsilon.
    \end{align}
\end{proposition}
\begin{IEEEproof}
    The proof is very similar to that of Proposition~\ref{prop:single_letter}. By Lemma~\ref{lem:cardinality_QC}, we can assume that  cardinalities of $|\cU_1|,|\cU_2|$ are bounded as $|\cU_j|\leq |\cX_1||\cX_2||\cY|$. Then, we have 
    \begin{align} \label{eq:single_letter-QC}
    R_j &\geq \frac{1}{n}I(E_j^n;U_j^{(n)})_{\tau'} \notag\\
    &= \frac{1}{n} \left[ H(E_j^n)_{\tau_{E_j}^{'\;\otimes n}}-H(E_{j}^n|U_j^{(n)})_{\tau'} \right]\notag\\
    &\overset{(a)}{=}\frac{1}{n}\left[ \sum_{i=1}^n H(E_{j,i})_{\tau'_{E_{j,i}}}-H(E_{j}^n|U_j^{(n)})_{\tau'} \right]\notag\\
    &=\sum_{i=1}^n \frac{1}{n} \left[ H(E_{j,i})_{\tau'_{E_{j,i}}}-H(E_{j,i}|E_{j,1}^{i-1},U_j^{(n)})_{\tau'} \right]\notag\\
    &\overset{(b)}{\geq} \sum_{i=1}^n \frac{1}{n} \left[ H(E_{j,i})_{\tau'_{E_{j,i}}}-H(E_{j,i}|U_{j}^{(n)})_{\tau'} \right] \notag\\
    &= \sum_{i=1}^n \frac{1}{n} I(E_{j,i};U_{j}^{(n)})_{\tau'}\notag\\
    &\overset{(c)}{=} \sum_{i=1}^n \frac{1}{n} I(E_{j,i};U_{j}^{(n)}|T=i)_{\tau'}\notag\\
    & = I(E_{j,T};U_{j}^{(n)}|T)_{\tau'} \notag\\
    &= H(U_{j}^{(n)}|T)_{\tau'}-H(U_{j}^{(n)}|E_{j,T},T)_{\tau'} \notag\\
    &\overset{(d)}{\geq} I(E_{j,T};U^{(n)}_{j})_{\pi_{T}\tau'},
    \end{align}
    where $(a)$ follows by iid distribution of $E_j$; $(b)$ holds  by the fact that conditioning reduces  entropy; $(c)$ follows by identifying the so-called time-sharing random variable $T$, uniformly distributed on $\{1,2,\ldots,n\}$ with p.m.f.  $\pi_T(i)=\frac{1}{n}$, and independent of $E_1^n,E_2^n,U_1^{(n)},U_2^{(n)},Y^n$; $(d)$ holds because $U_j^{(n)} \indep T$ and is classical, hence the fact that conditioning reduces entropy is used. 

    Now we define $E_j:=E_{j,T}$,$U_j:=U_{j}^{(n)}$, $X_j:=X_{j,T}$, $Y_j:=Y_{j,T}$  ($X_{j,T}, Y_{j,T}, E_{j,T} \indep T$). Notice that the distribution of the classical systems $X_1,X_2,U_2,U_2,Y$ is defined as (similar to \eqref{eq:identified_p})
    \begin{align}\label{eq:identified_p_QC}
    &p'_{X_{j,T}}=p_{X_j}, \; p'_{U_j^{(n)}|X_{j,T}}:= \sum_{i=1}^n \frac{1}{n} p'_{U_j^{(n)}|X_{j,i}}, \notag\\
    &p'_{X_{j,T},U_j^{(n)}}:= \sum_{i=1}^n \frac{1}{n} p_{X_j}p'_{U_j^{(n)}|X_{j,i}};  \mbox{ for }j=\{1,2\}  \notag \\
    &\mbox{ and } p'_{Y_T|U_1^{(n)},U_2^{(n)}}:= \sum_{i=1}^n \frac{1}{n} p'_{Y_i|U_1^{(n)},U_2^{(n)}}.
    \end{align} 
    and the overall state is
    \begin{align} \label{eq:tau_exp} 
    &\tau^{E_1X_1U_1E_2U_2Y}:= \notag\\
    &\sum_{\substack{{x_{1},x_{2}}\\{u_{1},u_{2},y}}} \left( \mathop{\bE}\limits_{T_1 \sim [1:n]} p'_{X_{1,T_1},U_{1}^{(n)}}(x_1,u_{1}) \ketbra{x_{1}}^{X_{1}} \otimes
     \ketbra{u_{1}}^{U_1} \right.  \notag\\
     & \quad \otimes \varphi_{x_{1}}^{E_1}\bigg) \otimes \left( \mathop{\bE}\limits_{T_2 \sim [1:n]} p'_{X_{2,T_2},U_{2}^{(n)}}(x_{2},u_{2}) \ketbra{x_{2}}^{X_{2}}  \otimes \right. \notag\\
     & \left. \ketbra{u_{2}}^{U_2}  \otimes \varphi_{x_{2}}^{E_2} \right)\hspace{-3 pt} \otimes   \hspace{-3 pt}\left(\mathop{\bE}\limits_{T_3 \sim [1:n]} p'_{Y_{T_3}|U_{1}^{(n)},U_{2}^{(n)}}(y|u_{1}u_{2})\ketbra{y}^Y\right) \notag\\
     &= \mathop{\bE}\limits_{T_1,T_2,T_3 \underset{\mathrm{iid}}{\sim} [1:n]} \Biggl[ \sum_{\substack{{x_{1},x_{2}}\\{u_{1},u_{2},y}}}  p'_{X_{1,T_1},U_{1}^{(n)}}(x_1,u_{1}) p'_{X_{2,T_2},U_{2}^{(n)}}(x_{2},u_{2}) \notag\\
     & \quad p'_{Y_{T_3}|U_{1}^{(n)},U_{2}^{(n)}}(y|u_{1}u_{2}) \ketbra{x_1}^{X_1} \otimes \ketbra{x_2}^{X_2} \otimes \ketbra{u_{1}}^{U_1}   \notag\\
     & \quad  
      \otimes \varphi_{x_{1}}^{E_1} \otimes   \ketbra{u_{2}}^{U_2} \otimes \varphi_{x_{2}}^{E_2}  \otimes \ketbra{y}^Y\Biggr].
    \end{align}
    The above identified state $\tau$ leads to the rate constraints 
    \begin{align} \label{eq:QC_rate_singleletter}
        R_j \geq I(E_j;U_j)_\tau; \mbox{ for }j\in\{1,2\}.
    \end{align}
   Since $\{p_{X_{j,i}}\}_{ i=1}^{n}$ for $j=1,2$ and $\{q_{Y_i|X_{1,i},X_{2,i}}\}_{i=1}^n$ are independent and identically distributed, we can write the target state to be simulated as:
    \begin{align}\label{eq:ideal_expectation}
        &\eta^{E_1E_2X_1X_2Y}= \notag\\
        &\sum_{x_1,x_2,y}  \left( \mathop{\bE}\limits_{T_1 \sim [1:n]} p_{X_{1,T_1}}(x_1) \ketbra{x_1}^{X_1} \otimes \varphi_{x_1}^{E_1}\right)  \notag\\
        & \quad \otimes \left( \mathop{\bE}\limits_{T_2 \sim [1:n]}p_{X_{2,T_2}}(x_2) \ketbra{x_2}^{X_2} \otimes \varphi_{x_2}^{E_2}\right) \notag\\
        & \quad \otimes \left( \mathop{\bE}\limits_{T_3 \sim [1:n]}p_{Y_{T_3}|X_{1,T_3},X_{2,T_3}}(y|x_1,x_2)  \ketbra{y}^Y\right) \notag\\
        &= \mathop{\bE}\limits_{T_1,T_2,T_3 \underset{\mathrm{iid}}{\sim} [1:n]} \left[ \sum_{x_1,x_2,y} p_{X_{1,T_1}}(x_1) p_{X_{2,T_2}}(x_2) \right. \notag\\
        & \quad \left. p_{Y_{T_3}|X_{1,T_3},X_{2,T_3}}(y|x_1,x_2) \ketbra{x_1}^{X_1} \otimes \varphi_{x_1}^{E_1} \otimes \ketbra{x_2}^{X_2}  \right. \notag\\
        & \quad \otimes \varphi_{x_2}^{E_2} \otimes   \ketbra{y}^Y   \bigg]:=\eta^{E_{1,T_1}E_{2,T_2}X_{1,T_1}X_{2,T_2}Y_{T_3}}.
    \end{align}
The above defined $\tau$ satisfies the simulation constraint since:
    \begin{align}\label{eq:final_tau}
    &\norm{\tau^{E_1E_2X_1X_2Y}-\eta^{E_1E_2X_1X_2Y}}_{tvd} \\
    &\overset{(i)}{=} \left\lVert \mathop{\bE}\limits_{T_1,T_2,T_3 \underset{\mathrm{iid}}{\sim} [1:n]} \left[ \Tr_{U_1,U_2} \tau^{E_{1,T_1}E_{2,T_2}U_1U_2X_{1,T_2}X_{2,T_2}Y_{T_3}} \right. \right. \notag\\
    & \qquad \qquad \qquad- \eta^{E_{1,T_1}E_{2,T_2}X_{1,T_2}X_{2,T_2}Y_{T_3}} \bigg]\bigg\rVert_{tvd}\notag\\
    &\overset{(ii)}{\leq} \mathop{\bE}\limits_{T_1,T_2,T_3 \underset{\mathrm{iid}}{\sim} [1:n]} \left\lVert \left[ \Tr_{U_1,U_2} \tau^{E_{1,T_1}E_{2,T_2}U_1U_2X_{1,T_2}X_{2,T_2}Y_{T_3}} \right. \right. \notag\\
    & \qquad \qquad \qquad \left. \left. -\eta^{E_{1,T_1}E_{2,T_2}X_{1,T_2}X_{2,T_2}Y_{T_3}} \right]\right\rVert_{tvd}\notag\\  
    & \leq \max_{t_1,t_2,t_3} \left\lVert \left[ \Tr_{U_1,U_2} \tau^{E_{1,t_1}E_{2,t_2}U_1U_2X_{1,t_2}X_{2,t_2}Y_{t_3}} \right. \right. \notag\\
    & \qquad \qquad \qquad \left. \left. -\eta^{E_{1,t_1}E_{2,t_2}X_{1,t_2}X_{2,t_2}Y_{t_3}} \right]\right\rVert_{tvd}\notag\\  
    &\overset{(iii)}{\leq} \norm{\Tr_{U_1,U_2}\tau^{\prime E_1^nE_2^nX_1^nU_1X_2^nU_2Y^n}-\eta^{\otimes n}_{E_1E_2X_1X_2Y}}_{tvd} \notag\\
    &= \norm{\tau^{\prime E_1^nE_2^nX_1^nX_2^nY^n}-\eta^{\otimes n}_{E_1E_2X_1X_2Y}}_{tvd} \notag\\
   & \overset{(iv)}{\leq} \epsilon,
    \end{align}
    where $(i)$ follows from \eqref{eq:tau_exp} and \eqref{eq:ideal_expectation}; $(ii)$ follows from convexity of $tvd$; $(iii)$ follows from the monotonicity of $tvd$ and $(iv)$ follows from \eqref{eq:sim_iid-QC} and \eqref{eq:sim-QC1}.\\
Hence, the state $\tau$ given in \eqref{eq:tau_exp} satisfies the simulation constraint and  \eqref{eq:QC_rate_singleletter} as desired.
\end{IEEEproof}
\subsection{\texorpdfstring{Continuity of the $\epsilon$-Rate Region}{Continuity of the epsilon-rate region}}
We start with the following fact about the continuity of mutual information $I(A;B)_{\tau^{AB}}$ with respect to $\tau_{A,B}$ having a fixed marginal $\tau^A$. When both the systems $A$ and $B$ are classical, the continuity bounds of \cite[Theorem~17.3.3]{Cover_Thomas} and \cite[Lemma~4]{Gohari_continuity}) suffice. However, an improved version of the continuity of mutual information for general quantum states was proven in \cite[Lemma~2]{Winter_continuity} known as the Alicki-Fannes-Winter continuity bound, which also applies to classical bipartite distributions. We state this result as the following fact:
\begin{fact}\cite[Lemma~2]{Winter_continuity} \label{fact:continuity_I}
Let $\rho^{'AB}$ and $\rho^{AB}$ be two quantum states on the joint Hilbert space $\cH^{AB}$ such that $\rho^A=\rho'^A$.  Then for any $\epsilon \in (0,1)$, it holds that:
\begin{align*}
    \norm{\rho^{AB}-\rho^{'AB}}_{tvd} &\leq \epsilon \\
    \Rightarrow |I(A;B)_\rho-I(A;B)_{\rho'}| &\leq 2\epsilon \log |\cA| + 2h_2\left( \frac{\epsilon}{1+\epsilon}\right).
\end{align*}
\end{fact}
The above fact also holds for classical probability distributions and hence we use this instead of a similar classical continuity bound. We now state and prove that the asymptotic extension of our outer bound, that is, $\cR_{outer}^{iid}(\epsilon)$ converges to $\cR^{iid}$ as mentioned in Corollary~\ref{cor:classical_IID}. Since this is a direct consequence of \cite[Lemma~6]{Gohari_continuity}, we state it as the following fact and include the similar proof for completeness. 
\begin{fact}\mbox{\cite[Lemma~6]{Gohari_continuity}}\label{fact:continuity}
Consider the asymptotic iid setting for simulating a MAC $q_{Y|X_1,X_2}$ with inputs $q_{\vec{X}}=q_{X_1} \times q_{X_2}$ (see  Definition~\ref{def:task_fixed_ip}) for any $\epsilon \in (0,1)$. 
Let an outer bound for this task be given by:
\begin{align*}
    \cR^{iid}_{outer}(\epsilon)&:= \bigg\{ (R_1,R_2): R_j \geq I(X_j;U_j)_p, \text{ for } j\in \{1,2\}, \\
    & \qquad \left. p_{\vec{X},\vec{U},Y} \;s.t.\;\mathop{\bE}\limits_{q_{\vec{X}}}\norm{p_{Y|\vec{X}}-q_{Y|\vec{X}}}_{tvd} \leq \epsilon \right\}
\end{align*}
with $|\cU_j| \leq |\cX_1||\cX_2||\cY|$ for $j \in \{1,2\}$ then:
\[
    \cR^{iid}=\bigcap_{\epsilon>0}\cR_{outer}^{iid}(\epsilon),
\]
where 
\begin{align*}
    &\cR^{iid}:= \left\{ (R_1,R_2): \right.\\
    & \left. R_j \geq I(X_j;U_j)_p, \text{ for } j\in \{1,2\},p_{\vec{X},\vec{U},Y} \mbox{ s.t. }p_{\vec{X},Y}=q_{\vec{X},Y} \right\}. 
\end{align*}
\end{fact}
\begin{IEEEproof}
    Since any rate pair $(R_1,R_2) \in \cR^{iid}$ lies in $\cR^{iid}_{outer}(\epsilon)$ for all $\epsilon>0$, therefore it holds that $\cR^{iid} \subseteq \bigcap\limits_{\epsilon>0} \cR^{iid}_{outer}(\epsilon)$.    The reverse direction  $\bigcap\limits_{\epsilon>0} \cR^{iid}_{outer}(\epsilon) \subseteq \cR^{iid}$ is quite non-trivial, for which we sketch the steps below. 
    Consider a sequence  $\{\epsilon_i\}_{i\geq 1}$ such that $\mathop{\lim}\limits_{i \to \infty} \epsilon_i=0$. Also, let
    \begin{align*}
        \cP(\vec{r})&:=\left\{p_{\vec{X},\vec{U},Y}: q_{X_1}q_{X_2}p_{U_1|X_1}p_{U_2|X_2}p_{Y|U_1,U_2} \right. \\
        & \qquad \left. \text{ and }p_{\vec{X},Y}=q_{\vec{X},Y} \text{ with } \{|\cU_j|\}_{j=1}^2 \leq |\cX_1||\cX_2||\cY|\right\}.\\
        \cP_{\epsilon}(\vec{r})&:=\left\{p_{\vec{X},\vec{U},Y}: q_{X_1}q_{X_2}p_{U_1|X_1}p_{U_2|X_2}p_{Y|U_1,U_2} \right.\\
        &\qquad \text{ and }\norm{p_{\vec{X},Y}-q_{\vec{X},Y}}_{tvd} \leq \epsilon \\
        & \qquad \left. \text{ with } \{|\cU_j|\}_{j=1}^2 \leq |\cX_1||\cX_2||\cY|\right\}.
    \end{align*}
    Now, we take a rate pair, say $\vec{R}^* = (R_1^*, R_2^*) \in \bigcap\limits_{\epsilon>0} \cR^{iid}_{outer}(\epsilon)$. There is a sequence of p.m.f. $p_i(\vec{x},\vec{u},y) \in \cP_{\epsilon_i}(\vec{r})$ corresponding to this rate pair. These p.m.f.s belong to the probability simplex $\cP$ of dimension $|\cX_1||\cX_2||\cU_1||\cU_2||\cY|$ and since the cardinalities of $|\cU_1|,|\cU_2|$ are bounded therefore the probability simplex is compact. Thus, there exists a subsequence $\{i_k\}_{k \geq 1}$ such that the subsequence of p.m.f. $\{p_{i_k}(\vec{x},\vec{u},y)\}_{k \geq 1}$ converges to some $p^*(\vec{x},\vec{u},y)$ in the probability simplex. \\
    The key point of the proof is that $p^*(\vec{x},\vec{u},y) \in \cP(\vec{r})$. 
    This can be proven by using the continuity of the total variation distance and the mutual information in the probability simplex. In particular, this follows from
    \begin{align*}
        &\norm{p^*(\vec{x},y)-q(\vec{x},y)}_1=\mathop{\lim}\limits_{k \to \infty}\norm{p_{i_k}(\vec{x},y)-q(\vec{x},y)}_1=0 \\
        &\Rightarrow p^*(\vec{x},y)=q(\vec{x},y)
    \end{align*}
    Furthermore, since $(X_{1i},U_{1i}) \indep (X_{2i},U_{2i})$ and the Markov condition $(X_{1i},X_{2i}) \to (U_{1i},U_{2i}) \to Y_i$ holds for all $i \geq 1$, the same   also holds in the limiting case. \\
    Finally it can also be shown that $\vec{R}^*$ is a point of $\cR^{iid}$ corresponding to the p.m.f. $p^*(\vec{x},\vec{u},y)$.  By using Fact~\ref{fact:continuity_I},  we get:
    \begin{align}\label{eq:continuity_region}
        & \norm{ p^*_{X_j,U_j}-p_{X_j,U_j} }_{tvd} \leq \epsilon \notag\\
        &\Rightarrow \left|I(X_j;U_j)_{p^*_{X_j,U_j}}- I(X_j;U_j)_{p_{X_j,U_j}} \right| \notag\\
        &\qquad \leq 2\epsilon \log (|\cX_j|) - 2h_2\left( \frac{\epsilon}{1+\epsilon}\right) := g(\epsilon).
    \end{align}
   Since $\mathop{\lim}\limits_{\epsilon \to 0} g(\epsilon) = 0$,  $\cR^{iid}_{outer}(\epsilon_{i_k})$ converges to $\cR^{iid}$. This completes the proof.
\end{IEEEproof}

\begin{proposition}\label{fact:continuity_QC}
Consider the asymptotic iid setting for simulating a CS-QC MAC with feedback denoted by $\cN^{A_1A_2 \to X_1X_2Y}$ with post-measurement inputs to the encoders as the state $\varphi_1^{E_1X_1X_1'} \otimes \varphi_2^{E_2X_2X_2'}$ (see  Definition~\ref{def:qtask}) for any $\epsilon \in (0,1)$. 
Let an outer bound for this task be given by:
\begin{align*}
    &\cR^{iid}_{outer}(\epsilon):= \left \{ (R_1,R_2): R_j \geq I(E_j;U_j)_\tau, \text{ for } j\in \{1,2\}, \right.\\
    & \qquad \quad\left. \;p_{\vec{X},\vec{U},Y} \;s.t.\;\norm{\tau^{E_1E_2X_1X_2Y}-\eta^{E_1E_2X_1X_2Y}}_{tvd} \leq \epsilon \right\}
\end{align*}
with $|\cU_j| \leq |\cX_1||\cX_2||\cY|$ for $j \in \{1,2\}$ then:
\[
    \cR^{iid}=\bigcap_{\epsilon>0}\cR_{outer}^{iid}(\epsilon),
\]
where 
\begin{align*}
    &\cR^{iid}:= \left\{ (R_1,R_2): \right.\\
    & \qquad \qquad \left. R_j \geq I(E_j;U_j)_\tau:\tau^{E_1E_2X_1X_2Y}=\eta^{E_1E_2X_1X_2Y} \right\}. 
\end{align*}
\end{proposition}
\begin{IEEEproof}
    The proof is similar to the proof of Fact~\ref{fact:continuity} owing to the block diagonal structure of the CQ state $\tau^{E_1E_2X_1X_2U_1U_2Y}$. We can essentially consider a sequence of states by keeping the basis (in which the target state $\eta$ is block diagonal) fixed and perturbing the distribution of the classical systems, see \eqref{eq:tau_prop3} and \eqref{eq:eta_prop3}. In this way the proof of Fact~\ref{fact:continuity} can be emulated for the CS-QC MAC simulation with feedback. We give a proof sketch for completeness. 
    
    Since any rate pair $(R_1,R_2) \in \cR^{iid}$ lies in $\cR^{iid}_{outer}(\epsilon)$ for all $\epsilon>0$, therefore it holds that $\cR^{iid} \subseteq \bigcap\limits_{\epsilon>0} \cR^{iid}_{outer}(\epsilon)$. \\
    To prove the reverse direction, that is,  $\bigcap\limits_{\epsilon>0} \cR^{iid}_{outer}(\epsilon) \subseteq \cR^{iid}$, consider a sequence  $\{\epsilon_i\}_{i\geq 1}$ such that $\mathop{\lim}\limits_{i \to \infty} \epsilon_i=0$. 
    Consider the target state $\eta^{E_1E_2X_1X_2Y}$ to be simulated and a CQ state $\tau$ with following block diagonal structure:
    \begin{align} \label{eq:tau_prop3}
        &\tau^{E_1X_1U_1E_2X_2U_2Y} := \sum_{\substack{{x_1,x_2}\\{u_1,u_2,y}}}  \bigg[ p_{X_1}(x_1) p_{X_2}(x_2) p_{U_1|X_1}(u_1|x_1) \notag\\
        & \quad p_{U_2|X_2}(u_2|x_2) p_{Y|U_1,U_2}(y|u_1,u_2) \varphi_{x_1}^{E_1} \otimes  \varphi_{x_2}^{E_2} \otimes  \ketbra{x_1}^{X_1} \notag \\
    &  \quad \otimes \ketbra{u_1}^{U_1} \otimes \ketbra{x_2}^{X_2} \otimes  \ketbra{u_2}^{U_2} \otimes \ketbra{y}^Y \bigg].
    \end{align} 
    \begin{align} \label{eq:eta_prop3}
    &\eta^{E_1E_2X_1X_2Y} := \sum_{{x_1,x_2,y}} \bigg[ p_{X_1}(x_1) p_{X_2}(x_2) q_{Y|X_1,X_2}(y|x_1,x_2) \notag\\
    & \quad \varphi_{x_1}^{E_1} \otimes  \varphi_{x_2}^{E_2} \otimes 
    \ketbra{x_1}^{X_1} \otimes \ketbra{x_2}^{X_2} \otimes \ketbra{y}^Y \bigg].
    \end{align}
    From \eqref{eq:tau_prop3} and \eqref{eq:eta_prop3} we have
    \begin{align}
        &\tau^{E_1E_2X_1X_2Y}= \eta^{E_1E_2X_1X_2Y} \Rightarrow p_{\vec{X},Y} = p_{\vec{X}}q_{Y|\vec{X}}; \mbox{ and } \label{eq:equiv_1}\\
        &\norm{\tau^{E_1E_2X_1X_2Y}- \eta^{E_1E_2X_1X_2Y}}_{tvd} \leq \epsilon \\
        &~\Rightarrow \norm{p_{\vec{X},Y}-p_{\vec{X}}q_{Y|\vec{X}}}_{tvd} \leq \epsilon, \label{eq:equiv_2}
    \end{align}
    where 
    \begin{align*}
    p_{X_1,X_2,Y} = \hspace{-3 pt}\sum_{u_1,u_2} \hspace{-3 pt}p_{X_1}p_{X_2}p_{U_1=u_j|X_1}p_{U_2=u_2|X_2} p_{Y|U_1=u_1,U_2=u_2}.
    \end{align*}
    Using this identification we can repeat the proof of Fact~\ref{fact:continuity} by defining the sets $\cP(\vec{r})$ and $\cP_\epsilon(\vec{r})$ and following the same line of argument mentioned therein. The analogous continuity of the mutual information (similar to \eqref{eq:continuity_region}) with respect to the total variation distance for CQ states is given by Fact~\ref{fact:continuity_I}. Finally, the cardinality bounds on $\cU_1$ and $\cU_2$ from Lemma~\ref{lem:cardinality_QC} can be used to imply $\cR^{iid}_{outer}(\epsilon_{i_k})$ converges to $\cR^{iid}$. This completes the proof.
\end{IEEEproof}


\section{Asymptotic equipartition property (AEP)}

\subsection{Asymptotic Expansion for Fixed Input}

We state the AEP for smoothed $\max$-mutual information for a classical distribution (see $(34)$ following Theorem~9 of \cite{AEP}) and a CQ state (equation~$(107)$ following Theorem~11 of \cite{AEP}).

\begin{fact}\cite[Equations~(34),(107)]{AEP}\cite[Equation~6]{Tomamichel_Hayashi }\label{fact:AEP_fixed_ip}
    Let $q_{X}p_{X|U} = p_{X,U} \in \cP$ be any joint distribution and $\tau^{EU} \in \cD(\cH^{EU})$ be any quantum state. Then for any $\epsilon \in (0,1)$, it holds that
    \begin{align}
    &\frac{1}{n} I_{\max}^\epsilon(X^n;U^n)_{p^{\otimes n}}
    = I(X;U)_{p} + \sqrt{\frac{V(X;U)}{n}} \Phi^{-1}(\epsilon) \notag\\
    & ~\qquad \qquad \qquad\qquad \qquad \qquad \quad+ O \left( \frac{\log n}{n}\right)\\
    &\Rightarrow \lim_{ n \to \infty} \frac{1}{n} I^\epsilon_{\max}(X:U)_{p_{X,U}^{\otimes n}} = I(X:U)_{p_{X,U}}, \mbox{and } \label{eq:AEP_classical} \\
    &\frac{1}{n}I_{\max}^{\epsilon}(E^n;U^n)_{{\tau}_{EU}^{\otimes n}} = 
    I(E;U)_\tau + \sqrt{\frac{V(E;U)}{n}} \Phi^{-1}(\epsilon^2) \notag\\
    &~\qquad \qquad \qquad\qquad \qquad \qquad \quad+ O\left( \frac{\log n}{n}\right)\\
   &\Rightarrow \lim_{ n \to \infty} \frac{1}{n} I^\epsilon_{\max}(E:U)_{\tau_{EU}^{\otimes n}} = I(E:U)_{\tau^{EU}}. \label{eq:AEP_quantum}
    \end{align}
    where  $V(X;U):= V(P_{X,U}||P_X \times P_U) := \bE \left[\log P_{X,U} - \log (P_X \times P_U) - \{D(P_{XU}||P_X \times P_U)\}^2 \right]$ is known as the mutual information variance between $p_{X,U}$ and $p_X \times p_U$ and $\Phi^{-1}(\cdot)$ is the inverse CDF of normal random variable with mean zero and unit variance.
\end{fact}

\subsection{Asymptotic Expansion for Universal Simulation}

We now state the AEP for smoothed $\max$-mutual information of the channel. This is obtained by taking the limit $\lim_{n \to \infty}$ in \cite[Corollary~9]{Berta_rej_sampling} resulting in the following fact: 
\begin{fact}\label{fact:AEP} \cite[Corollary~9]{Berta_rej_sampling}
For any point-to-point channel $p_{U|X}$ and $\epsilon \in (0,1)$, it holds that 
\[
    \lim\limits_{n \to  \infty} \frac{1}{n}I_{\max}^{\epsilon}(p^{\otimes n} _{U|X}) =  \max_{q_X} I(X;U)_{q_Xp_{U|X}}.
\]
\end{fact}

\section{Universal MAC simulation} \label{sec:universal}

\subsection{Proof of Lemma~\ref{lem:classical_achievability_universal}} \label{sec:achievability_universal}

We start by giving the achievability proof of Lemma~\ref{lem:classical_achievability_universal}.
\begin{IEEEproof}
Fix  $(\epsilon_1,\epsilon_2,\delta)$ satisfying the conditions of Lemma~\ref{lem:classical_achievability_universal} and choose auxiliary channels $p_{U_1|X_1}, p_{U_2|X_2}$ and the decoder $p_{Y|U_1,U_2}$ from the set $\cA^{inner}$, given in \eqref{eq:A_inner}. We need to show that for any $(R_1,R_2) \in  \cR_{\bU}^{inner}(\epsilon_1,\epsilon_2,\delta)$ (defined in \eqref{eq:universal_inner}), there exists an $(R_1,R_2,\epsilon)$ one-shot universal MAC simulation  protocol as mentioned in Definition \ref{def:task-univ}.  

     We will use the universal point-to-point channel simulation algorithm of Fact~\ref{lem:rej_sampling}-(ii) to simulate the auxiliary channels $p_{U_j|X_j}$ independently at each sender $j \in \{1,2\}$. 
    \begin{itemize}
         \item \textbf{Sender-$j$}: Let $s_{U_j}$ be a distribution with full support and choose $U_j \sim s_{U_j}$ as the shared randomness between the pair $(\cE_j,\cD)$. Using the rejection sampling algorithm mentioned in Fact~\ref{fact:accept-reject}, sender $j$ sends the appropriately chosen index of the shared randomness using $R_j$ bits to simulate the auxiliary channel $p_{U_j|X_j}$, irrespective of any particular input $q_{X_j}$.
         \item \textbf{Decoding:} After receiving the transmitted index of shared randomness from both the encoders, the decoder  first generates $(U_1,U_2)$ and applies the stochastic map $Y \sim p_{Y|U_1,U_2}$ to universally simulate $q_{Y|X_1,X_2}$.
         \item Hence, the output distribution of $U_j$ at $\cD$, denoted by $p_{U_j|X_j}^{algo}$ satisfies (from Fact~\ref{lem:rej_sampling}-(ii)): 
         \begin{align}\label{eq:sim_univ}
            \underset{x_j}{\max}\norm{p^{algo}_{U_j|X_j=x_j}-p_{U_j|X_j=x_j}}_{tvd} \leq \epsilon_j\;.
         \end{align}
         The amount of classical communication required to achieve this target distribution is given by the universal point-to-point channel simulation protocol from Fact~\ref{lem:rej_sampling}-(ii) as : 
         \begin{align*}
             R_j \geq I_{\max}^{\epsilon_j-\delta} (p_{U_j|X_j}) + \log \log \frac{1}{\delta}\;.
         \end{align*}
         \end{itemize}
         Thus, our algorithm generates the overall distribution
         \begin{equation} \label{eq:algo_output-univ}
         p^{algo}_{X_1,X_2,U_1,U_2,Y} = q_{X_1} \times q_{X_2} \times p^{algo}_{U_1|X_1} \times p^{algo}_{U_2|X_2} p_{Y|U_1,U_2}. 
         \end{equation}
         Note that the input distributions $q_{X_j}(x_j)$ above are arbitrary and play no role in the simulation criteria given by \eqref{eq:sim_protocoluniv}. \\
         To complete the proof, we finally need to show 
         \[
         \underset{x_1,x_2}{\max}\lVert p^{algo}_{Y|X_1=x_1,X_2=x_2}-q_{Y|X_1=x_1,X_2=x_2}\rVert_{tvd}\leq \epsilon_1+\epsilon_2.
         \]
         This follows by the following chain of inequalities:
         \begin{align*}
    &\underset{x_1, x_2}{\max}\lVert p^{\text{algo}}_{Y|X_1=x_1,X_2=x_2}-q_{Y|X_1=x_1,X_2=x_2}\rVert_{tvd}\\
    &\overset{(a)}{\leq}
     \underset{x_1, x_2}{\max}\lVert p^{\text{algo}}_{Y|X_1=x_1,X_2=x_2}-p_{Y|X_1=x_1,X_2=x_2}\rVert_{tvd} \\
     &\quad + \underset{x_1, x_2}{\max}\lVert p_{Y|X_1=x_1,X_2=x_2}-q_{Y|X_1=x_1,X_2=x_2}\rVert_{tvd}  \\
    & \overset{(b)}{=} \max_{x_1,x_2} \left\lVert \mathop{\Sigma}\limits_{u_1,u_2}p_{Y|U_1=u_1,U_2=u_2}\left(p^{\text{algo}}_{U_1|X_1}(u_1) \times \right.\right.\\
    &\quad \left. \left. p^{\text{algo}}_{U_2|X_2}(u_2)-p_{U_1|X_1}(u_1|x_1)p_{U_2|X_2}(u_2|x_2)\right) \right\rVert_{tvd}\\
    &= \frac{1}{2}\max_{x_1,x_2}\mathop{\Sigma} \limits_{u_1,u_2}\mathop{\Sigma}\limits_y p_{Y|U_1,U_2}(y|u_1,u_2)\left\lvert \left(p^{\text{algo}}_{U_1|X_1}(u_1) \times \right.\right.\\
    &\quad \left.\left.
    p^{\text{algo}}_{U_2|X_2}(u_2) -p_{U_1|X_1}(u_1|x_1) p_{U_2|X_2}(u_2|x_2) \right)\right\rvert\\
    & \overset{(c)}{\leq} \underset{x_1,x_2}{\max} \lVert p^{\text{algo}}_{U_1|X_1=x_1} p^{algo}_{U_2|X_2=x_2}-p^{algo}_{U_1|X_1=x_1} p_{U_2|X_2=x_2}\rVert_{tvd} \\
    &\quad +\underset{x_1,x_2}{\max}\lVert p^{\text{algo}}_{U_1|X_1=x_1} p_{U_2|X_2=x_2}-p_{U_1|X_1=x_1} p_{U_2|X_2=x_2}\rVert_{tvd}\\
    &= \underset{x_1}{\max}\lVert p^{\text{algo}}_{U_1|X_1=x_1}\rVert_{1} \underset{x_2}{\max}\lVert p^{algo}_{U_2|X_2=x_2}-p_{U_2|X_2=x_2}\rVert_{tvd} \\
     & \quad + \underset{x_2}{\max}\lVert p_{U_2|X_2=x_2}\rVert_1 \underset{x_1}{\max}\lVert p^{\text{algo}}_{U_1|X_1=x_1}-p_{U_1|X_1=x_1}\rVert_{tvd}\\
    & \overset{(d)}{\leq} \epsilon_1+\epsilon_2 \leq \epsilon,
         \end{align*}
         where $(a)$ and $(c)$ follow from the triangle inequality and the fact that maximum value of the sum of two non-negative functions is less than or equal to the sum of their individual maximum values; $(b)$ follows from the definition of distribution induced by the code in \eqref{eq:algo_output-univ} and $(d)$ follows from  \eqref{eq:sim_univ}.\\
     We have thus shown that there exists an $(R_1,R_2,\epsilon)$ code for universally simulating $q_{Y|X_1,X_2}$ which implies $\cR_{\bU}^{inner}(\epsilon_1,\epsilon_2, \delta) \subseteq \cR_{\bU}(\epsilon)$.
\end{IEEEproof}

\subsection{Proof of Lemma~\ref{lem:classical_converse_universal}}\label{sec:converse_universal}

We now prove Lemma~\ref{lem:classical_converse_universal}, the converse for the one-shot universal MAC simulation task.
\begin{IEEEproof} 
Let $(\epsilon_1,\epsilon_2)$ and $\epsilon$ satisfy the conditions of the lemma. We need to show that any $(R_1,R_2,\epsilon)$ MAC simulation protocol according to Definition \ref{def:task-univ} has $(R_1,R_2) \in \cR_{\bU}^{outer}(\epsilon_1,\epsilon_2,\epsilon)$ (defined in \eqref{eq:universal_outer}), which implies $\cR_{\bU}(\epsilon) \subseteq \cR_{\bU}^{outer}(\epsilon_1,\epsilon_2,\epsilon)$.

       Consider a MAC simulation protocol with any arbitrary input $q_{X_1} \times q_{X_2}$ and the overall distribution as
       \[
       \mathop{\bigotimes} \limits_{j=1}^2 \left( q_{X_j}q_{S_j}p'_{M_j|S_j,X_j} \right)p'_{Y|\vec{M},\vec{S}}.
       \]
       The encoders are specified by ${p'}_{M_1|X_1,S_1}$ and ${p'}_{M_2|X_2,S_2}$, and the decoder is specified by ${p'}_{Y|M_1,M_2,S_1,S_2}$. Since, the code is a faithful simulation code, we have from Definition~\ref{def:task-univ}:
     \begin{align} \label{eq:tilde_constraint}
        \mathop{\max}\limits_{x_1,x_2} \norm{{p'}_{Y|X_1=x_1,X_2=x_2} - q_{Y|X_1=x_1,X_2=x_2}}_{tvd} \leq \epsilon. 
     \end{align} 
     We now use a similar intuition as in Lemma~\ref{lem:classical_converse} to identify the auxiliary random variables for simulating $q_{Y|X_1,X_2}$.
  Define the following set for every vector $\vec{x}=(x_1,x_2)$
\begin{align} \label{eq:goodset_classical-uni}
    \bar{\cC}_{\vec{x}}:=\left\{(\vec{m},\vec{s}) : p'_{M_j|S_j,X_j}(m_j|s_j,x_j) \geq \frac{\epsilon_j}{|\cM_j|}, j=1,2 \right\}.
\end{align}
We henceforth denote the projection of $\cC_{\vec{x}}$ onto $(M_j,S_j,X_j)$ (or the $j^{th}$ user) as $\cC_{x_j}$ and we make the similar identification for their respective complements.\\ 
    Note that by union bound, we have
    \begin{align}
        \bP_{p'}(\cC_{\vec{x}}) &\leq  \sum_{j=1}^2\bP\left(\left\{p'_{M_j|S_j,X_j}(m_j|s_j,x_j) \leq \frac{\epsilon_j}{|\cM_j|}\right\}\right) \notag\\
        &\leq  \epsilon_1+\epsilon_2, \notag
    \end{align}
    where we have used:
    \begin{align} \label{eq:P(c_j)-univ}
     &\bP_{p'}\left(\left\{(m_j,s_j):p'_{M_j|S_j,X_j}(m_j|s_j,x_j) \leq \frac{\epsilon}{|\cM_j|}\right\}\right)\\
     &=\sum_{\substack{(m_j,s_j):\\p'_{M_j|S_j,X_j}(m_j|s_j,x_j)} \leq \frac{\epsilon_j}{|\cM_j|} } \hspace{-10 pt }p'_{S_j}(s_j)p'_{M_j|S_j,X_j}(m_j|s_j,x_j)  \notag \\
     &\Rightarrow \bP_{p'}(\cC_{x_j}) \leq \sum_{(m_j,s_j)}\frac{\epsilon_j}{|\cM_j|}q_{S_j} \leq \epsilon_j.
    \end{align}
    Hence, $\bP_{p'}(\bar \cC_{\vec{x}}) \geq 1-\epsilon_1 - \epsilon_2$, for all $\vec{x}$.\\     
    Consider the distribution defined as follows:
    \begin{align} \label{eq:p-univ'}
         &p_{M_j,S_j|X_j}(m_j,s_j|x_j):= \notag\\
         &\frac{p'_{S_j}(s_j)p'_{M_j|S_j,X_j}(m_j|s_j,x_j)}{\bP_{p'}(\bar \cC_{x_j})} \one_{\{(m_j,s_j) \in \bar \cC_{x_j}\}}
    \end{align}
    We have thus identified the auxiliary random variable $\{U_j\}_{j=1}^2$ for each $x_j$ as:    
    \begin{align*}
        &U_j:=(M_j,S_j) \one_{\bar \cC_{x_j}} \sim p_{U_j|X_j}(u_j|x_j)\\
        &p_{U_j|X_j}(u_j|x_j):=\frac{p'_{S_j}(s_j)p'_{M_j|S_j,X_j}(m_j|s_j,x_j) \one_{(m_j,s_j) \in \bar \cC_{x_j}}}{\bP_{p'}(\bar \cC_{x_j})} \;.
    \end{align*}
    Using this we identify the conditional distribution $p_{\vec{U},Y|\vec{X}}$ as:
    \begin{align} \label{eq:p-cond'}
         p_{\vec{U},Y|\vec{X}}(\vec{u},y|\vec{x})&:=
         \mathop{\bigotimes} \limits_{j=1}^2 \left[\frac{p'_{S_j}(s_j)p'_{M_j|S_j,X_j}(m_j|s_j,x_j)}{\bP_{p'}(\bar \cC_{x_j})}\right] \notag\\
         & \qquad \times  p'_{Y|\vec{S},\vec{M}}(y|\vec{s},\vec{m}) \one_{\{m_j,s_j \in \bar \cC_{x_j}\}}
    \end{align}
     Now, we identify the complete joint distribution $p$ defined as follows:
     \begin{align} \label{eq:p-joint-univ'}
         p_{\vec{X},\vec{U},Y}(\vec{x},\vec{u},y)&:= \mathop{\bigotimes} \limits_{j=1}^2 \left[ \frac{q_{X_j}(x_j)p'_{U_j|X_j}(u_j|x_j)}{\bP_{p'}(\bar \cC_{x_j})} \right] \notag\\
         & \qquad \times p'_{Y|U_1,U_2}(y|u_1,u_2) \one_{\{\vec{u} \in \bar{\cC}_{\vec{x}}\}}.
    \end{align}
    Note that \eqref{eq:P(c_j)-univ} also gives: 
    \begin{align}
        &\max_{\vec{x}} \norm{p_{Y|\vec{X}=\vec{x}}-p'_{Y|\vec{X}=\vec{x}}}_{tvd}  \notag\\
        &=\max_{\vec{x}}\norm{\mathop{\Sigma}\limits_{\vec{m},\vec{s}} \left( p_{\vec{M},\vec{S}|\vec{X}=\vec{x}}- p'_{\vec{M}\vec{S}|\vec{X}=\vec{x}}\right)p'_{Y|\vec{M}=\vec{m},\vec{S}=\vec{s}}}_{tvd} \notag\\
        &\leq \frac{\max\limits_{\vec{x}} \mathop{\sum}\limits_{(\vec{m},\vec{s}) \in \cC_{\vec{x}}}\left|{p(\vec{m},\vec{s}|\vec{x})-p'(\vec{m},\vec{s}|\vec{x})}\right|}{2} \notag\\
        &\quad + \frac{\max\limits_{\vec{x}} \mathop{\sum}\limits_{(\vec{m},\vec{s}) \in \bar{\cC}_{\vec{x}}}\left|{p(\vec{m},\vec{s}|\vec{x})-p'(\vec{m},\vec{s}|\vec{x})}\right|}{2} \notag \\
        &= \frac 12 \max_{\vec{x}} \bP_{p'}(\cC_{\vec{x}})+ \frac 12 \max_{\vec{x}}\bP_{p'}(\bar{\cC}_{x_1})\bP_{p'}(\bar{\cC}_{x_2}) \times \\
        & \quad \left( \frac{1}{\bP_{{p'}}(\bar{\cC}_{x_1}) \bP_{{p'}}(\bar{\cC}_{x_2})}-1\right) \notag \\
        & \leq \epsilon_1+\epsilon_2\; .\label{eq:tvddistbnd}
    \end{align}
Finally we define the following distribution on the random variable $U_j (=(M_j,S_j))$ that will be used to evaluate the quantity $I_{\max}^\epsilon(p_{U_j|X_j})$ for $j \in \{1,2\}$:
    \begin{align}
        r_{U_j}(u_j):= q_{S_j}(s_j)\frac{1}{|\cM_j|} 
    \end{align}
   These identifications leads to the following implications on the rate of the protocol:
    \begin{align}\label{eq:rate}
        2^{I_{\max}^{\epsilon_j}(p_{U_j|X_j})} & \overset{(a)}{\leq} 2^{D_{\max}(p'_{X_j,U_j}||p'_{X_j} \times r_{U_j})} \notag\\
         &=\max_{x_j} \max_{u_j}\frac{p'_{X_j,U_j}(x_j,u_j)}{p'_{X_j}(x_j)r_{U_j}(u_j)} \notag\\
        &\overset{(b)}{=} \max_{x_j} \max_{(m_j, s_j)} \frac{q_{S_j}(s_j)p'_{M_j|S_jX_j}(m_j|s_j,x_j)}{q_{S_j}(s_j)/|\cM_j|} \notag\\
        &\overset{(c)}{\leq } |\cM_j|,
    \end{align}
    where $(a)$ follows from the definition of channel smoothed $I_{\max}$ in Definition~\ref{def:smooth_Imax} and  observing that distribution $p_{U_j|X_j=x_j}=p_{M_j,S_j|X_j=x_j} \in \cB^{\epsilon_j}(p'_{M_j,S_j|X_j=x_j})$ because:
    \begin{align*}
        &\mathop{\max}\limits_{x_j} \norm{p_{U_j|X_j=x_j}-p'_{U_j|X_j=x_j}}_{tvd}\\
        &= \mathop{\max}\limits_{x_j} \frac{1}{2} \sum_{m_j,s_j} \left| p'_{M_j,S_j|X_j}(m_j,s_j|x_j)-p_{M_j,S_j|X_j}(m_j,s_j|x_j)\right|\\
        &=\mathop{\max}\limits_{x_j} \frac{1}{2} \left[ \sum_{(m_j,s_j) \in \bar \cC_{x_j}} p'_{M_j,S_j|X_j}(m_j,s_j|x_j) \left(\frac{1}{\bP_{p'}(\bar \cC_{x_j})}-1 \right)+ \right.\\
        & \quad \left. \sum_{(m_j,s_j) \in \cC_{x_j}} p'_{M_j,S_j|X_j}(m_j,s_j|x_j) \right] \\
        &=\bP_{p'}(\cC_{x_j}) \leq \epsilon_j \;(\mbox{ from \eqref{eq:P(c_j)-univ}}),
    \end{align*}
    $(b)$ follows from the identification of $U_j=(M_j,S_j)$ for all $p'$, the definition of $r_{U_j}$ and the Bayes rule and\\
    $(c)$ follows since $p'_{M_j,S_j|X_j}(m_j,s_j|x_j) \leq 1$.\\
    We thus have from \eqref{eq:rate}, the rate of the code is lower bounded by:
    \[
        R_j=\log|\cM_j| \geq I^{\epsilon_j}_{\max}(p_{U_j|X_j}) \; \text{ for }j\in\{1,2\}.
    \]
From \eqref{eq:tvddistbnd} we have that $p_{Y|\vec{X}=\vec{x}} \in \cB^{\epsilon_1+\epsilon_2}(p'_{Y|\vec{X}=\vec{x}})$.
This along with the simulation constraint of \eqref{eq:tilde_constraint} yields by the triangle inequality:
    \begin{align}
        \mathop{\max}\limits_{\vec{x}} \norm{p_{Y|\vec{X}=\vec{x}}-q_{Y|\vec{X}=\vec{x}}}_{tvd} \leq (\epsilon_1+\epsilon_2)+\epsilon \leq 2\epsilon.
    \end{align}
       We have thus identified the auxiliary channels $p_{U_1|X_1},p_{U_2|X_2}$ and a distribution $p_{Y|U_1,U_2}:=p'_{Y|U_1,U_2}$ (from the decoder of the simulation protocol), such that:
    \[
    (p_{U_1|X_1}, p_{U_2|X_2},p_{Y|U_1,U_2}) \in \cA_{\epsilon}^{outer},
    \]
   where the set $\cA_{\epsilon}^{outer}$ is given in \eqref{eq:A_univ} and the rate of any $(R_1,R_2,\epsilon)$-simulation code is bounded below by:
    \begin{align}
        R_j &\geq I^{\epsilon_j}_{\max}(p_{U_j|X_j}). \notag 
    \end{align}  
To complete the proof, we require a bound on the cardinalities of $\cU_1,\; \cU_2$, which is same as given in Lemma \ref{lem:cardinality-fixed_ip} and proven in Appendix \ref{sec:cardinality_proof}.
    \end{IEEEproof}


\subsection{Asymptotic Expansion}\label{sec:universal-iid}

We will extend the single-letter universal protocol to the asymptotic iid case and show that the rate region can still be single-letterized. We split the proof into two parts showing that the asymptotic inner and outer bounds converge to $\cR_{\bU}^{iid}$. Throughout the proof, we choose the parameters $\epsilon_j >0$ and $\delta \in (0, \min\{\epsilon_1,\epsilon_2, 1-\epsilon_1,1-\epsilon_2\})$.
    \begin{enumerate}
        \item \textbf{Asymptotic iid Universal Inner Bound: }
        We extend the universal one-shot inner bound to obtain the optimal universal asymptotic iid rate region. Let $(R_1,R_2) \in  \cR_{\bU}^{iid}$ (defined in \eqref{eq:universal_iid}) be such that for any $\eta>0$, 
        \begin{align} \label{eq:limit_univ}
        R_j \geq \mathop{\max}\limits_{q_{X_j}}I(X_j;U_j)_{q_{X_j}p_{U_j|X_j}} +\eta,
        \end{align}
        for some $p_{U_1,U_2,Y|X}=p_{U|X_1}p_{U_2|X_2}p_{Y|U_1,U_2}$ satisfying $ p(y|x_1,x_2) = q(y|x_1,x_2)$, for all $x_1,x_2$. Consider 
        \begin{align}
            &p_{U_1^{(n)}|X_1^n}=p_{U_1|X_1}^{\otimes n}\;, p_{U_2^{(n)}|X_2^n}= p_{U_2|X_2}^{\otimes n} \notag\\
            &\mbox{ and } p_{Y^n|U_1^{(n)},U_2^{(n)}}=p_{Y|U_1,U_2}^{\otimes n}. \label{eq:jointdist_univ}
        \end{align} 
        Note that the triple 
        \begin{align} \label{eq:A_inner_univ}(p_{U_1^{(n)}|X_1^n},p_{U_2^{(n)}|X_2^n},p_{Y^n|U_1^{(n)},U_2^{(n)}}) \in \cA_{inner}^{(n)},
        \end{align}
        where the set $\cA_{inner}^{(n)}$ is the $n^{th}$ extension of the set $\cA_{inner}$ defined in \eqref{eq:A_inner} of Theorem~\ref{thm:one-shot_universal}.
        The AEP from Fact~\ref{fact:AEP} for the channel smoothed $\max$-mutual information gives 
            \begin{align}
        &\lim_{n \to \infty} \frac{1}{n}\left[n\cdot \max_{q_{X_j}} I(X_j;U_j)_{q_{X_j} p_{U_j|X_j}} + \log \log \frac{1}{\delta} \right] \notag\\
        &= \max_{q_{X_j}} I(X_j;U_j)_{q_{X_j} p_{U_j|X_j}}. \notag 
            \end{align}  
            which from \eqref{eq:limit_univ} means that 
    \begin{align} \label{eq:univ-inner_rate}
        nR_j \geq I_{\max}^{\epsilon_j-\delta}(p^{\otimes n}_{U_j|X_j}) + \log \log  \frac{1}{\delta}, 
    \end{align}
    for all sufficiently large $n$ (depending on $\eta$). 
    Hence, \eqref{eq:A_inner_univ} and \eqref{eq:univ-inner_rate} together, in the asymptotic limit $n \to \infty$ and $\epsilon_j \to 0$ imply 
$\cR_{\bU}^{iid} \subseteq \cR_{\bU,inner}^{(n)}(\epsilon_1,\epsilon_2)$, where
\begin{align}
 &\cR_{\bU,inner}^{(n)}(\epsilon_1,\epsilon_2) \notag\\
 &=\left\{(R_1,R_2):  nR_j \geq I_{\max}^{\epsilon_j-\delta} (p_{U_j^{(n)}|X_j^n}) +  \log \log \frac{1}{\delta}\right\}.  
\end{align}                
        \item \textbf{Asymptotic iid Universal Outer Bound: }
        In order to prove that the asymptotic extension of the universal one-shot outer bound is the region $\cR^{iid}_{\bU}$, for any $\epsilon \in (0,1)$ we first define the following $\epsilon$-approximate universal iid region as follows:
\begin{align} \label{eq:delta_iid_converse-univ}
    &\cR_{\bU}^{iid}(\epsilon):= \Biggl\{ (R_1,R_2): R_j \geq \max_{q_{X_j}}I(X_j;U_j)_{q_{X_j}p_{U_j|X_j}}, \notag\\
    & \quad \left. \forall \; p_{X_1,X_2,U_1,U_2,Y}=q_{X_1}q_{X_2}p_{U_1|X_1}p_{U_2|X_2}p_{Y|U_1,U_2} \mbox{ s.t.} \right. \notag\\
    &\quad \mathop{\max}\limits_{x_1,x_2}\left\lVert \mathop{\sum}\limits_{u_1,u_2} p(u_1|x_1)p(u_2|x_2)p_{Y|U_1=u_1,U_2=u_2} \right. \notag\\
    & \quad \left. -q_{Y|X_1=x_{1},X_2=x_{2}}\right\rVert_{tvd} \leq 2\epsilon \Biggr\}.
\end{align}
We will now use the converse of Theorem~\ref{thm:one-shot_universal} to an $n$-letter universal simulation block code with the distribution $ p_{U_1^{(n)}|X_1^n}p_{U_2^{(n)}|X_2^n}p_{Y^n|U_1^{(n)}, U_2^{(n)}}$. For any $\epsilon \in (0,1)$ and $\epsilon_1, \epsilon_2 >0 $ such that $\max\{\epsilon_1,\epsilon_2 \} \leq \epsilon/2$, let $\cR_{\bU,outer}^{(n)}(\epsilon_1,\epsilon_2, \epsilon)$ be the $n$-fold extension of the region $\cR_{\bU}^{outer}(\epsilon_1,\epsilon_2, \epsilon)$ with respect to the input and auxiliary random variables   $(X_j^n,U_j^{(n)}) \sim q^{\otimes n}_{X_j} p_{U_j^{(n)}|X_j^n}$. Suppose $(nR_1,nR_2) \in \cR_{\bU,outer}^{(n)}(\epsilon_1,\epsilon_2, \epsilon)$ with $p_{X_1^n,U_1^{(n)},X_2^n,U_2^{(n)},Y^n}:=q_{X_1}^{\otimes n} q_{X_2}^{\otimes n}p_{U_1^{(n)}|X_1^n}p_{U_2^{(n)}|X_2^n}p_{Y^n|U_1^{(n)},U_2^{(n)}}$ being the distribution induced by any $n$-fold simulation code satisfying
\begin{align}
    &\mathop{\max} \limits_{x_1^n,x_2^n}\bigg\lVert \mathop{\sum}\limits_{u_1^{(n)},u_2^{(n)}} p(u_1^{(n)}|x_1^n)p(u_2^{(n)}|x_2^n)  \notag\\
    &~~ p_{Y^n|U_1^{(n)}=u_1^{(n)},U_2^{(n)}=u_2^{(n)}}- q^{\otimes n}_{Y|X_1=x_1,X_2=x_{2}}\bigg\rVert_{tvd} \leq 2\epsilon.
\end{align}
Then, we have
     \begin{align*}
         &nR_j \\
         &\geq    I_{\max}^{\epsilon_j} (p_{U_j^{(n)}|X_j^n}) \notag \\
            & \overset{(a)}{=} \max_{q_{X_j}^{\otimes n}}\min_{\substack{{\bar p_{U_j^{(n)}|X_j^n}} \\ \in \cB^{\epsilon_j}(p_{U_j^{(n)}|X_j^n})}} \min_{r_{U_j^{(n)}}}D_{\max} \left({q_{X_j}^{\otimes n}\bar p_{U_j^{(n)}|X_j^n}}\bigg\lVert \right. \notag\\
            &\qquad \qquad \qquad \qquad \qquad \qquad \qquad \qquad \left. q^{\otimes n}_{X_j} \times r_{U_j^{(n)}}\right)\\
            & \geq \max_{q_{X_j}^{\otimes n}}\min_{\substack{{\bar p_{U_j^{(n)}|X_j^n}} \\ \in \cB^{\epsilon_j}(p_{U_j^{(n)}|X_j^n})}} \min_{r_{U_j^{(n)}}}D \left({q_{X_j}^{\otimes n}\bar p_{U_j^{(n)}|X_j^n}}\bigg\lVert\right. \notag\\
            & \qquad \qquad \qquad \qquad \qquad \qquad \qquad \qquad \left. q^{\otimes n}_{X_j} \times r_{U_j^{(n)}}\right)\\
            & \overset{(b)}{=} \max_{q_{X_j}^{\otimes n}}\min_{\substack{{\bar p_{U_j^{(n)}|X_j^n}} \\ \in \cB^{\epsilon_j}(p_{U_j^{(n)}|X_j^n})}} \min_{r_{U_j^{(n)}}}D \left({q_{X_j}^{\otimes n}\bar p_{U_j^{(n)}|X_j^n}}\bigg\lVert \right. \notag\\
            & \qquad \qquad \qquad \left. q^{\otimes n}_{X_j} \times \left\{\sum_{x_j^n}q_{X_j}^{\otimes n}(x_j) \bar p_{U_j^{(n)}|X_j^n=x_j^n}\right\}\right)\\
            &=\max_{q_{X_j}^{\otimes n}}\min_{\bar{p} \in \cB^{\epsilon_j}(p_{U_j^{(n)}|X_j^n})} I(X_j^n;U_j^{(n)})_{\bar p}\\
            & \overset{(c)}{\geq} \max_{q_{X_j}^{\otimes n}} \left[ I(X_j^n;U_j^{(n)})_{q_{X_j^{\otimes n}}p_{U_j^{(n)}|X_j^n}}- \right.\\
            & \qquad \left. \qquad 2h_2\left(\frac{\epsilon_j}{1+\epsilon_j}\right)-2n\epsilon_j \log (|\cX_j|)\right]\\
            & \overset{(d)}{\geq} \max_{q_{X_j}} \left[ n I(X_{j};U_{j})_{q_{X_j}p_{U_{j}|X_{j}}}- \right. \notag\\
            & \qquad \qquad \left.2h_2\left(\frac{\epsilon_j}{1+\epsilon_j}\right)-2n\epsilon_j \log (|\cX_j|)\right]
            \end{align*}
        where $(a)$ follows since the smoothed channel $\max$-information is independent of the input distribution from \eqref{eq:def_channelImax} $(b)$ follows from the fact that the minimum in $D\left(q_{X_j}^{\otimes n}\bar{p}_{U_j^{(n)}|X^n_j}\bigg\lVert q_{X_j}^{\otimes n} \times r_{U_j^{(n)}}\right)$ is achieved at $r_{U_j^{(n)}}=\sum_{x_j^n} q^{\otimes n}_{X_j}(x_j)\bar{p}_{U_j^{(n)}|X_j^n}$; $(c)$ follows from Fact~\ref{fact:continuity_I} and cardinality bounds on $\cU_j$ from Lemma~\ref{lem:cardinality-fixed_ip}; and $(d)$ follows by a similar analysis to that of  Proposition~\ref{prop:single_letter} (by maximizing over inputs $q_{\vec{X}}$) for some $p_{U_1,U_2,Y|X_1,X_2}=p_{U_1|X_1}p_{U_2|X_2}p_{Y|U_1,U_2}$ satisfying $\max\limits_{\vec{x}}\norm{p_{Y|\vec{X}=\vec{x}}-q_{Y|\vec{X}=\vec{x}}}_{tvd} \leq 2\epsilon$ due to monotonicity of trace distance.

        By taking the limits $\mathop{\lim}\limits_{\epsilon_j \to 0} \mathop{\lim}\limits_{n \to \infty}$, we get
        \begin{align*}
          R_j &\geq \lim_{\epsilon_j \to 0} \lim_{n\to \infty} \max_{q_{X_j}} \bigg[ I(X_{j};U_{j})_{q_{X_j}p_{U_{j}|X_{j}}} \\
          & \qquad \frac{2h_2\left(\frac{\epsilon_j}{1+\epsilon_j}\right)}{n}-2\epsilon_j \log (|\cX_j|)\bigg]\\
            &= \max_{q_{X_j}} \lim_{\epsilon_j \to 0} \lim_{n\to \infty}  \bigg[ I(X_{j};U_{j})_{q_{X_j}p_{U_{j}|X_{j}}} \\
            &\qquad  -\frac{2h_2\left(\frac{\epsilon_j}{1+\epsilon_j}\right)}{n}-2\epsilon_j \log (|\cX_j|)\bigg]\\
            &=\max_{q_{X_j}} I(X_{j};U_{j})_{q_{X_j}p_{U_{j}|X_{j}}}.
        \end{align*}
         Hence, we have shown that in the asymptotic iid limit:
    \begin{equation} \label{eq:R_outsubset_R-univ}
    \lim_{\epsilon_1,\epsilon_2 \to 0}\lim_{n \to \infty}\cR_{\bU,outer}^{(n)}(\epsilon_1,\epsilon_2,\epsilon) \subseteq \cR_{\bU}^{iid}(\epsilon).
    \end{equation} 
     Since, in our setting we have bounded cardinalities of the auxiliary random variables, we can directly apply \cite[Lemma~6]{Gohari_continuity} in our case (see Fact~\ref{fact:continuity-universal} for exact statement) to obtain 
     \begin{align*}
       \mathop{\lim}\limits_{\epsilon \to 0} \mathop{\lim}\limits_{n \to \infty} \cR_{\bU, outer}^{(n)}(\epsilon_1,\epsilon_2,\epsilon)&:= \cR_{\bU,outer}^{iid}\\
       &\subseteq \cR_{\bU}^{iid}=\mathop{\lim}\limits_{\epsilon \to 0} \cR_{\bU}^{iid}(\epsilon).
     \end{align*}
\end{enumerate}
     Thus we have shown that in the asymptotic iid limit:
     \begin{align*}
    &  \cR_{\bU}^{outer} \subseteq \cR_{\bU}^{iid} \subseteq  \lim_{n \to \infty}\cR_{\bU, inner}^{(n)} \subseteq \cR^{outer}_{\bU},\\
     \Rightarrow  & \cR_{\bU}^{inner}:=\lim_{\epsilon_1,\epsilon_2 \to 0}\lim_{n \to \infty}\cR_{\bU,inner}^{(n)}(\epsilon_1,\epsilon_2) = \cR_{\bU}^{iid} = \cR_{\bU}^{ outer}\;.
     \end{align*}
The following fact states that the region $\lim_{\epsilon \to 0} \cR^{iid}_{\bU}(\epsilon)=\cR_{\bU}^{iid} $.
\begin{fact}\label{fact:continuity-universal}
Consider the asymptotic iid setting for universally simulating a MAC $q_{Y|X_1,X_2}$ (see  Definition~\ref{def:task-univ}) for any $\epsilon \in (0,1)$. 
Let an outer bound on the cost region for this task be
\begin{align*}
    &\cR^{iid}_{\bU}(\epsilon):= \Bigg\{ (R_1,R_2): R_j \geq \max_{p_{X_j}}I(X_j;U_j)_{p_{X_j}p_{U_j|X_j}}, \\
    & \qquad \qquad\qquad \mbox{ and } \mathop{\max}\limits_{\vec{x}}\norm{p_{Y|\vec{X}=\vec{x}}-q_{Y|\vec{X}=\vec{x}}}_{tvd} \leq \epsilon \Bigg\},
\end{align*}
with $|\cU_j| \leq |\cX_1||\cX_2||\cY|$ for $j \in \{1,2\}$ then:
\[
    \cR^{iid}_{\bU} = \bigcap_{\epsilon>0}\cR^{iid}_{\bU}(\epsilon),
\]
where 
\begin{align*}
    &\cR_{\bU}^{iid}:= \Bigg\{(R_1,R_2): R_j \geq \max_{p_{X_j}}I(X_j;U_j)_{p_{X_j}p_{U_j|X_j}},\\
    & \qquad \qquad \qquad \text{ and } p_{Y|\vec{X}=\vec{x}}=q_{Y|\vec{X}=\vec{x}} \Bigg\}. 
\end{align*}
\end{fact}
The proof of the above fact is almost the same as that of Fact~\ref{fact:continuity} using the continuity of the function\\ $\max_{p_{X_j}}I(U_j;X_j)_{p_{X_j}p_{U_j|X_j}}$ with respect to the argument $p_{U_j|X_j}$ (see \cite[Lemma~12 and Corollary~13]{capacity_continuity} by considering $p_{U_j|X_j}$ as channels).

\section{One-shot Measurement Compression}\label{sec:QCMAC_meas_comp}

\subsection{Task and Achievability}

We first recall one of the most fundamental theorems of quantum information theory that is used to prove the existence of an isometry which can serve as either encoder or a decoder. This is the following widely used Uhlmann's theorem. 
\begin{fact}\mbox{Uhlmann's Theorem~\cite{Uhlmann}} \label{fact:Uhlmann}
   Consider  (finite dimensional) density matrices $\rho^A, \sigma^A$. Let $\ket{\psi_\rho}^{AB}$ be a purification of $\rho^A$, and let $\ket{\phi_\sigma}^{AC}$ be a purification of $\sigma^A$. Then there exists an isometry $V^{C \to B}$ such that,
   \[
        F((I^A \otimes V)\ketbra{\phi_\sigma}(I^A \otimes V^\dagger),\ketbra{\psi_\rho}) = F(\rho^A, \sigma^A),
    \]
    where $F(\rho,\sigma):=\norm{\sqrt{\rho}\sqrt{\sigma}}_1$.
\end{fact}
We now state a version of Uhlmann's theorem for CQ states, which we shall be using in our achievability proof of Lemma~\ref{lem:CS_QC_MAC}.
\begin{fact} \label{fact:CQ_Uhlmann}\cite[Claim~4]{Anshu_compression}
     Let $\varphi^{EE'U},\tau^{EIU}$ be two CQ states of the form 
     \begin{align*}
        \varphi^{EE'U}&:= \sum_{u} p(u)\ketbra{u}^U \otimes \ketbra{\varphi_u}^{EE'} \mbox{ and } \\
        \tau^{EIU}&:= \sum_{u} q(u)\ketbra{u}^U \otimes \ketbra{\tau_u}^{EI}. 
     \end{align*}
     Then there exists a set of isometries indexed by the contents of the classical register $U$, denoted by $\{V_u^{E' \to I}\}$ such that 
     \begin{align*}
        &F\left( \varphi^{EU}, \tau^{EU} \right) = F\bigg( \left\{ \sum_u \ketbra{u}^U \otimes I^E \otimes V_u \right\} \varphi^{EE'U} \\
        & \qquad \left\{ \sum_u \ketbra{u}^U \otimes I^E \otimes V_u^\dagger\right\}, \tau^{EIU} \bigg).
    \end{align*}
\end{fact}

We now define the task of one-shot measurement compression with feedback which was studied in \cite{Anshu_compression}.
\begin{definition}\cite[Definition~1]{Anshu_compression} \label{def:meas_comp}
    Let $\Lambda^{A \to X}$ be a quantum measurement described as
         \[
          \eta^{EX}:=  \Lambda^{A \to X}(\psi^{EA}):= \mathop{\Sigma}\limits_x \Tr[{\Lambda_x \rho^A}]\ketbra{x}^X \otimes  \psi_x^E \;
         \]
         and let  $\ket{\psi}^{EA}$ be  any purification of $\rho^A$. Then for any given $\epsilon \in (0,1)$, a purification $\ket{\eta}^{E E' X X'}$ of $\eta^{E X}$, with $\cX \cong \cX'$,  an $(R,\epsilon)$-quantum measurement compression protocol with feedback consists of:
         \begin{itemize}
            \item A pre-shared random state $S^{A'A''}$ between the sender and the receiver
            \item A rate limited noiseless classical channel of rate $R$;
            \item Encoder  $\cE_{\textrm{meas. comp.}}:\cH^{X'} \otimes \cH^{S}\to [1:2^R]$ and a Decoder  $\cD_{\textrm{meas. comp.}}:[1:2^R] \otimes \cH^{S} \to \cH^{X}$ such that
            \end{itemize}
            \begin{align*}
                &\norm{\cD_{\textrm{meas. comp.}} \circ \cE_{\textrm{meas. comp.}}(\eta^{E E'XX'} \otimes S^{A'A''}) - \eta^{EE'XX'}}_{tvd} \\
                &\leq \epsilon.
            \end{align*}
 Note that  the amount of classical communication required for the above task is $R$ bits. 
\end{definition}
An achievable rate for the above measurement compression task was given in \cite[Theorem 1]{Anshu_compression}. To obtain a one-shot achievable rate for simulation of CS-QC MAC with feedback, we will employ the protocol of \cite{Anshu_compression} individually for each sender. However, we characterize the required  rate in terms of our definition of $I_{\max}^{\epsilon}$ (see \eqref{eq:quantum_Imax}) by modifying the analysis slightly. The exact statement of the convex split lemma for CQ states with its characterization in terms of $I_{\max}^{\epsilon}$ according to \eqref{eq:quantum_Imax} is given in Fact~\ref{fact:convex_split}.

 We further recall that the task of quantum measurement compression with feedback is very similar to that of quantum state splitting \cite{Berta_splitting}. Both these tasks, in turn use convex split lemma of Fact~\ref{fact:convex_split} to quantify the rate. We also use Fact~\ref{fact:convex_split} to derive our CS-QC MAC with feedback simulation rate region and hence the minute distinction between measurement compression with feedback and quantum state splitting is not useful for our purpose. In fact we essentially prove the achievability for the task of Definition~\ref{def:meas_comp} (similar to that of \cite[Theorem~1]{Anshu_compression}, but with a slightly different definition of $I_{\max}^{\epsilon}$) to derive $\cR_{inner}^{QC-fb}(\epsilon_1, \epsilon_2, \delta)$ given by \eqref{eq:fixed_inner_QC} in Definition~\ref{def:QC_in_out}. This is because we can equivalently see our encoding as the simulation of the post-measurement states $\ket{\varphi_j}^{E_jE_j'X_jX_j''U_j \tilde U_j}$ by compressing $\tilde U_j$ and allowing the receiver to reconstruct it, so that the overall joint-state shared with the receiver is still close to $\ket{\varphi_j}^{E_jE_j'X_jX_j''U_j \tilde U_j}$ and receiver holds the register $\tilde U_j$ relabeled as $\bar U_j$ in Lemma~\ref{lem:CS_QC_MAC} (see \eqref{eq:tilde_phi}).


\subsection{Proof of Lemma~\ref{lem:CS_QC_MAC}} \label{appendix:CS-QC_achievability}

We give a self-contained proof of the achievable rates mentioned in Lemma~\ref{lem:CS_QC_MAC}. This follows by the direct application of the convex split lemma to the states $\ket{\varphi_j}^{E_j E_j'X_jX_j'' U_j \tilde U_j}$ with the shared randomness as $n_j$-fold iid copy of $S_j^{U'_jU''_j}:=\sum_{u_j} \tilde p_{U_j}(u_j) \ketbra{u_j}^{U'_j} \otimes \ketbra{u_j}^{U''_j}$,
where the registers $U_{j,1}', \ldots , U_{j,n_j}'$ are held by the sender and $U''_{j,1}, \ldots, U_{j,n_j}''$ are held by the receiver. Further, the distribution $\tilde p_{U_j}$ of the shared randomness is the  optimizing distribution on $U_j$ in the definition $I_{\max}^{\epsilon_j-\delta}(E_j;U_j)_{\varphi_j}$ (see \eqref{eq:quantum_Imax}). Recall that the CQ state $\ket{\varphi_j}$ from \eqref{eq:tilde_phi} and its reduced state $\varphi_j^{E_j \tilde U_j}$ from \eqref{eq:phi-EU} are given as:
\begin{align*}
  \ket{\varphi_j}^{E_j E_j' X_jX_j''U_j \tilde U_j}&=\sum_{x_j,u_j} \sqrt{p_{X_j,U_j}(x_j,u_j)} \ket{x_jx_j}^{X_jX_j''} \otimes \\
  & \qquad \ket{u_ju_j}^{U_j \tilde U_j} \otimes \ket{\varphi_{x_j}}^{E_jE_j'} \mbox{ and } \\
  \varphi_j^{E_j \tilde U_j}&= \sum_{u_j} p_{U_j}(u_j) \ketbra{u_j}^{\tilde U_j} \otimes \tilde \varphi_{u_j}^{E_j}, \\
  \qquad \mbox{ with } \tilde \varphi_{u_j}^{E_j}&=\sum_{x_j} p_{X_j|U_j}(x_j|u_j) \varphi_{x_j}^{E_j}.
\end{align*}
We will thus use the convex split lemma from Fact~\ref{fact:convex_split} for transmitting $\tilde U_j$ to the receiver such that the correlation of $U_j$ with the (untouched) environment $E_j$ is (almost) preserved.  
We define the following convex split state and its CQ extension with the quantum system of the CQ state being pure, for each sender $j$:
\begin{align}\label{eq:split_CQ}
    & \mu_j^{E_jU'_{j,1},\ldots,U'_{j,n_j}} := \frac{1}{n_j} \sum_{i_j=1}^{n_j} {\varphi_j}^{E_jU'_{j,i_j}} \mathop{\bigotimes}\limits_{k \neq i_j} S_j^{U'_{j,k}}, \notag \\
     &\mu_j^{E_j I_j X_j X_j'' U'_{j,1} U''_{j,1} \ldots U'_{j,n_j} U''_{j,n_j}} := \notag\\
     &\sum_{u_1,\ldots,u_{n_j}} \hspace{-9 pt} \tilde p_{U_1,\ldots,U_{n_j}}(u_1,\ldots,u_{n_j})\hspace{-3 pt} \ketbra{u_1 \ldots u_{n_j}}^{U'_{j,1} \ldots U_{j,n_j'}} \notag\\
     & \qquad \otimes \ketbra{u_1 \ldots u_{n_j}}^{U''_{j,1} \ldots U''_{j,n_j}}  \otimes \notag\\
    & \qquad \left\{\left( \sum_{i_j=1}^{n_j} \frac{1}{\sqrt{n_j}} \ket{i_j}^{I_j} \ket{\varphi_{u_{i_j}}}^{E_jE_j'X_jX_j''} \right) \right. \cdot \notag\\
    &\qquad \left. \left( \sum_{i_j=1}^{n_j} \frac{1}{\sqrt{n_j}} \bra{i_j}^{I_j}\bra{\varphi_{u_{i_j}}}^{E_jE_j'X_jX_j''} \right)\right\},
\end{align}
where
\begin{align*}
&\ket{\varphi_{u_{i_j}}}^{E_jE_j'X_jX_j''}\\
&=\sum_{x_j} \sqrt{p_{X_j|U_{j}}(x_j|u_{i_j})} \ket{\phi_{x_j}}^{E_jE_j'} \ket{x_jx_j}^{X_j X_j''}.
\end{align*}
Similarly, we have the following CQ extension with the quantum system being pure, of the state $\varphi_{j}^{E_j} \otimes \left( \mathop{\otimes} \limits_{i_j=1}^{n_j} S_j^{U_{j,i_j}'} \right)$:
\begin{align} \label{eq:varphi_extension}
&\varphi_{j}^{E_jE_j' X_j X_j'' U_j \tilde U_j U_{j,1}'U_{j,1}'' \ldots U_{j,n_j}' U_{j,n_j}''}= \sum_{u_{1},\ldots,u_{n_j}} \tilde p_{\vec{U}_{j}}(\vec{u}_j)\notag\\
& \ketbra{u_1,\ldots,u_{n_j}}^{U_{j,1}' \ldots U_{j,n_j}'} \hspace{-2 pt}\otimes \hspace{-2 pt} \ket{u_1,\ldots,u_{n_j}}^{U_{j,1}'' \ldots U_{j,n_j}''} \notag \\
& \bra{u_1,\ldots,u_{n_j}}^{U_{j,1}'' \ldots U_{j,n_j}''} \otimes \ketbra{\varphi_j}^{E_jE_j'X_jX_j'' U_j \tilde U_j}.
\end{align}
Convex split lemma from Fact~\ref{fact:convex_split} implies that for $\log n_j \geq I_{\max}^{\epsilon_j - \delta}(E_j;U_j) + 2 \log \frac{1}{\delta}$, it holds that:
\begin{align} \label{eq:convex_split_closeness}
    \norm{ \varphi_{j}^{E_j}\mathop{\bigotimes}\limits_{k=1}^{n_j} S_j^{U'_{j,k}}-\mu_j^{E_jU'_{j,1},\ldots,U'_{j,n_j}}}_{tvd} \leq \epsilon_j\;.
\end{align}
We can now apply the CQ Uhlmann's theorem from Fact~\ref{fact:CQ_Uhlmann} due to the desired structure of the CQ extensions of $\mu_j$ and $\varphi_{u_j}$ in equations \eqref{eq:split_CQ} and \eqref{eq:varphi_extension} , respectively. Thus, from Fact~\ref{fact:CQ_Uhlmann} there exists a conditional isometry $V_{u_1,\ldots,u_{n_j}}^{E_j'X_jX_j'' U_j \tilde U_j \to E_j'X_jX_j'' I_j}$ such that, for  
\begin{align} \label{eq:encoded_ip}
    &\tilde \nu_j^{E_jI_jU'_{j,1}U''_{j,1} \ldots U'_{j,n_j} U''_{j,n_j}}:= \notag\\
    &\left(\sum_{u_1, \ldots, u_{n_j}} \ketbra{u_1 \ldots u_{n_j}}^{U_{j,1}' \ldots U_{j,n_j}'} \otimes V_{u_1, \ldots, u_{n_j}} \right) \notag\\
    & \qquad \qquad \quad \left(\ketbra{\varphi_{j}} \otimes S_j \right) \notag\\
    &= \sum_{u_1,\ldots,u_{n_j}} \tilde{p}_{U_{j,1},\ldots, U_{j,n_j}}(u_1, \ldots, u_{n_j}) \ket{u_1 \ldots u_{n_j}}^{U_{j,1}' \ldots U_{j,n_j}'} \notag\\
    &~~ \bra{u_1 \ldots u_{n_j}}^{U_{j,1}' \ldots U_{j,n_j}'} \otimes \ketbra{u_1 \ldots u_{n_j}}^{U_{j,1}'' \ldots U_{j,n_j}''} \notag \\
    & \quad \otimes 
    \ketbra{\tilde \varphi_j}^{E_jX_jX_j''I_j},
\end{align}
where
\begin{align*}
    &\ket{\tilde \varphi_j}^{E_jE_j'X_jX_j''I_j}:= \\
    &\sum_{x_j} \sqrt{\tilde{p}_{X_j|U_j}(x_j|u_{i_j})} \ket{\varphi_{x_j}}^{E_jE_j'} \ket{x_j}^{X_j} \otimes \ket{x_j}^{X_j''} \otimes \ket{i_j}^{I_j}, 
\end{align*}
$i_j$ depends on the contents of shared randomness $u_1, \ldots, u_{n_j}$ and the classical registers $U_j, \tilde U_j$ (since $U_j$ is classically correlated with $E_j$ via $X_j$, the action of the isometry above holds without any loss of generality), we get:
\begin{align} \label{eq:convex_split-main}
    &\bigg\lVert \tilde \nu_j^{E_jE_j'X_jX_j''I_jU'_{j,1}U''_{j,1} \ldots U'_{j,n} U''_{j,n_j}}- \notag\\
    & \quad \mu_j^{E_jE_j'X_jX_j'' I_jU'_{j,1} U''_{j,1} \ldots U'_{j,n_j} U''_{j,n_j}}\bigg\rVert_{tvd} \leq \epsilon_j.
\end{align}
$\bullet$ We now give a  protocol that allows the receiver to recover the state $\tilde \varphi_j^{E_jX_j \bar U_j } \widesim[2]{\epsilon_j} \varphi_j^{E_jX_j \tilde{U}_j}$ (by recovering $\bar U_j$ from $\tilde U_j$):
\begin{enumerate}
    \item $\cE_j$ has the state $\ketbra{\varphi_j}^{E_jE_j'X_jX_j''U_j \tilde{U}_j}$ as input with access to the  random state $\mathop{\bigotimes}\limits_{k=1}^{n_j} S_j^{U'_{j,k}U''_{j,k}}$ shared with the receiver. $\cE_j$ applies the conditional isometry $V_{u_1,\ldots, u_{n_j}}$, conditioned on the contents of the shared randomness register from \eqref{eq:encoded_ip}, to the input and obtains the state $\tilde \nu_j$. This creates the necessary correlation between $E_j$ and the shared random state $S_j$ and is recorded in the register $I_j$. 
    \item $\cE_j$ then measures the register $I_j$ and sends the classical message $i_j$ using $\log n_j$ bits to the receiver. Thus, the rate of the protocol is (from  Fact~\ref{fact:convex_split}):
    \begin{align}
        R_j:=\log n_j \geq I_{\max}^{\epsilon_j-\delta}(E_j;U_j)_{{\tau}}+2 \left( \log \frac{1}{\delta} \right) \;.
    \end{align}
    Note that $\log n_j$ here is the rate $R_j$ of the main CS-QC MAC simulation protocol for Lemma~\ref{lem:CS_QC_MAC}.
    \item  
    The final overall state is  $\tilde \varphi_j^{E_j X_j \bar U_{j}}$. Using \eqref{eq:convex_split-main} the encoder $\cE_j$ can pretend as if step 2 is applied on the state $\mu_j$ as input,  which would have resulted in the overall output state of the protocol  with $U_j \sim p_{U_j}$ at the receiver's end. Thus, the encoder for our CS-QC MAC simulation protocol 
    is
    \begin{align*}
        &\cE_{j,meas.comp}= \{\ketbra{i_j}\}^{I_j} \circ \\
        &\quad \sum_{u_1, \ldots, u_{n_j}} \ketbra{u_1 \ldots u_{n_j}}^{U'_{j,1} \ldots U'_{j,n_j}} \otimes \\
        & \qquad V_{u_1 \ldots u_{n_j}}^{E_j'X_jX_j''U_j \tilde U_j \to E_j'X_jX_j'' I_j},
    \end{align*}
    where $\{\ketbra{i_j}\}^{I_j}$ denotes the measurement in the computational basis $\ketbra{i_j}$ for the $j^{th}$-sender.
    The receiver picks up the shared random register $U''_{j,i_j}$ given in \eqref{eq:encoded_ip} and relabels it to $\bar U_j$, as its finally recovered state.
    \item Let the step $2$ of the encoder $\cE_j$ measuring $I_j$ and transmitting the measurement outcome $i_j$  and the step $3$ of the receiver recovering $\bar U_j$ from  $U''_{j,i_j}$ register be represented as a quantum operation $\cO_j$.\\ 
    Thus we have:
    \begin{align} \label{eq:QC_achievable}
        &\norm{\varphi_j^{E_jX_j  U_j}-\tilde \varphi_j^{E_jX_j \bar U_j}}_{tvd}  \notag\\
        &\overset{(a)}{=} \norm{ \mu_j^{E_jX_j  U_j} - \cO_j  \left( \ketbra{\varphi_j} \mathop{\bigotimes}\limits_{k=1}^{n_j} \ketbra{S_j} \right) }_{tvd} 
        \overset{(b)}{\leq} \epsilon_j\;,
    \end{align}
    where $(a)$ follows from steps $2$ and $3$ of the protocol above defining $\cO_j$; $(b)$ follows from the equation \eqref{eq:convex_split-main} and the monotonicity of the total variation distance.
    Hence, from equation~\eqref{eq:QC_achievable} we get that:
    \begin{align} \label{eq:QC_main}
        \norm{p_{X_j,U_j}-\tilde p_{X_j,U_j}}_{tvd} \leq \epsilon_j.
    \end{align}
    To finish the protocol, the decoder of the CS-QC MAC simulation achievability protocol use these states $\tilde \varphi_j^{\bar U_j}$ to generate $Y \sim p_{Y|\bar U_1,\bar U_2}$. Note that the quantum operation $\cO_j =\cD_{j,\mathrm{meas. comp.}} \circ \cE_{j,\mathrm{meas. comp.}}$ is the encoder-decoder operation for each sender, before the final decoding of $U_1,U_2$ to obtain the desired output $Y$.
\end{enumerate}
\begin{remark}
    We note that the encoding above is essentially the same as that of \cite[Theorem~1]{Anshu_compression}. The only difference is that the aforementioned reference proves the achievability for one-shot measurement compression with feedback using a different definition of the smoothed $\max$-mutual information than our Definition~\ref{def:smooth_Imax_quantum}. If we employ the achievability of \cite[Theorem~1]{Anshu_compression} with the definition of $I_{\max}^{\epsilon}$ considered therein,  we get different  slack factors. A direct comparison of the different definitions of smoothed $\max$-mutual information (including our Definition~\ref{def:smooth_Imax_quantum}) can be found in \cite{Ciganovic}. 
\end{remark}
\bibliographystyle{IEEEtran}
\bibliography{Final_IEEE_TIT}

\begin{IEEEbiographynophoto}{Aditya Nema}
the Ph.D. degree from STCS, Tata Institute of Fundamental Research in Mumbai, India 2020. Currently, he is an Assistant Professor in the department of Electrical Engineering at IIT Gandhinagar, India. His prior appointments include a postdoc position from 2023-2025 in the Institute of Quantum Information group at RWTH Aachen University, Germany succeeding a Specially Appointed Assistant Professor in the Department of Mathematical Informatics at Nagoya University, Japan, from 2021-2023. His research interests include topics in information and communication theory, mathematical analysis, theoretical computer science, and quantum
information science.
\end{IEEEbiographynophoto}

\vspace{11pt}

\begin{IEEEbiographynophoto}{Sreejith Sreekumar}
received the Ph.D. degree in Electrical engineering from
 Imperial College London in 2019. Currently, he is a CNRS researcher at the Laboratoire Des Signaux Et Syst\'emes, CentraleSup\'elec,
University Paris-Saclay. Prior to this, he was a Post-Doctoral Associate with the Institute for Quantum Information, RWTH Aachen University, Germany,
and the School of Electrical and Computer Engineering, Cornell University, Ithaca. His research interests include topics in information and communication theory, mathematical statistics, theoretical machine learning, and quantum
information science.
\end{IEEEbiographynophoto}

\vspace{11pt}

\begin{IEEEbiographynophoto}{Mario Berta (Associate Member, IEEE)}
received the Ph.D. degree in theoretical physics from ETH Zurich in 2013. He is currently a Professor in physics ¨
with the Institute for Quantum Information, RWTH Aachen University; and a Visiting Reader with the Department of Computing, Imperial College London.
Previously, he was a Senior Research Scientist with the Amazon Web Services Center for Quantum Computing and a Post-Doctoral Researcher with the Institute for Quantum Information and Matter Caltech. His research interests
include mathematical questions in quantum information science, focusing on quantum communication theory, theoretical quantum cryptography, and the
theory of quantum algorithms.
\end{IEEEbiographynophoto}

\vfill 

\end{document}